\documentclass[longauth]{aa}

\usepackage{graphicx}
\usepackage{txfonts}
\usepackage{lipsum}
\usepackage{subcaption}         
\usepackage{lscape}             
\usepackage{placeins}           

\def\approxinf{%
  \def\p{%
    \setbox0=\vbox{\hbox{$<$}}%
    \ht0=0.6ex \box0 }%
  \def\s{%
    \vbox{\hbox{$\sim$}}%
  }%
  \mathrel{\raisebox{0.7ex}{%
      \mbox{$\underset{\s}{\p}$}%
    }}%
}

\def\approxsup{%
  \def\p{%
    \setbox0=\vbox{\hbox{$>$}}%
    \ht0=0.6ex \box0 }%
  \def\s{%
    \vbox{\hbox{$\sim$}}%
  }%
  \mathrel{\raisebox{0.7ex}{%
      \mbox{$\underset{\s}{\p}$}%
    }}%
}

\usepackage{hyperref}
\hypersetup{colorlinks=true,
            linkcolor = blue,
            urlcolor  = blue,
            citecolor = blue}

\newcommand{\massb}{6.75\,$\pm$\,1.25\,M$_{\oplus}$}
\newcommand{\radiusb}{2.234\,$\pm$\,0.074\,R$_{\oplus}$}
\newcommand{\teqb}{368\,$\pm$\,9\,K}

\begin{document}

\title{A temperate sub-Neptune transiting the M4 Dwarf TOI-210 identified by NIRPS and TESS}
\subtitle{Uncovering hidden M-dwarf planetary systems in the near-infrared}

\titlerunning{The TOI-210 system}
\authorrunning{C. Cadieux, et al.}


\author{C.~Cadieux\inst{1,2} \thanks{Corresponding author: \email{charles.cadieux@unige.ch}} 
\and B.\,N.~Skinner\inst{3,4}
\and F.~Bouchy\inst{1}
\and E.\,D.~Gillis\inst{3}
\and V.~Bourrier\inst{1}
\and R.~Doyon\inst{2,5}
\and L.\,Y.~Messamah\inst{1}
\and B.\,L.~Canto~Martins\inst{6}
\and A.~L'Heureux\inst{2}
\and Y.\,S.~Messias\inst{2,6}
\and J.\,M.~Almenara\inst{7}
\and A.\,K.~Stefanov\inst{8,9}
\and X.~Bonfils\inst{7}
\and K.\,A.~Collins\inst{10}
\and H.\,M.~Relles\inst{10}
\and C.~Rodr\'iguez\inst{1}
\and P.-A.~Roy\inst{17}
\and A.~Shporer\inst{11}
\and G.~Srdoc\inst{12}
\and S.~Taylor\inst{13}
\and C.~Ziegler\inst{14}
\and R.~Allart\inst{2}
\and K.~Al~Moulla\inst{15,1}
\and \'E.~Artigau\inst{2,5}
\and F.~Baron\inst{2,5}
\and S.\,C.\,C.~Barros\inst{15,16}
\and F.~Bélanger\inst{2}
\and B.~Benneke\inst{17,2}
\and M.~Bryan\inst{18}
\and T.~Ciolak\inst{1}
\and R.~Cloutier\inst{3}
\and N.\,J.~Cook\inst{2}
\and N.\,B.~Cowan\inst{19,20}
\and E.~Cristo\inst{15}
\and L.~Dang\inst{21}
\and J.\,R.~De~Medeiros\inst{6}
\and X.~Delfosse\inst{7}
\and E.~Delgado-Mena\inst{22,15}
\and X.~Dumusque\inst{1}
\and D.~Ehrenreich\inst{1,23}
\and D.\,O.~Fontinele\inst{6}
\and T.~Forveille\inst{7}
\and J.~Gagn\'e\inst{24,2}
\and M.\,J.~Hobson\inst{7,3}
\and P.~Lamontagne\inst{25,2}
\and I.\,C.~Le\~ao\inst{6}
\and R.~de~Lima~Gomes\inst{2,6}
\and C.~Lovis\inst{1}
\and L.~Malo\inst{2,5}
\and C.~Melo\inst{26}
\and L.~Mignon\inst{7}
\and C.~Mordasini\inst{27}
\and N.~Nari\inst{28,8,9}
\and L.~Parc\inst{1}
\and R.~Rebolo\inst{8,9,29}
\and R.~Rosener\inst{2}
\and J.~Rowe\inst{30}
\and N.\,C.~Santos\inst{15,16}
\and D.~S\'egransan\inst{1}
\and A.~Srivastava\inst{2}
\and A.~Su\'arez~Mascare\~no\inst{8,9}
\and S.~Udry\inst{1}
\and D.~Valencia\inst{18}
\and G.~Wade\inst{31,32}
\and J.\,P.~Wardenier\inst{27,2}
\and D.~Weisserman\inst{3}
}

\institute{
\inst{1}Observatoire de Gen\`eve, D\'epartement d’Astronomie, Universit\'e de Gen\`eve, Chemin Pegasi 51, 1290 Versoix, Switzerland\\
\inst{2}Institut Trottier de recherche sur les exoplan\`etes, D\'epartement de Physique, Universit\'e de Montr\'eal, Montr\'eal, Qu\'ebec, Canada\\
\inst{3}Department of Physics \& Astronomy, McMaster University, 1280 Main St W, Hamilton, ON, L8S 4L8, Canada\\
\inst{4}Origins Institute, McMaster University, 1280 Main St W, Hamilton, ON, L8S 4L8, Canada\\
\inst{5}Observatoire du Mont-M\'egantic, Qu\'ebec, Canada\\
\inst{6}Departamento de F\'isica Te\'orica e Experimental, Universidade Federal do Rio Grande do Norte, Campus Universit\'ario, Natal, RN, 59072-970, Brazil\\
\inst{7}Univ. Grenoble Alpes, CNRS, IPAG, 38000 Grenoble, France\\
\inst{8}Instituto de Astrof\'isica de Canarias (IAC), Calle V\'ia L\'actea s/n, 38205 La Laguna, Tenerife, Spain\\
\inst{9}Departamento de Astrof\'isica, Universidad de La Laguna (ULL), 38206 La Laguna, Tenerife, Spain\\
\inst{10}Center for astrophysics $\vert$ Harvard \& Smithsonian, 60 Garden Street, Cambridge, MA 02138, USA\\
\inst{11}Department of Physics and Kavli Institute for Astrophysics and Space Research, Massachusetts Institute of Technology, Cambridge, MA 02139, USA\\
\inst{12}Kotizarovci Observatory, Sarsoni 90, 51216 Viskovo, Croatia\\
\inst{13}Western Colorado University, 1 Western Way, Gunnison, CO 81230, USA\\
\inst{14}Department of Physics, Engineering and Astronomy, Stephen F. Austin State University, 1936 North St, Nacogdoches, TX 75962, USA\\
\inst{15}Instituto de Astrof\'isica e Ci\^encias do Espa\c{c}o, Universidade do Porto, CAUP, Rua das Estrelas, 4150-762 Porto, Portugal\\
\inst{16}Departamento de F\'isica e Astronomia, Faculdade de Ci\^encias, Universidade do Porto, Rua do Campo Alegre, 4169-007 Porto, Portugal\\
\inst{17}Department of Earth, Planetary, and Space Sciences, University of California, Los Angeles, CA 90095, USA\\
\inst{18}Department of Physics, University of Toronto, Toronto, ON M5S 3H4, Canada\\
\inst{19}Department of Physics, McGill University, 3600 rue University, Montr\'eal, QC, H3A 2T8, Canada\\
\inst{20}Department of Earth \& Planetary Sciences, McGill University, 3450 rue University, Montr\'eal, QC, H3A 0E8, Canada\\
\inst{21}Department of Physics and Astronomy, University of Waterloo, 200 University W, Waterloo, ON N2L 3G1, Canada\\
\inst{22}Centro de Astrobiolog\'ia (CAB), CSIC-INTA, Camino Bajo del Castillo s/n, 28692 Villanueva de la Ca\~nada, Madrid, Spain\\
\inst{23}Centre Vie dans l’Univers, Facult\'e des sciences de l’Universit\'e de Gen\`eve, Quai Ernest-Ansermet 30, 1205 Geneva, Switzerland\\
\inst{24}Plan\'etarium de Montr\'eal, Espace pour la Vie, 4801 av. Pierre-de Coubertin, Montr\'eal, Qu\'ebec, Canada\\
\inst{25}Instituto de Astrof\'isica de Andaluc\'ia (IAA-CSIC), Glorieta de la Astronom\'ia s/n, 18008 Granada, Spain\\
\inst{26}European Southern Observatory (ESO), Karl-Schwarzschild-Str. 2, 85748 Garching bei München, Germany\\
\inst{27}Space Research and Planetary Sciences, Physics Institute, University of Bern, Gesellschaftsstrasse 6, 3012 Bern, Switzerland\\
\inst{28}Light Bridges S.L., Observatorio del Teide, Carretera del Observatorio s/n, 38500 G\"uimar, Tenerife, Spain\\
\inst{29}Consejo Superior de Investigaciones Cient\'ificas (CSIC), 28006 Madrid, Spain\\
\inst{30}Bishop's University, Dept of Physics and Astronomy, Johnson-104E, 2600 College Street, Sherbrooke, QC, Canada, J1M 1Z7, Canada\\
\inst{31}Department of Physics, Engineering Physics, and Astronomy, Queen’s University, 99 University Avenue, Kingston, ON K7L 3N6, Canada\\
\inst{32}Department of Physics and Space Science, Royal Military College of Canada, 13 General Crerar Cres., Kingston, ON K7P 2M3, Canada
}

  \abstract
   {Super-Earths and sub-Neptunes dominate the exoplanet population, yet their compositions, formation mechanisms, and capability of retaining atmospheres remain poorly understood. M-type stars offer a unique opportunity to investigate these questions, thanks to their favorable planet-to-star radius and mass ratios and the diversity of planetary systems they host.}
   {We aim to characterise the planetary system of the M4 dwarf TOI-210, in which TESS  during its primary mission identified one transiting exoplanet candidate at $P= 9.01$\,days.
   } 
   {We combine multi-technique observations, including transits from 40 sectors of TESS and ground-based follow-up with LCOGT and ExTrA, and radial velocities (RV) from NIRPS obtained as part of its Guaranteed Time Observations program. These data were analysed using a Bayesian framework to constrain stellar, planetary, and orbital parameters.}
   {We confirm TOI-210\,b, a temperate sub-Neptune ($T_{\rm eq} =368$\,$\pm$\,9\,K) with a mass of \massb\ and a radius of \radiusb. Its bulk density implies a volatile-rich composition, either as an extended H/He atmosphere with mass fraction of approximately 1\% (gas dwarf), a substantial water reservoir comprising at least 29\% of the mass (2$\sigma$ lower limit; water world), or a mixture of both. The NIRPS RVs show moderate evidence ($\Delta \ln \mathcal{Z} = 5.9$) for at least one non-transiting exoplanet interior to TOI-210\,b. In a targeted search, we identify two Keplerian signals at 2.15 and 3.76\,days, with minimum masses of 3.2$\pm$0.8\,M$_{\oplus}$ and 4.4$\pm$1.0\,M$_{\oplus}$, respectively, that provide a plausible explanation for the observed RV variations. If these signals are planets, the absence of transits in the TESS photometry implies mutual inclinations with TOI-210\,b above $2.3$--3.5$^{\circ}$.
   }
   {The discovery of TOI-210\,b using near-infrared spectroscopy opens a new observational window on faint M dwarfs ($V > 14$\,mag, $H > 9$\,mag) hosting low-mass planets that are largely inaccessible to optical spectrographs. TOI-210\,b closely resembles TOI-270\,d in mass, radius, and temperature, making it a prime JWST target to assess whether this temperate sub-Neptune also has a relatively clear metal-rich atmosphere.}


   \maketitle
   \nolinenumbers

\section{Introduction}

A major unresolved question in exoplanetary science concerns the nature of the most common types of planets in the Galaxy: super-Earths and sub-Neptunes. While the majority of planets with $R_{\rm p} < 1.5$\,R$_{\oplus}$ have an Earth-like composition (e.g., \citealt{Rogers_2015, Plotnykov_2020}), the lower densities observed in sub-Neptunes imply that they are not mostly made of rocks and metals; they either retained a primary atmosphere of H/He (e.g., \citealt{Owen_2017, Lee_2022}) or formed water-rich beyond the ice line and migrated inwards (e.g., \citealt{Venturini_2020, Burn_2024}). Distinguishing between these formation pathways remains challenging because of inherent degeneracy in internal structure models: planets with similar masses and radii can have radically different compositions. Breaking this degeneracy requires precise density measurements and, in some cases, atmospheric characterisation to distinguish between competing models (e.g., \citealt{Cadieux_2024b}).

M-type stars offer a unique opportunity to investigate the nature of sub-Neptunes. Thanks to their smaller radii and lower masses compared to Sun-like stars, they produce deeper transits and larger Doppler signals, allowing in-depth characterisation of their planetary systems. In recent years, planetary synthesis studies have revealed that M dwarfs may preferentially form water-rich planets \citep{Burn_2021, Venturini_2024}. This theoretical prediction is increasingly supported by both individual discoveries of potentially water-rich planets around M dwarfs (e.g., \citealt{Cadieux_2022, Piaulet_2023, Cherubim_2023, Cadieux_2025}) and population-level studies (e.g., \citealt{Luque_2022, Parc_2024, Weisserman_2026, Hobson_2026}). Intriguingly, the transition between super-Earths and sub-Neptunes around mid-to-late M dwarfs may differ from that observed around FGK and even early M stars. Instead of exhibiting a pronounced radius valley \citep{Fulton_2017,Cloutier_2020}, these systems show an unimodal radius distribution peaking at $R_{\rm p}=1.25$\,R$_{\oplus}$ \citep{Gillis_2026}, potentially reflecting different formation or atmospheric evolution pathways across stellar types.

Temperate sub-Neptunes ($T_{\rm eq} < 400$\,K) are especially valuable in this context. Because they receive relatively modest stellar irradiation, they are more likely to have preserved a significant fraction of their primordial volatile inventory and atmospheric composition, whether H/He or H$_2$O. As such, they offer a window into planet formation near or beyond the snow line. Around M dwarfs, the temperate zone corresponds to relatively short orbital periods (10--30\,days), meaning that most currently known temperate sub-Neptunes orbit such a star. Recent JWST observations have revealed diversity in their atmospheric properties, ranging from clear atmospheres with strong molecular signatures \citep{Madhusudhan_2023, Benneke_2024, Rigby_2025} to haze-dominated atmospheres with muted transmission spectra \citep{Roy_2026}. This diversity suggests that formation and evolution history may be imprinted in the transmission data, motivating a more systematic exploration of temperate sub-Neptunes across different systems.

The Near-InfraRed Planet Searcher (NIRPS) is a high-precision near-infrared (970--1900\,nm) spectrograph installed on the ESO 3.6-m telescope at La Silla Observatory \citep{Bouchy_2025} with approximately one third of its Guaranteed Time Observations (NIRPS-GTO, PI: Bouchy \& Doyon) dedicated to characterise transiting systems around M dwarfs. Thus far, the NIRPS-GTO has produced results across a broad range of planetary regimes, from ultra-short-period rocky worlds \citep{Srivastava_2026a}, super-Earths/sub-Neptunes \citep{Parc_2025, Lamontagne_2026, Weisserman_2026} to hot super-Neptunes \citep{Osborn_2026} and giant planets \citep{Frensch_2026}. A `Temperate' subprogram within the GTO aims to characterise small exoplanets ($R_{\rm p} < 3$\,R$_{\oplus}$) around M dwarfs in a low-irradiated regime ($200 < T_{\rm eq} < 400$\,K). NIRPS is well-suited for the radial-velocity follow-up of such targets, achieving sub-meter per second stability and precision over timescales of months \citep{Suarez_2025}. A first highlight of this subprogram was the confirmation of a two-planet system around TOI-406 by \cite{Lacedelli_2024}. Here, we present the discovery and characterisation of the planetary system around TOI-210.

The paper is organised as follows. Sect.~\ref{sec:observations} describes all the observations of TOI-210. Sect.~\ref{sec:stellar_char} presents a characterisation of the star TOI-210, including chemical abundance measurements derived from the NIRPS spectrum. Sect.~\ref{sec:analysis_results} presents our transit and radial velocity (RV) data analysis and results, followed by a discussion about our findings in Sect.~\ref{sec:discussion}. Finally, we summarise our results and draw our conclusions in Sect.~\ref{sec:conclusions}.

\section{Observations} \label{sec:observations}

\subsection{TESS photometry} \label{sec:tess}

TOI-210 (TIC~141608198) was observed semi-continuously by TESS \citep{Ricker_2015} in a total of 40 sectors over seven years, from Sector 1 in July 2018 to Sector 98 in January 2026. The most recent Sectors 97 and 98 were `double' sectors, the first since the start of the mission to last twice as long, for approximately 50 consecutive days. Identified early in the mission as a high-priority M dwarf target \citep{Muirhead_2018}, TOI-210 was observed at the short 2-min cadence.  A preliminary search of the Sector 1--3 data using an adaptive, wavelet-based matched filter (\citealt{Jenkins_2002, Jenkins_2010, Jenkins_2020}) first revealed transit-like signatures with depth of 0.41\%. The TESS Science Office subsequently announced the candidate planet TOI-210.01 in May 2019 \citep{Guerrero_2021}. In the latest Data Validation Reports (DVR; \citealt{Twicken_2018, Li_2019}) from sectors 1--69, the signal is modeled using a limb-darkened transit profile, yielding a signal-to-noise ratio (S/R) of 51.6, an orbital period of $9.01053\pm0.00001$\,days, a time of inferior conjunction of $1329.8256\pm0.0005$\,(BJD$- 257000$), and a preliminary planetary radius of $2.28 \pm 0.12$\,R$_\oplus$. Our transit analysis presented later in Sect.~\ref{sec:transit_analysis} yields more precise transit parameters that remain consistent with those from the DVR.

We collected from the Mikulski Archive for Space Telescopes the \texttt{PDCSAP} photometry \citep{Smith_2012,Stumpe_2012,Stumpe_2014} issued by the TESS Science Processing Operations Center (SPOC; \citealt{Jenkins_2016}). The \texttt{PDCSAP} data includes a correction for instrumental systematics and flux dilution from nearby sources within a few TESS pixels (21$\arcsec$ per pixel). The field of view is not crowded and the dilution factor ($F_{\rm c} / F_\star$) remained below 0.185 in all sectors.


We normalised all transits to a common baseline flux by removing correlated structures in the out-of-transit light curve using a Gaussian process (GP), following the methodology of \cite{Cadieux_2025}. We first isolated the in-transit photometric points using the period and phase available on ExoFOP ($P = 9.010541$\,days, $t_0 = 2459338.993$\,BJD). The GP was implemented with the \texttt{celerite2} package \citep{celerite1_2017,celerite2_2018}, using a critically damped simple harmonic oscillator (SHO) kernel. For each sector, we fit for a baseline flux $f_0$, a photometric amplitude $\sigma_{\rm phot}$, a coherence timescale $\tau$, and an excess white noise term $\sigma_{\rm jitter}$ added in quadrature to the covariance matrix. The parameters were optimised within a Bayesian framework using the \texttt{emcee} sampler \citep{Foreman-Mackey_2013}, with broad uninformed log-uniform priors: $\mathcal{LU}(0.5, 1.5)$ for $f_0$, $\mathcal{LU}(1, 10^6)$ ppm for $\sigma_{\rm phot}$, $\mathcal{LU}(0.2, 20)$ days for $\tau$, and $\mathcal{LU}(1, 10^6)$ ppm for $\sigma_{\rm jitter}$. The sampler used 100 walkers, ran for 3000 steps, and discarded the first 500 steps as burn-in. The TESS light curves for all sectors, along with the resulting GP models, are shown in Fig.~\ref{fig:tess_gp}. After dividing the \texttt{PDCSAP} fluxes by the mean GP prediction, we retained the in-transit photometric points (highlighted in red in Fig.~\ref{fig:tess_gp}) for the transit analysis.

\subsection{LCOGT transit photometry} \label{sec:lco}

We observed two full transits of TOI-210.01, on UTC 2018 December 30 in Sloan $i'$ band and UTC 2019 January 17 in Sloan $g'$ band, from the Las Cumbres Observatory Global Telescope (LCOGT; \citealp{Brown:2013}) 1.0\,m network node at Siding Spring Observatory near Coonabarabran, Australia. The 1-m telescopes are equipped with $4096\times4096$ SINISTRO cameras having an image scale of $0\farcs389$ per pixel, resulting in a $26\arcmin\times26\arcmin$ field of view. The images were calibrated by the standard LCOGT {\tt BANZAI} pipeline \citep{McCully:2018} and differential photometric data were extracted using {\tt AstroImageJ} \citep{Collins:2017}. We used photometric apertures having $3\arcsec$ radii that excluded most of the flux from the nearest known neighbor \textit{Gaia} DR3 4650160713423321728 ($\Delta G = 3.71$\,mag), which is $4\farcs4$ southwest of TOI-210. We detected the transit in the target star photometric aperture in both bands, which confirms that the TESS detected event is indeed occurring in TOI-210, and that strong transit depth chromaticity is ruled out. The LCOGT transits are displayed in Fig.~\ref{fig:transits+ground_based_photometry}.

\subsection{ExTrA transit photometry} \label{sec:ExTrA}

We observed nine transits of TOI-210.01 with ExTrA \citep{Bonfils_2015}, a nIR (0.85--1.55\,$\mu$m) multi-object spectrophotometer installed at La Silla Observatory, Chile. The ExTrA instrument can be fed by up to three 60-cm telescopes, each equipped with five field-unit fibers: one to collect light from the science target and four from nearby comparison stars. The observations were acquired with an exposure time of 60\,s using all three telescopes on UT2020-11-23, 2022-12-20, 2023-02-03, and with two of them on UT2022-12-11, 2022-12-29,  2023-01-07, 2023-01-16, 2023-02-21, 2025-02-20. The star TOI-210 was observed with an 8$\arcsec$ fiber using the low-resolution mode ($R \sim 20$) of ExTrA. Comparison stars were selected to have similar $J$ magnitude and spectral type to TOI-210. The raw data were processed with a custom pipeline outlined in \cite{Cointepas_2021}, producing in total 21 light curves (3 transits $\times$ 3 telescopes + 6 transits $\times$ 2 telescopes). The raw light curves were subsequently corrected for correlated noise using a Matérn-3/2 GP following the methodology of \cite{Cointepas_2021}. The resulting products are detrended broadband light curves (0.85--1.55\,$\mu$m).

\begin{figure*}
    \centering
    \minipage{0.295\textwidth}
    \includegraphics[width=1\linewidth]{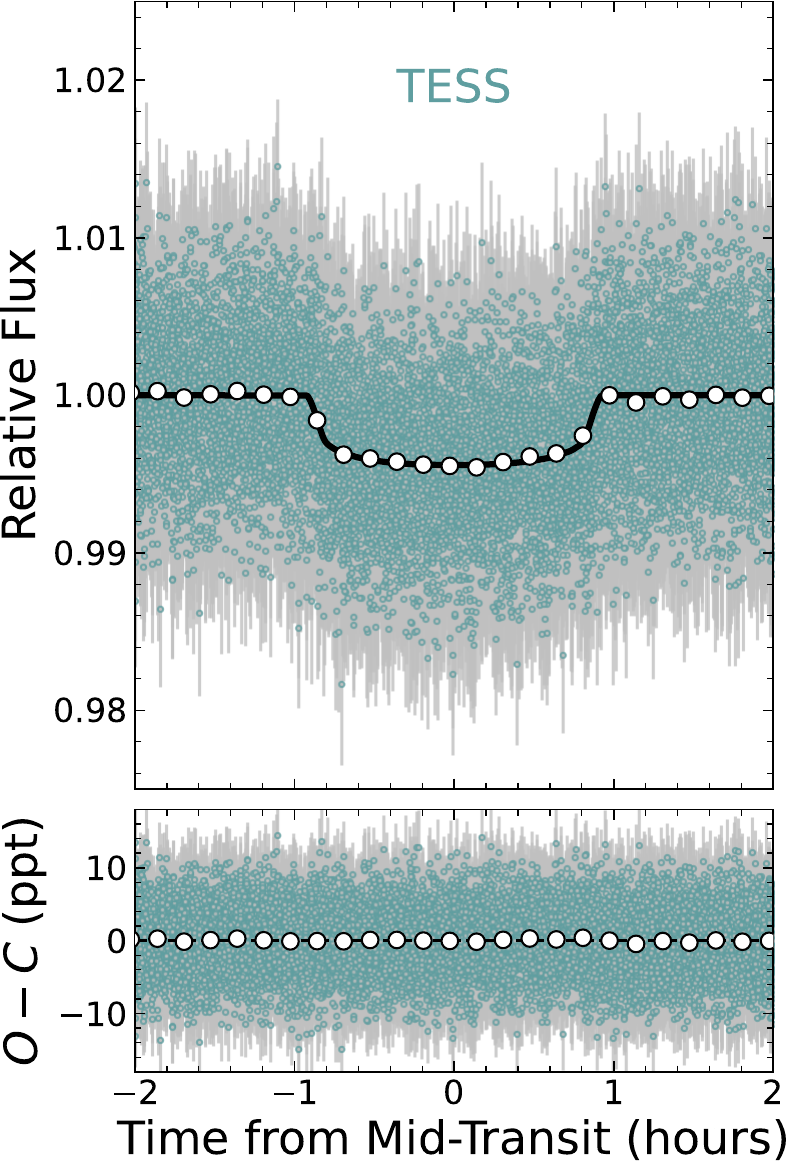}
    \endminipage
    \hfill
    \minipage{0.295\textwidth}
    \includegraphics[width=1\linewidth]{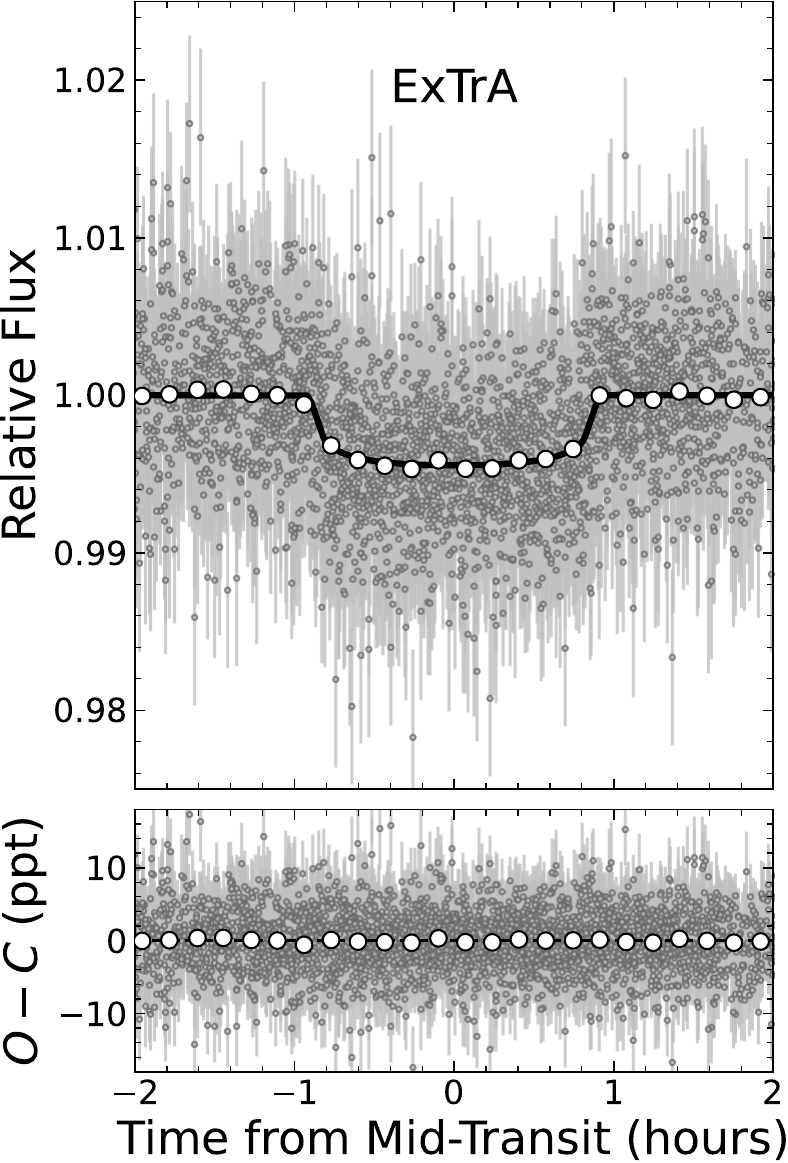}
    \endminipage
    \hfill
    \minipage{0.38\textwidth}
    \includegraphics[width=1\linewidth]{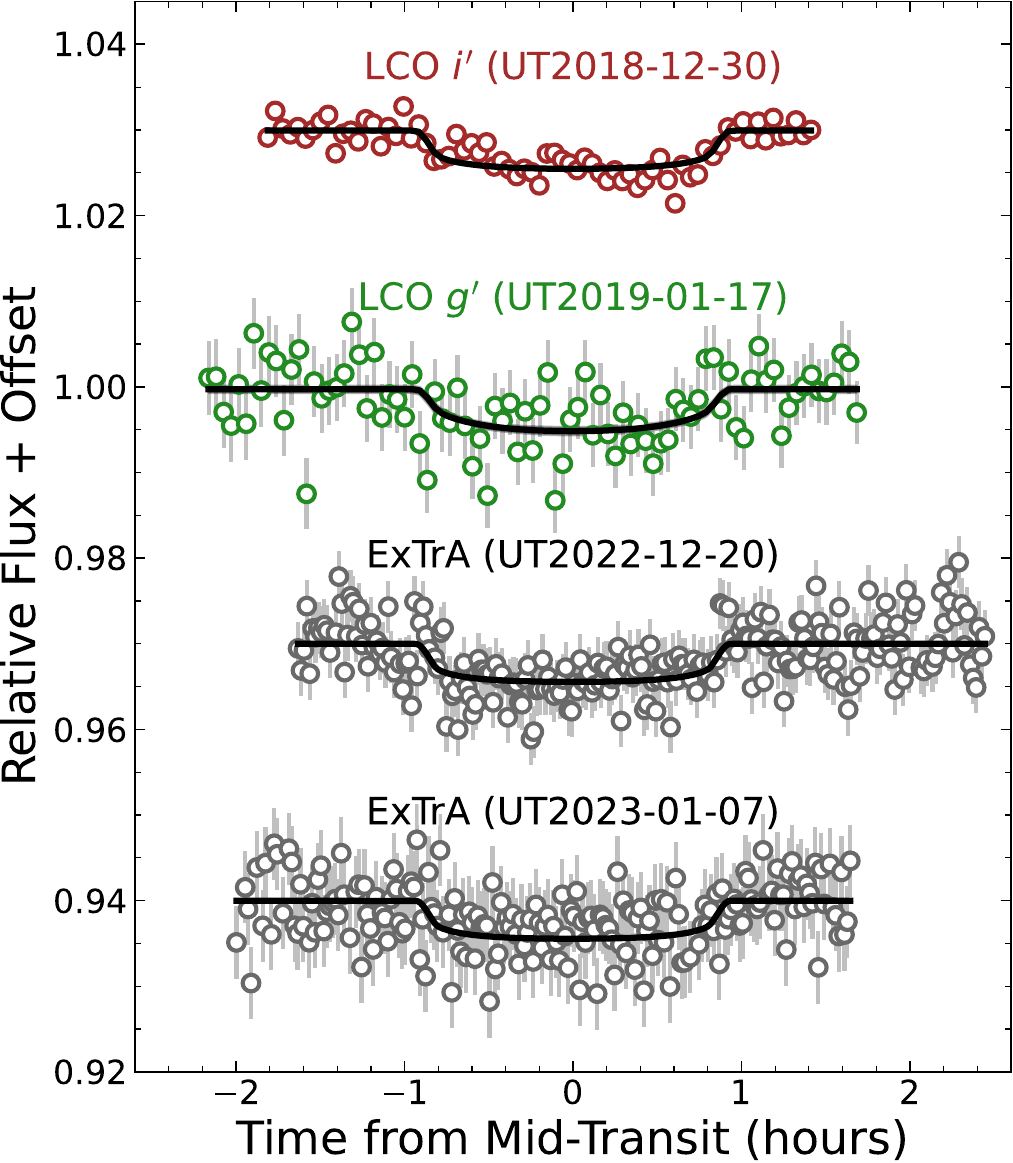}
    \endminipage
    \caption{Transit of TOI-210\,b. \textit{Middle and Left panels}:~Phase-folded transits from TESS and ExTrA with bins of 10\,min (in phase) shown with white circle. \textit{Right panel}:~Ground-based transit photometry from LCOGT and ExTrA stacked with an arbitrary vertical offset and ordered chronologically from top to bottom. We show two representative transits of ExTrA to illustrate the range of data quality across a total of 16 transits, displaying respectively the most- and least-significant detections according to our statistical analysis (Sect.~\ref{sec:ExTrA}).}
    \label{fig:transits+ground_based_photometry}
\end{figure*}

We further curated the ExTrA dataset based on the overall quality of the transits using a simple statistical test. For each light curve, we compared the Bayesian Predictive Information Criterion Simplified (BPICS) metric between a flat line and a transit model. The BPICS, introduced by \cite{Ando_2011} and later defined by \cite{Thorngren_2026}, is a model comparison metric similar to the Bayesian Information Criterion (BIC; \citealt{Schwarz_1978}) with the only difference being that it incorporates the average log-likelihood along the posterior samples instead of the maximum value, which yields a more robust indicator of the predictive power of the model. We retained only light curves with $\Delta\mathrm {BPICS} > 40$, a threshold chosen to exclude the lowest-quality transits. This criterion removes five light curves from the low end of the BPICS distribution (mean 77.3, maximum 160.3), leaving a final set of 16 ExTrA transits for the joint analysis. A phase-fold of the complete dataset and two representative light curves (lowest and highest retained BPICS) are shown in Fig.~\ref{fig:transits+ground_based_photometry}.

\subsection{SOAR high-resolution imaging} \label{sec:soar}

We obtained high-resolution imaging of the star TOI-210 using the High Resolution Camera (HRCam) speckle imager installed on the 4.1-m Southern Astrophysical Research (SOAR) telescope \citep{Tokovinin_2024}. The observations were taken on UT2019-02-18 with the $I$ filter (879\,nm, bandwidth 289\,nm) to search for nearby blended sources that could dilute the TESS photometry or affect the interpretation of the transit origin (SOAR TESS Survey; \citealt{Ziegler_2020}). HRCam provides a pixel scale of $0\farcs01575$ and the estimated point-spread function width of the star was $0\farcs07683$ (5\,pixels). From the image sequence, we constructed a 5$\sigma$ contrast curve (Fig.~\ref{fig:imaging}) and rule out any source with a contrast ratio $\Delta I < 4.00$\,mag at $0\farcs2$.

\subsection{NIRPS and HARPS radial velocity}

The NIRPS spectrograph can operate simultaneously with HARPS (380--690\,nm; \citealt{Mayor_2003}). We acquired 263 spectroscopic observations of TOI-210 with NIRPS and 122 with HARPS over 129 nights between UT2023-11-15 and UT2026-01-27 (Program IDs 112.25N, 112.25NS, 112.25NS, and 116.2985; PI: Bouchy \& Doyon). Operating behind an adaptive optics (AO) system, the NIRPS data were acquired in the High Efficiency mode (HE) providing a $0\farcs9$ on-sky fiber entrance and a spectral resolution $R \approx 75\,000$. HARPS observations are seeing-limited and were obtained primarily in its high-throughput EGGS mode ($R \approx 80\,000$, $1\farcs4$ fiber), with some exposures taken in high-accuracy HAM mode ($R \approx 115\,000$, $1\farcs0$ fiber). For both instruments, the simultaneous calibration fiber was set to observe the sky. The default observing setup consisted of two consecutive exposures of 900\,s with NIRPS per night, while HARPS integrated for the full 1800\,s sequence. For some nights, a single spectrum or up to four spectra were obtained with NIRPS due to the observing conditions that required exposures to be repeated or aborted. Overall, the observations were conducted under median airmass of 1.44 and seeing conditions of $0\farcs9$.

NIRPS benefits from having two independent data reduction softwares (DRS) capable of processing raw detector frames into science-ready, telluric-corrected, order-by-order spectra. The TOI-210 observations were reduced with version 3.3.12 of the \texttt{NIRPS-DRS} \citep{Bouchy_2025}, an adaptation of the ESPRESSO pipeline \citep{Pepe_2021}, and with version 0.7.294 of \texttt{APERO} \citep{Cook_2022}. In the \texttt{NIRPS-DRS}, telluric correction follows the method of \cite{Allart_2022} with a new module for treating OH emission lines making use of the simultaneous sky observations from the calibration fiber \citep{Srivastava_2026b}. The \texttt{APERO} side, the tellurics both in absorption and in emission are corrected using a combination of TAPAS atmospheric models \citep{Bertaux_2014} and a library of (featureless) fast-rotating hot stars  observed with NIRPS under various conditions (airmass and water column density). The extracted NIRPS spectra have a median S/R per pixel of 32.9 in the middle of the $H$ band. The HARPS data were processed with the original DRS version 3.5 \citep{Lovis_2007}. The median S/R per pixel in the redder orders of HARPS only reaches approximately 3 for this optically faint star ($V = 15.366$\,mag).

We measured RVs using a line-by-line (LBL) approach \citep{Dumusque_2018} with the publicly available code \texttt{LBL} compatible with both NIRPS and HARPS \citep{Artigau_2022}. The \texttt{LBL} code matches each spectrum to the wavelength-derivative of a high-S/R template following the method of \cite{Bouchy_2001}. For NIRPS, we used the template of GJ~682 (M4V) and for HARPS, the template of GJ~3090 (M3V), both from NIRPS-GTO observations of brighter stars with similar spectral type to TOI-210. This template-matching is performed on $\sim$13\,000 spectral lines, building a distribution from which an outlier-resilient weighted average RV is derived from a statistical model (Appendix B of \citealt{Artigau_2022}). The LBL framework also provides ancillary spectroscopic quantities that trace stellar activity. One indicator is the differential line width (dLW; \citealt{Zechmeister_2018}), akin to the full-width at half maximum of the cross-correlation function (mean line profile) in the approximation of Gaussian lines. Another novel indicator is the differential stellar temperature (d\textit{Temp}; \citealt{Artigau_2024}). d\textit{Temp} time series have emerged as a powerful tracer of stellar rotation, demonstrating a strong correlation with simultaneous TESS photometry in NIRPS observations of the M dwarf GJ 3090 \citep{Lamontagne_2026}.

The NIRPS RVs have a median uncertainty of 5.2\,m\,s$^{-1}$ and a scatter of 10.6\,m\,s$^{-1}$ (nightly bins). For comparison, the HARPS LBL RVs yield a median error of 23\,m\,s$^{-1}$ and a dispersion of 21\,m\,s$^{-1}$, about four times less precise than NIRPS for an equivalent exposure time. For this reason, we excluded the HARPS measurements from the analysis. The RV analysis presented in Sect.~\ref{sec:rv_analysis} is performed on individual NIRPS exposures. 

We discarded ten low-S/R nights, primarily affected by AO-loop openings during the sequence, telescope vibrations and oscillations affecting pointing stability, or high-airmass observations, as recorded in the observing logs. In addition, we removed 15 exposures identified via a 3.5$\sigma$ clipping in the d\textit{Temp}, RV, and dLW data. The resulting time series are provided in Table~\ref{table:rv} for \texttt{NIRPS-DRS}, with a comparison with \texttt{APERO} shown in Fig.~\ref{fig:drs_comps}. The two pipelines are in good agreement, with median absolute deviations of 0.69$\sigma$ in d\textit{Temp}, 0.65$\sigma$ in RV, and 0.81$\sigma$ in dLW. For the remainder of the analysis, we adopted the \texttt{NIRPS-DRS} products as our default dataset, while systematically verifying that all results are consistent across pipelines.

With a systemic velocity of $15.0$\,km\,s$^{-1}$ and a small yearly Barycentric Earth Radial Velocity (BERV) variation of $\Delta \mathrm{BERV} = 7.6$\,km\,s$^{-1}$, TOI-210 avoids the BERV-crossing systematics reported in other NIRPS datasets (e.g., \citealt{Parc_2025, Frensch_2026, Osborn_2026, Srivastava_2026a}). These systematics occur when the BERV matches the systemic velocity, causing stellar lines and telluric residuals from common species (e.g., H$_2$O, OH) to overlap. Despite avoiding this regime, we still detect residual correlations between d\textit{Temp}, RV, and dLW with BERV in both pipelines. These effects were mitigated in the analysis by detrending with a second-order polynomial in BERV, following the approach of \cite{Suarez_2025} for NIRPS observations of Proxima.

\section{Stellar characterisation} \label{sec:stellar_char}

\subsection{Basic stellar properties}

TOI-210 (2MASS J05555049-7359046, TIC 141608198) is a Southern Hemisphere field star \citep{Gagne_2018,Gagne_2026} at a distance of $42.58 \pm 0.03$\,pc \citep{Gaia_2023}. The analysis of the NIRPS d\textit{Temp} and dLW activity indicators revealed a 71 $\pm$ 6\,days rotation period (Sect.~\ref{sec:rv_analysis}). M-type stars lose angular momentum over time, which slows down their rotation rate, allowing a calibration of their age. Based on \cite{Engle_2023}, TOI-210's slow rotation is consistent with an age of $\sim$4.5\,Gyr. TOI-210 has a \textit{Gaia} DR3 RUWE metric of 1.221, below the threshold of 1.4 which would indicate excess astrometric noise indicative of unresolved companions \citep{Ziegler_2020}.

Using the $G - G_{\rm RP}$ to spectral type relation of \cite{Kiman_2019}, we get a spectral type of M$3.5 \pm 0.5$V for TOI-210. The $T_{\rm eff} = 3200 \pm 50$\,K we derived in Sect.~\ref{sec:sed} from the spectral energy distribution (SED) of the star is consistent with an M4V spectral type according to Table 5 of \cite{Pecaut_2013}. Based on these two estimates, we assigned a spectral type of M4V for TOI-210.

We next estimated the stellar mass and radius using the empirical relations of \cite{Mann_2015,Mann_2019} with absolute $K_s$ magnitude. For TOI-210, $K_{\rm s} = 9.995 \pm 0.023$ from 2MASS \citep{Skrutskie_2006} and using the distance from \textit{Gaia} DR3, we infer a $M_{\star} = 0.310 \pm 0.008$ M$_\odot$ and a $R_{\star} = 0.329 \pm 0.010$ R$_\odot$ with the uncertainty estimated through a Monte Carlo approach that accounts for the intrinsic scatter in the \cite{Mann_2015,Mann_2019} relations. From $M_{\star}$ and $R_{\star}$, we derived the surface gravity ($\log g$) and the stellar density ($\rho_\star$). We summarise the properties of the star TOI-210 in Table~\ref{table:stellar_params}.

\begin{table}
\caption{\label{table:stellar_params}Summary of TOI-210 Stellar Properties}
\centering
\renewcommand{\arraystretch}{1.25} 
\begin{tabular}{lcc}
\hline\hline
Parameter & Value & Ref.\\
\hline
\multicolumn{3}{c}{\textit{Identifiers}}\\
TOI & 210 & 1\\
TIC & 141608198 & 1\\
2MASS & J05555049-7359046 & 2\\
Gaia DR3 & 4650160717726370816 & 3\\
\hline
\multicolumn{3}{c}{\textit{Astrometry}}\\
RA, $\alpha$ (J2016.0) & 05$^{\mathrm{h}}$55$^{\mathrm{m}}$50$\fs$85 & 3\\
Dec, $\delta$ (J2016.0) & $-73^{\circ}59\arcmin07\farcs1$ & 3\\
$\mu_{\alpha} \cos \delta$ (mas\,yr$^{-1}$) & $+$80.388 $\pm$ 0.016 & 3\\
$\mu_{\delta}$ (mas\,yr$^{-1}$) & $-139.150$ $\pm$ 0.020 & 3\\
$\pi$ (mas) & 23.4846 $\pm$ 0.0156 & 3\\
$d$ (pc) & 42.58 $\pm$ 0.03 & 3\\
\hline
\multicolumn{3}{c}{\textit{Stellar parameters}}\\
Spectral Type & M4V & 4\\
$T_{\rm eff}$ (K) & 3200 $\pm$ 50 & 4\\
$L_{\star}$ (L$_{\odot}$) & 0.01141 $\pm$ 0.00005 & 4\\
$M_{\star}$ (M$_{\odot}$) & 0.310 $\pm$ 0.008 & 4\\
$R_{\star}$ (R$_{\odot}$) & 0.329 $\pm$ 0.010 & 4\\
$\rho_{\star}$ (g\,cm$^{-3}$) & 12.3 $\pm$ 1.2 & 4\\
log $g$ (cgs) & 4.90 $\pm$ 0.03 & 4\\
$\left[ {\rm M/H} \right]$ & $0.15 \pm 0.11$ & 4\\
$P_{\rm rot}$ (days) & 71 $\pm$ 6 & 4\\
Age (Gyr) & $\sim4.5$ &  4\\
\hline
\multicolumn{3}{c}{\textit{Photometry}}\\
$B$ & 16.674 $\pm$ 0.014 & 5\\
$V$ & 15.366 $\pm$ 0.149 & 5\\
$G$ & 13.828 $\pm$ 0.003 & 3\\
$G_{\rm BP}$ & 15.435 $\pm$ 0.004 & 3\\
$G_{\rm RP}$ & 12.606 $\pm$ 0.004 & 3\\
$T$ & 12.540 $\pm$ 0.008 & 1\\
$g$ & 15.797 $\pm$ 0.029 & 5\\
$r$ & 14.551 $\pm$ 0.021 & 5\\
$i$ & 13.023 $\pm$ 0.046& 5\\
$J$ & 10.873 $\pm$ 0.023 & 2\\
$H$ & 10.258 $\pm$ 0.024 & 2\\
$K_{\rm s}$ & 9.995 $\pm$ 0.023 & 2\\
$W1$ & 9.817 $\pm$ 0.023 & 6\\
$W2$ & 9.680 $\pm$ 0.020 & 6\\
$W3$ & 9.538 $\pm$ 0.030 & 6\\
\hline
\end{tabular}
\tablebib{(1) TICv8.2 \citep{Stassun_2019}. (2) 2MASS \citep{Skrutskie_2006}. (3) 
\textit{Gaia} DR3 \citep{Gaia_2023}. (4) This work. (5) APASS \citep{Henden_2015}. (6) WISE \citep{WISE_2010}.}
\end{table}

\subsection{Spectral energy distribution fit} \label{sec:sed}

\begin{figure}
    \centering
    \includegraphics[width=1\linewidth]{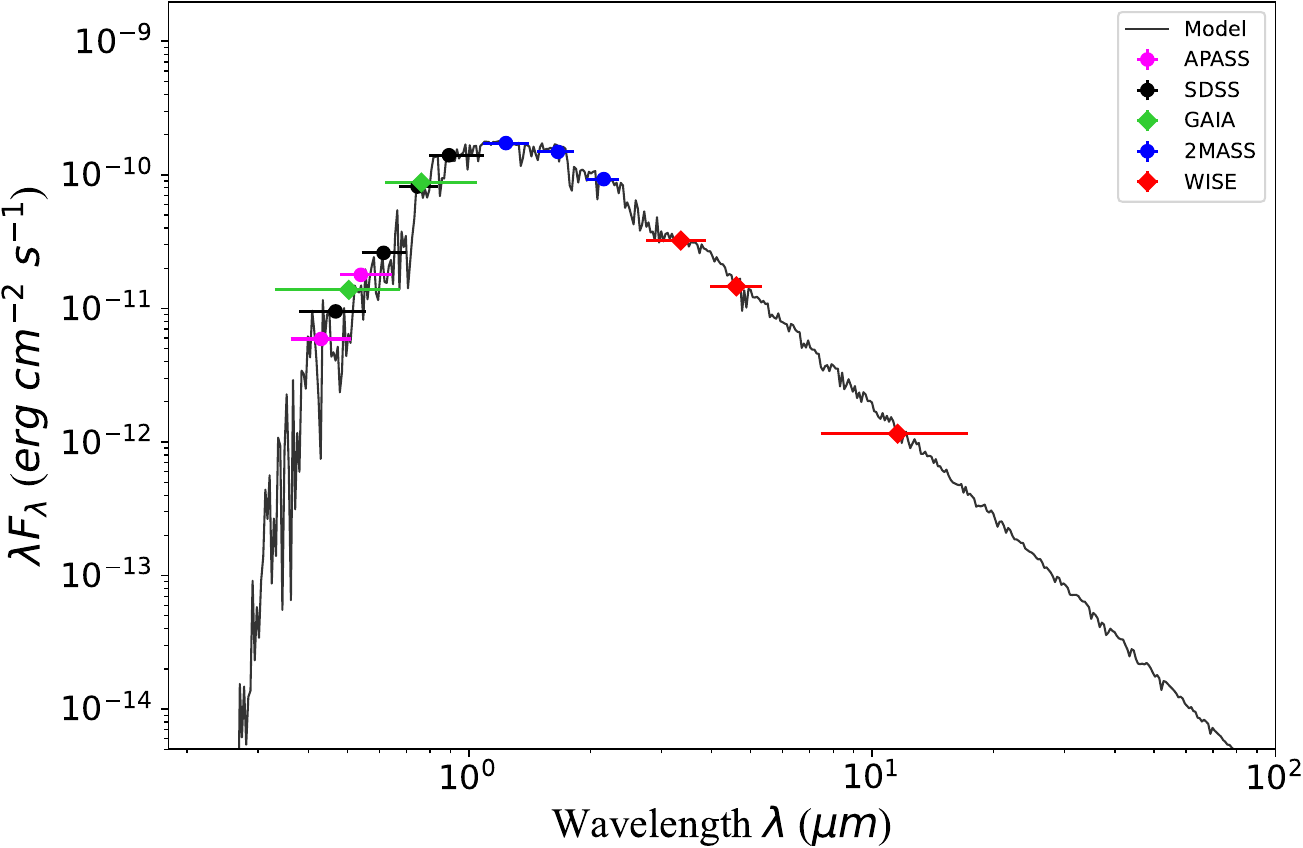}
    \caption{Spectral energy distribution (SED) of TOI-210 fitted with a BT-Settl stellar atmosphere model \citep{Allard_2012, Allard_2013} with $T_{\rm eff} = 3200$\,K, [M/H]~$= 0.3$\,dex, and $\log g = 4.5$ (cgs). Photometric measurements obtained in different bandpasses are shown as colored points, with horizontal bars indicating the effective widths of the corresponding filters.}
    \label{fig:sed_fit}
\end{figure}

We independently determined the stellar parameters from an analysis of the SED using available broadband photometry spanning 0.4--12~$\mu$m (see Fig.~\ref{fig:sed_fit}). For TOI-210, we compiled photometric measurements from the following catalogs: $G$-$G_{\rm BP}$-$G_{\rm RP}$ from \textit{Gaia} DR3, $BV$ from APASS \citep{Henden_2015}, $gri$ from SDSS system \citep{Alam2015} collected by APASS, $JHK_{\rm s}$ from 2MASS, and $W1$--$W3$ from WISE \citep{WISE_2010}.

The SED fitting was performed using the Virtual Observatory Spectral Analyzer (VOSA; \citealt{Bayo_2008}). For M dwarfs, VOSA employs a grid of BT-Settl stellar atmosphere models \citep{Allard_2012, Allard_2013} with free parameters $T_{\rm eff}$, [M/H], and $\log g$. We assumed negligible extinction ($A_V = 0$\,mag) given the proximity to TOI-210. The model minimising the $\chi^2$ yields $T_{\rm eff} = 3200 \pm 50$\,K, [M/H]~$= 0.3 \pm 0.1$\,dex, and $\log g = 4.5 \pm 0.25$ (cgs), where the uncertainties correspond to half the spacing of the model grid. By integrating the best-fit SED model and with the \textit{Gaia} distance, we obtain a bolometric luminosity of $L_{\star} = 0.01141 \pm 0.00005$\,L$_{\odot}$. Using the Stefan-Boltzmann law, we derive a stellar radius of $R_{\star} = 0.348 \pm 0.011$\,R$_\odot$. This radius is consistent at the 1.3\,$\sigma$ level with the value obtained from the empirical relation of \citealt{Mann_2015}.

\subsection{Spectroscopic chemical abundances determination} \label{sec:chemical_abundances}

The NIRPS template spectrum of TOI-210 was used to measure the chemical abundances of several elements. For this task, we adopt the methodology developed by \cite{Jahandar_2024, Jahandar_2025} for near-infrared spectroscopy with SPIRou \citep{Donati_2020}, adapted and previously applied for NIRPS observations of the cool dwarfs LHS~1140 (M4.5V; \citealt{Cadieux_2024a}), TOI-406 (M3V; \citealt{Lacedelli_2024}), TOI-756 (M1V; \citealt{Parc_2025}), TOI-270 (M3V; \citealt{Coulombe_2025}), GJ~3090 (M2V; \citealt{Lamontagne_2026}), TOI-4666 (M2.5V; \citealt{Frensch_2026}), and TOI-4336~A/TOI-4342 (M3.5V/M0V, \citealt{Parc_2026}). We briefly summarise the method here.

The co-added spectrum of TOI-210 is compared to a grid of PHOENIX-ACES stellar models \citep{Allard_2012, Husser_2013} generated at a fixed log\,$g = 5$ with interpolation steps of 100\,K in $T_{\rm eff}$ and 0.1\,dex in [M/H]. The synthetic spectra were computed at high spectral resolution and then convolved with a Gaussian kernel to match the resolving power of NIRPS. Strong spectral lines in the $YJH$ bands were fitted to obtain an average effective temperature and its 1$\sigma$ dispersion in $T_{\rm eff}$ for [M/H] in the interval [$-1$, 1]\,dex. The [M/H] with the smallest dispersion in $T_{\rm eff}$ is considered the best-fit solution. While this method spectroscopically derives  $T_{\rm eff}$, the lack of $K$ band coverage by NIRPS often results in an overestimated temperature (here $T_{\rm eff} = 3577 \pm 37$\,K), as discussed by \cite{Coulombe_2025}. To avoid potentially propagating this bias into the abundance analysis, we instead fixed the stellar effective temperature to the SED-derived value of $T_{\rm eff} = 3200$\,K. Next, spectral lines from known chemical species were fitted individually while allowing the metallicity to vary freely. To account for the uncertainty in effective temperature, we repeated the analysis with fixed $T_{\rm eff}$ corresponding to the two extremes of its 1$\sigma$ confidence interval, using the same set of spectral lines as the $T_{\rm eff}=3200$\,K case, following \citet{Parc_2026}. The resulting abundances and their 1$\sigma$ dispersions are reported in Table~\ref{table:stellar_abundances}, yielding a global metallicity of [M/H] $= 0.15 \pm 0.11$.

\begin{table}[h]
\caption{\label{table:stellar_abundances}Chemical abundances of TOI-210 obtained with NIRPS}
\centering
\renewcommand{\arraystretch}{1.25} 
\begin{tabular}{lcc}
\hline\hline
Element$^*$ & [X/H] & \# of lines\\
\hline
Fe I &  $-0.13$ $\pm$ 0.31 & 8 \\
Mg I &  $0.73$ $\pm$ 0.27 & 3 \\
Si I &  $0.26$ $\pm$ 0.33 & 2 \\
Ca I & $-0.10$ $\pm$ 0.28 & 1 \\
Ti I &  $-0.13$ $\pm$ 0.19 & 10 \\
Al I &  $0.60$ $\pm$ 0.31 & 2 \\
Na I &  0.49 $\pm$ 0.13 & 3 \\
C I  &  $-0.19$ $\pm$ 0.38 & 1 \\
K I  & 0.01 $\pm$ 0.32 & 2 \\
O I$^{\dag}$ & 0.05 $\pm$ 0.09 & 21 \\ 
\hline
[M/H] & 0.15 $\pm$ 0.11 & --- \\
\hline
\end{tabular}
\vspace{0.1cm}
\caption*{\footnotesize {\bf Notes.} $^*$Elements other than Fe, Ti, and O are based on 3 or fewer lines and should be treated with caution. $^\dag$The oxygen abundance is inferred from OH lines.
}
\end{table}

\section{Data analysis and results} \label{sec:analysis_results}

\subsection{Transit analysis}\label{sec:transit_analysis}

We jointly analysed the transit photometry from TESS, LCOGT, and ExTrA, and then used the resulting constraints on the orbital period and phase of TOI-210\,b as priors for the RV analysis. We chose this sequential approach to reduce computational costs, given the large combined transit data set and the relative complexity of the RV activity modelling.

The multi-instrument transit fit was performed within the \texttt{juliet} framework \citep{Espinoza_2019}, a Bayesian inference tool that combines transit modelling with \texttt{batman} \citep{Kreidberg_2015} and nested sampling, here implemented with \texttt{dynesty} \citep{Speagle_2020}. The \texttt{dynesty} sampler is based on the dynamic nested sampling algorithm \citep{Higson_2019} which progressively allocates more live points in high-likelihood regions to improve accuracy and Bayesian evidence estimates.

We assumed a circular orbit ($e = 0$, $\omega = 90^{\circ}$) for TOI-210\,b, as expected for a short-period planet subject to tidal circularisation. This choice is further supported by the RV analysis (Sect.~\ref{sec:rv_analysis}), which does not provide evidence for a non-zero eccentricity. The parameters to describe the orbit and transit are the orbital period $P$, the time of inferior conjunction $t_{0}$, the planet-to-star radius ratio $R_{\rm p} / R_{\star}$, the scaled semi-major axis $a/R_{\star}$, and the orbital inclination $i$. We placed uniform priors on $P$ and $t_{0}$ centered on the ExoFOP ephemeris and with a half-width of 0.1\,days. We reparameterised $a / R_{\star}$ by fitting instead the log stellar density $\ln \rho_{\star}$ as suggested by \cite{Gilbert_2022} to avoid biases at high impact parameter, with a prior using $\rho_{\star}$ in Table~\ref{table:stellar_params} ($\ln \rho_{\star}$\,$\sim$\,$\mathcal{N}\left(9.41, 0.11\right)$, SI units). For efficient exploration of $R_{\rm p}/R_{\star}$ and the transit impact parameter ($b = a/R_{\star}\cos i$), we sampled instead the parameters $r_1$ and $r_2$ defined by \citet{Espinoza_2019}, which ensure physically plausible transiting configurations for values between 0 and 1.

Stellar limb darkening in each instrument bandpass is modeled using the quadratic parameters $q_1$ and $q_2$ which can take values between 0 and 1 \citep{Kipping_2013}. The photometric precision of the TOI-210 data is insufficient to constrain second-order limb-darkening effects. We therefore adopted broad Gaussian priors on $q_1$ and $q_2$ for each bandpass, centered on theoretical predictions computed with \texttt{ExoTic-LD} \citep{Grant_Wakeford_2024} using a PHOENIX stellar atmosphere model \citep{Husser_2013} with $T_{\rm eff}=3200$\,K, $\log g = 5$, and [M/H] = 0.0.

For each instrument, we modeled the out-of-transit flux as $f_{0} = 1/(1 + DM)$, where $D$ is a dilution factor and $M$ is a baseline flux offset \citep{Espinoza_2019}. We fixed $D = 1$ (i.e., no dilution) and fitted for $M$ using a broad Gaussian prior, $\mathcal{N}(0, 0.01)$. Finally, we included an additional photometric jitter term $\sigma$ for each instrument to account for excess white noise beyond the nominal uncertainties.

The nested sampling was done with $100\times N_{\rm param}$ live points, with $N_{\rm param} = 21$ the number of free parameters. As recommended by \cite{Speagle_2020} for dimensionality above 20, we used the random slice sampling (`rslice') in \texttt{dynesty} to propose new live points. The priors and posterior results from the joint transit fit are reported in Table~\ref{table:transit_params}. The uncertainties are given from the 16$^{\rm th}$ and 84$^{\rm th}$ percentiles of the posterior distributions. The best-fit transit model for TOI-210\,b in TESS, LCOGT, and ExTrA photometry can be seen in Fig.~\ref{fig:transits+ground_based_photometry}.

Our transit data set spans a wide range of bandpasses, allowing us to validate the achromatic nature of the transit signal. We tested this by fitting the transits for each instrument individually, with the same priors as the joint analysis, with the exception of $P$ and $t_{0}$, which are constrained with normal priors informed by the combined fit. The resulting measurements of $R_{\rm p}/R_\star$ are shown in Fig.~\ref{fig:transit_chromaticity}. We find no evidence for important chromaticity, with all $R_{\rm p}/R_\star$ consistent within $2\sigma$.

\subsection{Radial velocity analysis} \label{sec:rv_analysis}

We now present the analysis of the NIRPS RVs of TOI-210. We modeled Keplerian signals with \texttt{radvel} \citep{Fulton_2018}, stellar activity using a multidimensional GP (multi-GP) with \texttt{tinygp} \citep{Foreman-Mackey_2024}, and detrended the data against the BERV.

We used normal priors for the $P$ and $t_0$ of TOI-210\,b based on the median and 1$\sigma$ results of the transit fit (Table~\ref{table:transit_params}). The semi-amplitude $K$ followed a uniform prior $\mathcal{U}\left(0, 10\right)$\,m\,s$^{-1}$. The eccentricity $e$ and argument of periastron $\omega$ are either fixed to 0 and 90$^{\circ}$ for circular orbit models or, when allowing for an eccentric orbit, we instead fitted for $\sqrt{e} \cos \omega$ and $\sqrt{e} \sin \omega$ between $-1$ and $1$, forcing $e < 0.99$ to avoid numerical errors.

The stellar activity modelling follows the multi-GP framework of \cite{Rajpaul_2015}, itself motivated by the \cite{Aigrain_2012} formalism that activity-induced RV variations from brightness heterogeneities and convective blueshift suppresion from active regions can be approximated by the product of the stellar flux and its time derivative ($FF^{\prime}$). Our multi-GP models three time series, corresponding to the observable $k \in \{\mathrm{d\textit{Temp}, RV, dLW}\}$. Each observable is described as a linear combination of a shared Gaussian process $\mathcal{GP}$ and its time derivative:
\begin{equation}
\label{eq:multi-GP}
\Delta y_k = \alpha_{k}\cdot\mathcal{GP} +\beta_{k}\cdot\mathcal{GP}^{\prime}
\end{equation}
where $\beta_{\textrm{d}Temp}$ is fixed to 0, allowing d\textit{Temp} to serve as a photometry proxy, as previously established for the mid-M dwarf GJ~3090 \citep{Lamontagne_2026}. For $\mathcal{GP}$, we adopted the quasi-periodic covariance function (\citealt{Haywood_2014}; \citealt{Rajpaul_2015}):
\begin{equation}
k_{i,j} = \exp \left[ -\frac{|t_i - t_j|^2}{2 \ell^2} - \Gamma \sin^2 \left( \frac{\pi | t_i - t_j|}{P_{\rm rot}} \right) \right]
\end{equation}
where $|t_i - t_j|$ is the time lag between data $i$ and $j$, $\ell$ is the coherence timescale usually scaling with the lifetime of active regions, $\Gamma$ sets the number of inflection points within one recurrence period $P_{\rm rot}$, associated with stellar rotation. Note here the $\alpha_{k}$ and $\beta_{k}$ in Eq.\ (\ref{eq:multi-GP}) serve as amplitude terms.

A periodogram analysis of the d\textit{Temp} and dLW time series revealed significant peaks with false alarm probability (FAP) below 0.1\% at 49, 58, 99, and 140 days. We therefore adopted a broad prior $\mathcal{LU}(30, 200)$\,days that covers a wide range of plausible rotation periods, with the upper bound approximately corresponding to the longest periods known for fully convective stars \citep{Newton_2018}. The priors for $\ell$ and $\Gamma$ were motivated by \cite{Stock_2023} when $P_{\rm rot}$ is unknown, i.e., $\ell$ is at least one stellar rotation $\mathcal{LU}\left(30, 1000\right)$\,d and $\Gamma$ is constrained within $\mathcal{LU}\left(0.01, 10\right)$ for a plausible level of harmonic complexity.

Residuals of the telluric correction can leave imprints in spectroscopic measurements. Like \cite{Suarez_2025}, we detrended the time series with a quadratic term against BERV:
\begin{equation}
\Delta y_k = a_{k}\cdot\mathrm{BERV}^2 + b_{k}\cdot\mathrm{BERV}
\end{equation}
We adopted zero-centered Gaussian priors $\mathcal{N}(0, 0.5)$ for $a_{k}$ and $b_{k}$ (in their respective units), which remain weakly informative while disfavoring large values. Finally, we fit for constant offsets $c_{\mathrm{d}Temp}$, $c_{\mathrm{RV}}$, and $c_{\mathrm{dLW}}$, and for any excess white noise $s_{\mathrm{d}Temp}$, $s_{\mathrm{RV}}$, and $s_{\mathrm{dLW}}$, using broad, uninformative priors.

We explored the posterior distribution using the \texttt{jaxns} nested sampling code \citep{Albert_2020, Albert_2023}. Both \texttt{tinygp} and \texttt{jaxns} are written in \texttt{JAX}\footnote{\href{https://github.com/jax-ml/jax}{\texttt{github.com/jax-ml/jax}}}, enabling automatic differentiation, just-in-time compilation, and efficient parallelisation. This substantially reduced the computational cost of likelihood evaluations, which number in the millions for this high-dimensional inference problem. Based on the benchmarking results of \citet{Albert_2023}, we adopted $c = 40 \times N_{\rm param}$ Markov chains, a slice factor of $s=6$, and $k=5$ `phantom' samples for accurate Bayesian evidence calculations at large $N_{\rm param}$. The phantom-powered nested sampling algorithm implemented in \texttt{jaxns} incorporates a subset of intermediate samples (`phantom') generated during slice sampling into the evidence and posterior estimation, rather than discarding them as in standard nested sampling, thereby reducing the total number of likelihood evaluations by a factor of 5 or more \citep{Albert_2023}. This setup generates approximately $200 \times N_{\rm param}$ live points, providing robust exploration of the prior volume.

\begin{figure*}
    \centering
    \minipage{0.78\textwidth}
    \includegraphics[width=1\linewidth]{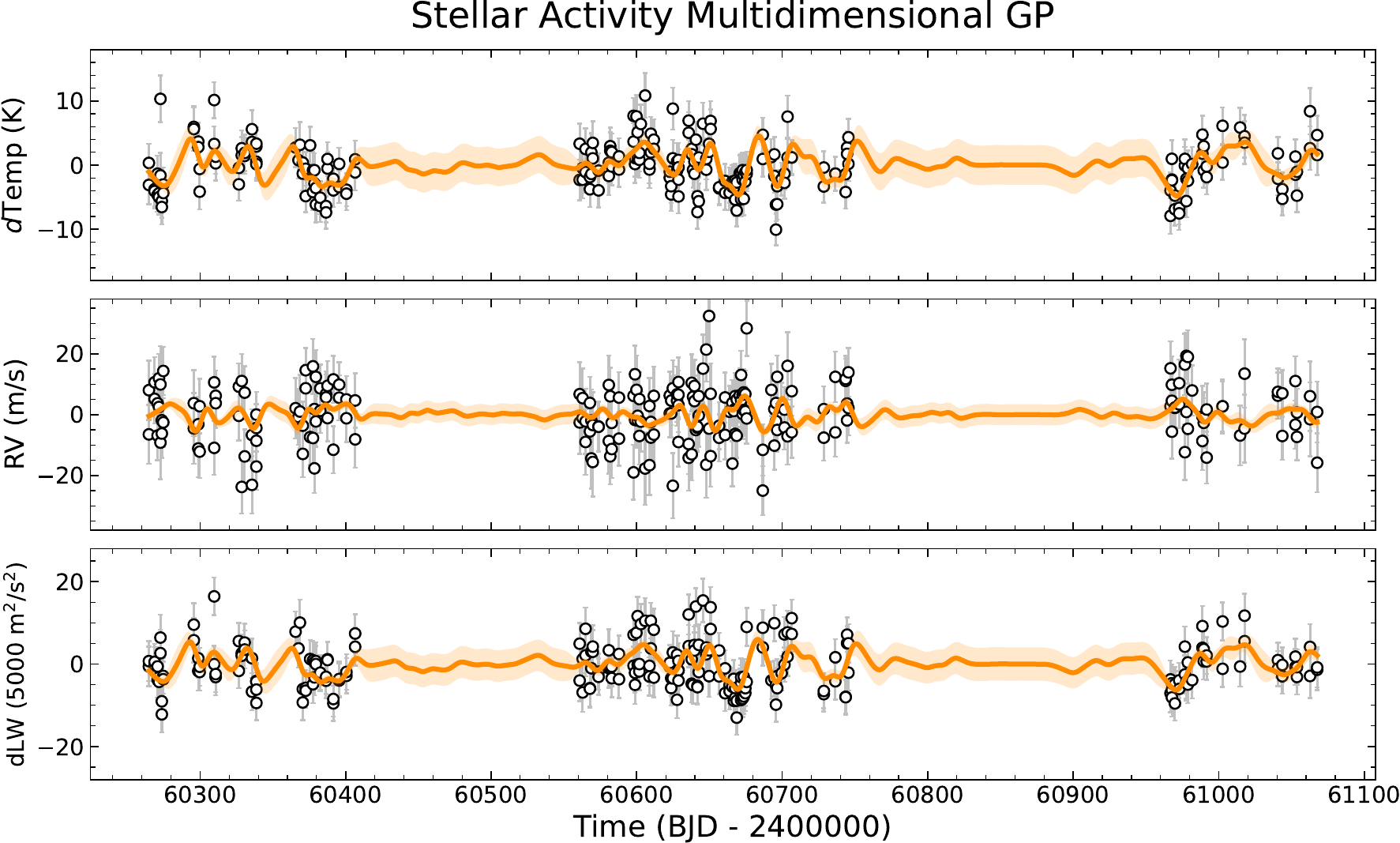}
    \endminipage
    \hfill
    \minipage{0.193\textwidth}
    \vspace{-0.1cm}
    \includegraphics[width=1\linewidth]{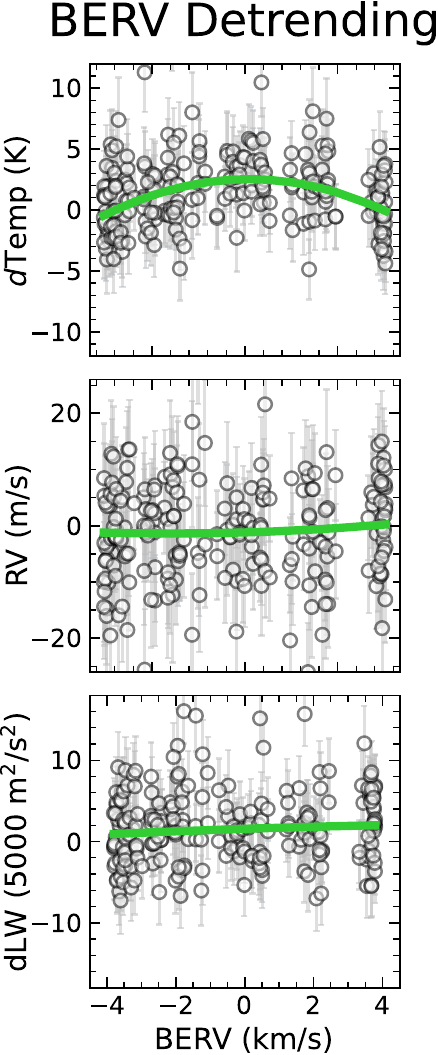}
    \endminipage\\[0.3cm]
    \minipage{0.33\textwidth}
    \includegraphics[width=1\linewidth]{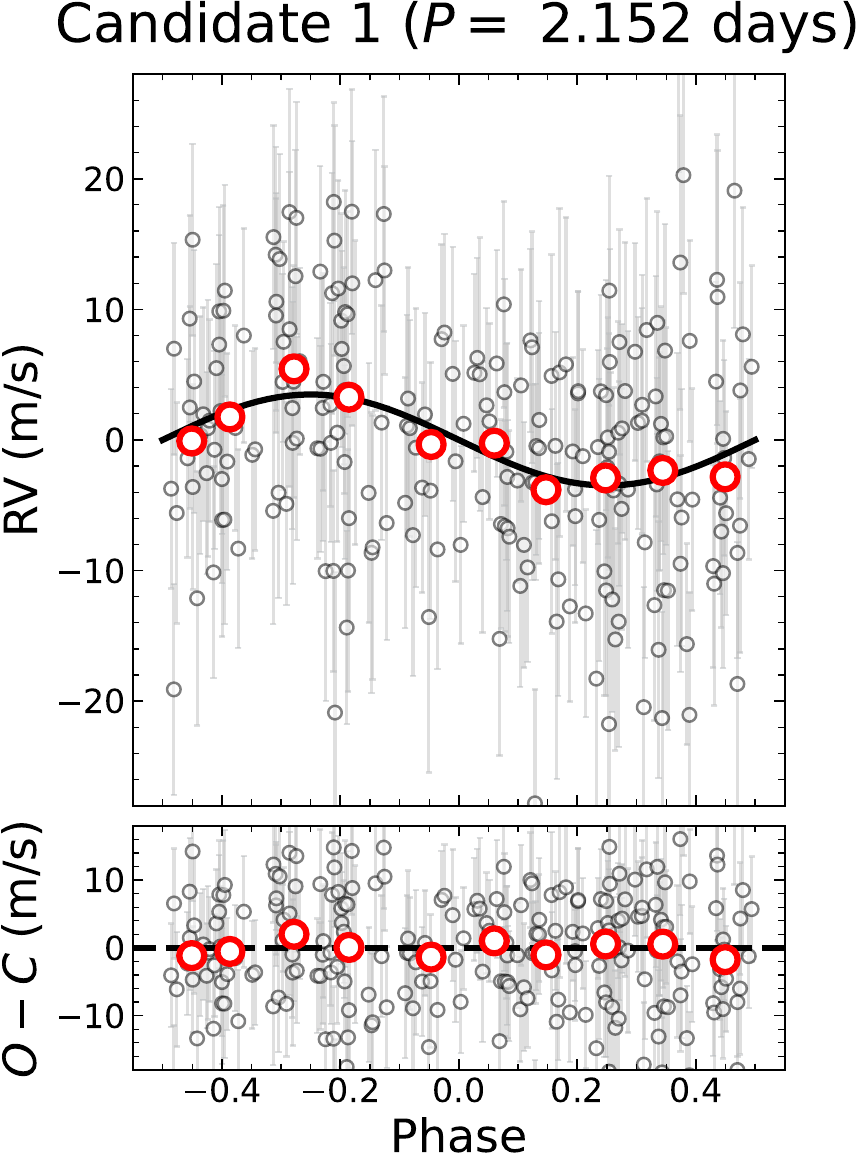}
    \endminipage
    \minipage{0.33\textwidth}
    \hfill
    \includegraphics[width=1\linewidth]{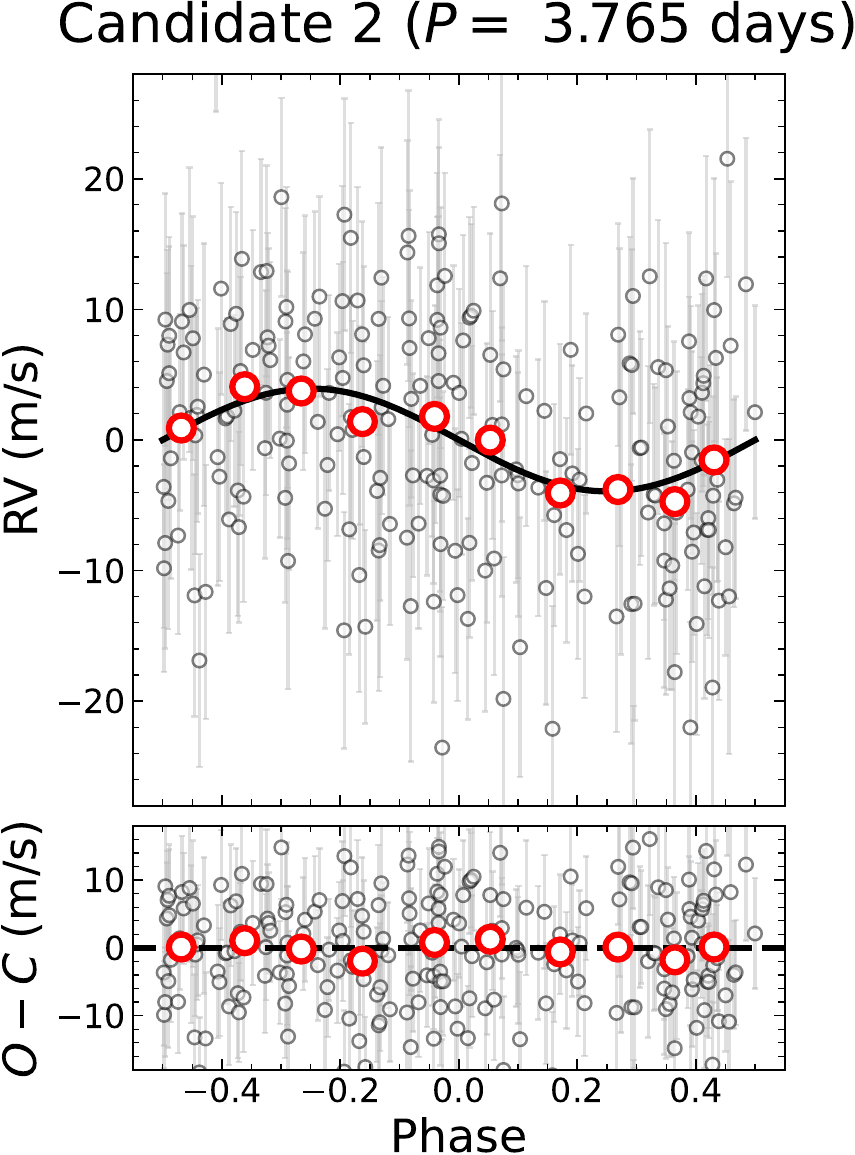}
    \endminipage
    \hfill
    \minipage{0.33\textwidth}
    \includegraphics[width=1\linewidth]{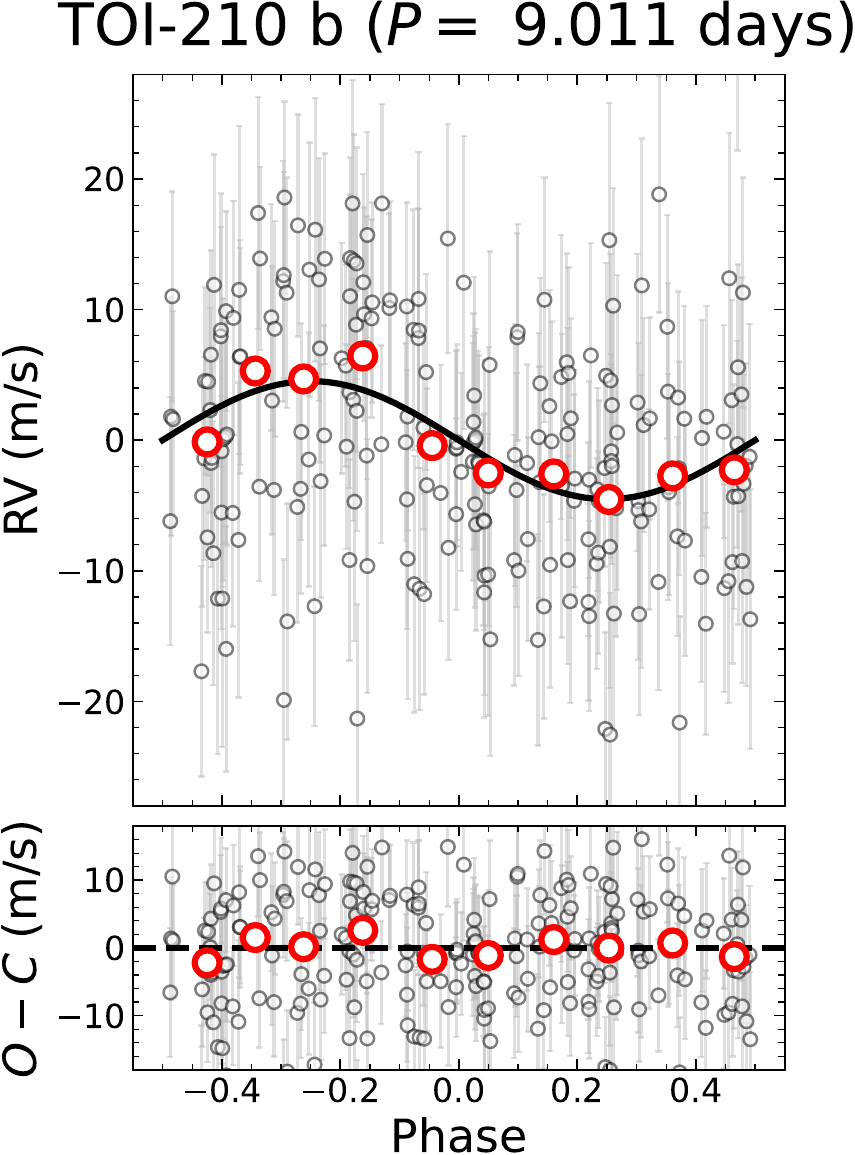}
    \endminipage
    \hfill
    \caption{Summary of the radial-velocity analysis. The best-fit model includes a multidimensional Gaussian process (GP) to account for stellar activity ($P_{\rm rot} = 71^{+6}_{-3}$\,days), three Keplerian signals at 2.15\,d ($3.49 \pm 0.87$\,m\,s$^{-1}$), 3.76\,d ($3.93 \pm 0.93$\,m\,s$^{-1}$) and 9.01\,d ($4.52 \pm 0.84$\,m\,s$^{-1}$), and a polynomial detrending with respect to BERV (see Sect.~\ref{sec:rv_analysis} for details). In each self-titled panel, we show the corresponding component of the full model, with the other components subtracted.}
    \label{fig:rv_analysis}
\end{figure*}

We compared the Bayesian evidence ($\mathcal{Z}$) of an activity-only model (0p) and a model including TOI-210\,b on a circular orbit (1cp). The transiting planet is robustly detected by NIRPS ($K_{\rm b} = 4.43 \pm 0.90$\,m\,s$^{-1}$), increasing the log-Bayesian evidence by $\Delta \ln \mathcal{Z}=10.3$. Allowing for an eccentric orbit (1ep) was marginally disfavored compared to 1cp ($\Delta \ln \mathcal{Z}=-0.5$), providing no evidence for a non-zero eccentricity. From this model, we derived a $2\sigma$ upper limit of $e < 0.38$ for TOI-210\,b. The periodogram of the residual RVs revealed peaks at tentative periods of 2.15\,d and 3.76\,d, along with their respective 1-d aliases (Fig.~\ref{fig:periodogram_residuals}). We explored additional models to assess whether these signals could be attributed to additional planetary companions.

We first performed a blind search for either an ultra-short period (USP) planet with $P \sim \mathcal{LU}(0.3, 1)$\,days or an inner planet with $P \sim \mathcal{LU}(1, 9)$\,days. Here, we parameterised the orbital phase as $t_0 = P \times \phi + 2460630$\,BJD, with $\phi \in [-0.5, 0.5]$, to avoid multi-modal solutions in $t_0$. The USP search converged to a primary period 0.44\,d and a secondary peak at 0.79\,d (both 1-d alias of 3.76\,days), but is disfavored relative to the single-planet model ($\Delta \ln \mathcal{Z} = -0.7$). The inner planet search, on the other hand, marginally increased $\ln \mathcal{Z}$ by 0.7, with the highest peak at 3.76\,d and secondary mode at 2.15\,d. Based on these results, we conclude that the NIRPS data cannot robustly confirm an additional planet in the TOI-210 system. 

We next performed targeted searches around the promising peaks 2.15\,d and 3.76\,d with normal priors $\mathcal{N}(P_{\rm peak}, 0.1)$\,d. Here, we sample $t_0$ uniformly between 2460630 and 2460630 + $1.2 \times P_{\rm peak}$\,BJD. The two-planet model with 2.15\,d increased $\ln \mathcal{Z}$ by $+1.7$, while the 3.76\,d case increased it by $+3.3$. A three-planet model (2.15, 3.76, 9.01\,d) provides the highest Bayesian evidence of all models explored ($\Delta \ln \mathcal{Z} = +5.9$). The Bayesian evidence values for each model are reported in Table~\ref{table:RV_model_comparison}. Based on these numbers, we consider that the NIRPS data provide moderate evidence for at least one Keplerian signal with a period below that of TOI-210\,b. We refer to the signals at 2.15\,d and 3.76\,d as Candidate 1 and 2, respectively, as none of these additional planet models increase the log-Bayesian evidence by the nominal value of 6 required to robustly claim a detection \citep{Trotta_2008, Thorngren_2026}. These candidate signals correspond to minimum masses of $3.2 \pm 0.8$\,M$_{\oplus}$ and $4.4 \pm 1.0$\,M$_{\oplus}$, for Candidate 1 and 2, respectively. 

We adopted the three-planet model (3cp) with the highest $\ln \mathcal{Z}$ to derive the planetary parameters of TOI-210\,b, while noting that all models yield fully consistent mass estimates. We provide the full prior and posterior values in Table~\ref{table:multi_dim_gp}. Figure~\ref{fig:rv_analysis} shows the full model, with panels representing the stellar activity modelling, candidate signals, the detection of TOI-210\,b, and the BERV detrending. The periodogram of the NIRPS RVs highlighting the candidate periods and aliases is available in Fig.~\ref{fig:periodogram_residuals}.

All models considered converged to a well-constrained rotation period $P_{\rm rot} = 71^{+6}_{-3}$\,days. We measure statistically significant $\alpha_{\rm RV} = -2.4^{+0.9}_{-1.0}$\,m\,s$^{-1}$ and $\beta_{\rm RV} = 14.0^{+5.5}_{-4.9}$\,m\,s$^{-1}$\,d$^{-1}$, showing sensitivity to both $F$ and $F^{\prime}$ components in the RV constrained principally by d\textit{Temp}. In addition, $\alpha_{\rm dLW}$ is positive like $a_{\textrm{d}Temp}$, with $\beta_{\rm dLW}$ consistent with 0. This could either indicate that we lack precision to detect the $F^{\prime}$ component in dLW or that dLW is principally a photometric proxy for TOI-210. The activity components accounts for 30--40\% of variance in dLW and d\textit{Temp}, compared to only 8\% in RV, highlighting how the multi-GP approach can capture correlated stellar noise that would otherwise be difficult to model in RVs alone.

To further test the planetary interpretation of the candidate signals, we carried out several complementary analyses. A systematic search for transit signals that may have been missed by the TESS pipeline is presented in Appendix~\ref{sec:transit_search}. We find no evidence that either Candidates 1 or 2 transits TOI-210. If these candidates are genuine planets, avoiding a transiting geometry requires mutual inclinations of $\Delta i > 2.3^{\circ}$ and $\Delta i > 3.5^{\circ}$ relative to TOI-210\,b for Candidates 2 and 1, respectively. We further investigated whether the gravitational perturbations induced by these candidate planets could generate detectable transit timing variations (TTVs) for TOI-210\,b (Appendix~\ref{sec:ttv_search}). The predicted TTV amplitudes remain below the timing precision, and we find no evidence for deviations from a linear ephemeris. Finally, we performed injection--recovery tests to quantify the sensitivity of the NIRPS data to planetary signals and to assess any attenuation or amplification introduced by the activity GP over periods ranging from 0.3 to 1000\,days. These tests presented in Appendix~\ref{sec:sensitivity_map} confirm that the NIRPS observations are sensitive to signals down to the 3\,m\,s$^{-1}$ level.

\begin{table}[h]
\caption{\label{table:RV_model_comparison}Bayesian model comparison of the NIRPS RV analysis of TOI-210}
\centering
\renewcommand{\arraystretch}{1.25} 
\begin{tabular}{lcccc}
\hline\hline
Model & $N_{\rm param}$ & $\ln \mathcal{Z}$ & $\Delta \ln \mathcal{Z}$\\
\hline
{[0p]} Activity GP only & 20 & $-2139.5$ & $-10.3$\\
{[1cp]} 9.01d & 23 & $-2129.2$ & 0\\
{[1ep]} 9.01d & 25 & $-2129.7$ & $-0.5$\\
{[2cp]} $x_{\rm USP}$d + 9.01d & 26 & $-2129.9$ & $-0.7$\\
{[2cp]} $x_{\rm inner}$d + 9.01d & 26 & $-2128.5$ & $+0.7$\\
{[2cp]} 2.15d + 9.01d & 26 & $-2127.5$ & $+1.7$\\
{[2cp]} 3.76d + 9.01d & 26 & $-2126.4$ & $+3.3$\\
{[3cp]} 2.15d + 3.76d + 9.01d & 29 & $-2123.8$ & $+5.9$\\
\hline
\end{tabular}
\vspace{0.1cm}
\caption*{\footnotesize {\bf Notes.} The uncertainty in $\ln \mathcal{Z}$ is approximately 0.2, based on repeated runs.}
\end{table}

\section{Discussion} \label{sec:discussion}

\subsection{Planetary parameters and bulk composition of TOI-210\,b}

We report a radius of \radiusb\ and a mass of \massb\ for TOI-210\,b, corresponding to respective relative precisions of 3\% and 18\%. Together, these measurements yield a bulk density of $\rho = 3.3 \pm 0.7$\,g\,cm$^{-3}$. The precision in density enables a first investigation of the planet’s internal structure \citep{Plotnykov_2024}, which we present in Sect.~\ref{sec:internal_structure}. TOI-210\,b receives about three times the bolometric instellation of the Earth ($S = 3.1 \pm 0.4$\,S$_{\oplus}$). Assuming a zero Bond albedo and homogeneous heat redistribution, this irradiation level corresponds to an equilibrium temperature of \teqb. The orbital, transit, and physical parameters of TOI-210\,b derived in this work are reported in Table~\ref{table:derived_params}.

\begin{table}[h]
\caption{\label{table:derived_params}Orbital, transit, and physical parameters of TOI-210\,b}
\centering
\renewcommand{\arraystretch}{1.25} 
\begin{tabular}{lc}
\hline\hline
Parameter & TOI-210\,b\\
\hline
\multicolumn{2}{c}{\textit{Orbital parameters}}\\[0.05cm]
$P$ (days) &  9.0105563 $\pm$ 0.0000028\\
$t_{\rm 0}$ (BJD\,$-$\,2457000) &  2339.00523 $\pm$ 0.00024\\
$a$ (au) &  0.0579 $\pm$ 0.0027\\
$i$ ($^{\circ}$) &  89.54$^{+0.25}_{-0.18}$\\
$e$ &  $<0.38$\,(2$\sigma$)\\[0.05cm]
\multicolumn{2}{c}{\textit{Transit parameters}}\\[0.05cm]
$b$ &  0.30$^{+0.10}_{-0.16}$\\
$\delta$ (ppt) &   3.83 $\pm$ 0.10\\
$t_{14}$ (hours) &  1.85 $\pm$ 0.08\\[0.05cm]
\multicolumn{2}{c}{\textit{Physical parameters}}\\[0.05cm]
$R_{\rm p}$ (R$_{\oplus}$) &  2.234 $\pm$ 0.074\\
$M_{\rm p}$ (M$_{\oplus}$) & 6.75 $\pm$ 1.25\\
$\rho$ (g\,$\cdot$\,cm$^{-3}$) &  3.3 $\pm$ 0.7\\
$S$ (S$_{\oplus}$) &  3.1 $\pm$ 0.4\\
$T_{\textrm{eq},A_{\rm B} = 0}$ (K) & 368 $\pm$ 9\\
\hline
\end{tabular}
\end{table}

\begin{figure}[h]
    \centering
    \includegraphics[width=1\linewidth]{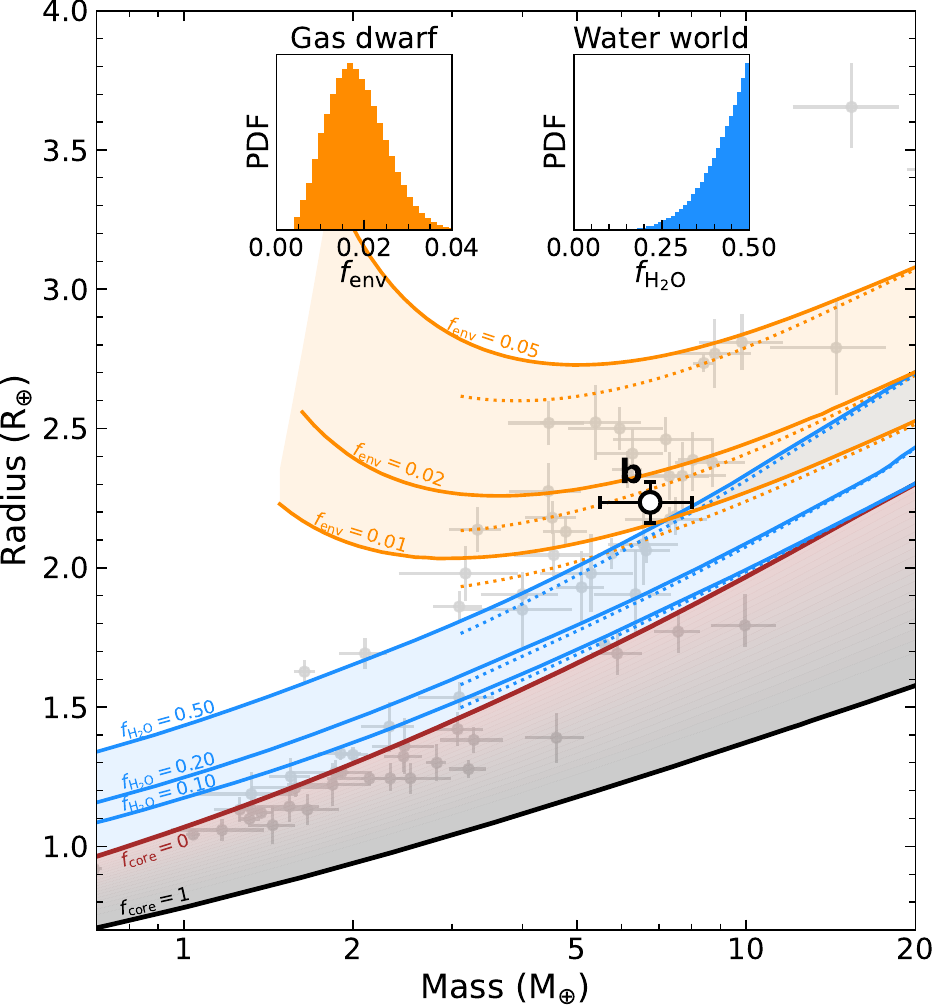}
    \caption{Mass--radius diagram of well-characterised small exoplanets around M dwarfs from the \textit{PlanetS} catalog \citep{Otegi_2020, Parc_2024}. The solid lines show planetary composition curves from \cite{Skinner_2026}. The H$_2$O-rich and H/He-rich models assume an Earth-like solid interior ($f_{\rm core}=0.325$) and are computed for a temperature of 500\,K and a Bond albedo of 0.3. Dotted curves show equivalent water mass fractions ($f_{\rm H_{2}O}$) and envelope mass fractions ($f_{\rm env}$) from a custom grid computed at 376\,K. The shaded regions delineate the mass--radius regimes corresponding to rocky super-Earths (black--red gradient), water worlds (blue), and gas dwarfs (orange). The mass and radius of TOI-210\,b are best explained by a $f_{\rm env} = 0.018^{+0.007}_{-0.006}$ gas dwarf. A water world scenario requires $f_{\rm H_{2}O} > 0.29$ (2$\sigma$) to explain the low density.}
    \label{fig:MR}
\end{figure}

We place TOI-210\,b in a mass--radius diagram (Fig.~\ref{fig:MR}) populated with entries of M-dwarf transiting exoplanets from the \textit{PlanetS} catalog \citep{Otegi_2020, Parc_2024} with mass and radius precisions better than 25\% and 8\%, respectively. The population of small exoplanets around M dwarfs is compared to theoretical models from \cite{Skinner_2026} for rocky, \hbox{water-,} and gas-rich compositions, computed at $T_{\rm eq}= 500$\,K. Here, the models output is a `transit radius' corresponding to the photospheric radius ($\tau = 2/3$ optical depth for a transiting ray of light), enabling a direct comparison with $R_{\rm p}$ inferred from transit observations \citep{Haldemann_2024,Skinner_2026}.

In Fig.~\ref{fig:MR}, we delineate three regions of parameter space corresponding to rocky planets (black--red gradient), water worlds (blue), and gas dwarfs (orange). These categories are intended as interpretative regimes for planetary compositions rather than strict physical boundaries. The limit between rocky and water-rich corresponds to the pure silicate line ($f_{\rm core} = 0$), above which the lower density cannot be explained by rocks and metals only. Water enrichment in the condensed material of protoplanetary discs is not expected to exceed 50\% by mass (e.g., \citealt{Marboeuf_2014}), an expectation borne out by Solar System bodies \citep{Lodders_2003}. Thus, planets with radii above that anticipated for water mass fractions $f_{\rm H_{2}O} = 0.50$ are expected to have an extended H/He envelope. Gas accretion models typically predict primordial envelopes with mass fractions of $f_{\rm env} \gtrsim 1\%$ \citep{Ginzburg_2016}, which we adopted as the lower end of the gas-dwarf parameter space.

The position of TOI-210\,b in Fig.~\ref{fig:MR} suggests two possible alternatives for the nature of this temperate sub-Neptune: (1) a gas dwarf with percent-level hydrogen and helium by mass, and (2) a water-rich planet with a large water mass fraction, comparable to or exceeding that of icy Solar System bodies ($\sim$50\% H$_2$O). It is possible that reality lies between these two extremes, with a well-mixed supercritical envelope containing significant fractions of both H/He and H$_2$O. In this case, the inferred envelope mass fraction is an upper limit, reflecting a scenario in which the planet formed completely dry inside the snow line or all primordial water was chemically destroyed. 

Other explanations have been proposed in recent years for the low densities of small planets orbiting M dwarfs. For example, a substantial fraction of a planet's water inventory may be incorporated into its interior rather than residing in a distinct outer volatile layer \citep{Luo_2024,Weisserman_2026}. More exotic scenarios have also been suggested, including carbon-rich `soot planets' \citep{Li_2026} and planets whose thin H/He envelopes become partially miscible with a molten silicate interior \citep{Young_2026}. Given the relatively modest 18\% precision on the mass of TOI-210\,b, distinguishing among these possibilities is not currently feasible.

\subsection{Internal structure analysis} \label{sec:internal_structure}

To interpret the bulk density of TOI-210\,b, we investigated the gas-dwarf and water-world scenarios discussed above, assuming a differentiated internal structure in both cases. Our goal is to constrain three key interior parameters: the core mass fraction ($f_{\rm core}$), the water mass fraction ($f_{\rm H_{2}O}$), and the H/He envelope mass fraction ($f_{\rm env}$). For this exercise, we coupled a Markov chain Monte Carlo (MCMC) sampler to a precomputed grid of planetary models, with grid spacings of 5\% in $f_{\rm core}$ and $f_{\rm H_{2}O}$ and 0.5\% in $f_{\rm env}$. The grid was generated following the description in \cite{Skinner_2026}, with parameters tailored to TOI-210\,b, setting a zero-albedo $T_{\rm eq}$ of 376 K and a rotation period of 9.01\,days (under the assumption of tidal locking). All other input parameters are either varied or set following Table 1 in \cite{Skinner_2026}. The planet is modelled with three layers: a predominantly iron core, a silicate mantle, and either a water layer or a H/He layer with a solar He fraction. For a given mass and set of compositional fractions ($f_{\rm core}$, $f_{\rm H_{2}O}$, $f_{\rm env}$), the model returns the corresponding transit radius. Example models from the custom grid are shown as dotted curves in Fig.~\ref{fig:MR}.

Water on TOI-210\,b is expected to be predominantly in the steam phase at low pressures and in the supercritical state at high pressures, according to 3D planetary climate models \citep{Turbet_2023}. Interior models that assume fully condensed water will therefore tend to overestimate its $f_{\rm H_{2}O}$ (e.g., \citealt{Zeng_2019, Plotnykov_2020}), while models that neglect the contraction of planetary atmospheres as they cool will underestimate it (e.g., \citealt{Aguichine_2021}). By incorporating both steam/supercritical water and atmospheric contraction, our models \citep{Skinner_2026} provide a more physically consistent estimate of $f_{\rm H_{2}O}$. We further compared the results below with the recent models of \cite{Aguichine_2025}, which include these effects, and find good agreement.

We performed the interior inference using the \texttt{smint} interpolation framework \citep{Piaulet_2021, Piaulet_2023}, adapted to incorporate our custom grid of planetary structure models. The planetary radius is treated as a datum with the uncertainty incorporated in the Gaussian likelihood calculations. We apply a Gaussian prior on $M_{\rm p}$ of $\mathcal{N}\left(6.75, 1.25\right)$\,M$_{\oplus}$. In \texttt{smint}, the sampled parameter is $f^{\prime}_{\rm core}$, the core mass fraction of the core+mantle interior only, excluding water or atmospheric layer. The total planetary core mass fraction is given by $f_{\rm core}=f^{\prime}_{\rm core}(1-f_{\rm env})$ for a gas dwarf or $f_{\rm core}=f^{\prime}_{\rm core}(1-f_{\rm H_{2}O})$ for a water world. We considered two prior cases for $f^{\prime}_{\rm core}$: (1) a free chemistry exploration with $\mathcal{U}(0, 0.8)$ with the upper bound reflecting the maximum iron-enrichment found in a sample of rocky super-Earths \citep{Plotnykov_2020}, and (2) a stellar chemistry prior informed by the abundance measurements of the host star. Recent studies have reported empirical correlations between stellar and planetary compositions \citep{Brinkman_2024, Brinkman_2025, Behmard_2025, Plotnykov_2026}, motivating the use of stellar abundances as an additional constraint on the planet’s interior. We converted the NIRPS abundance measurements (Table~\ref{table:stellar_abundances}) into an equivalent planetary core mass fraction using \texttt{exopie} \citep{Plotnykov_2024}. Because the Mg abundance was derived from only three spectral lines, resulting in a relatively uncertain measurement, we instead adopted Ti as a proxy for this $\alpha$ element, following \citet{Weisserman_2026}. We obtained a stellar chemisty prior $f^{\prime}_{\rm core} = 0.27 \pm 0.16$, compatible with Earth’s value of $f^{\prime}_{\rm core} = 0.325\pm0.003$ \citep{Wang_2018}. The priors on $f_{\rm H_{2}O}$ and $f_{\rm env}$ are uniform between 0 and 0.5, and between 0.0 and 0.06, respectively. The free parameters are $M_{\rm p}$, $f^{\prime}_{\rm core}$, and either $f_{\rm H_{2}O}$ or $f_{\rm env}$ (3 free parameters). For each model, the MCMC exploration used 100 walkers for 10\,000 steps, discarding the first 2\,000 samples as burn-in. The posterior distributions on $f_{\rm env}$ (gas dwarf) and $f_{\rm H_{2}O}$ (water world) for the stellar chemistry prior cases are shown in Fig.~\ref{fig:MR}.

For the gas-dwarf experiment, we obtained $f_{\rm env} = 0.022^{+0.010}_{-0.009}$ in the free chemistry case, and $f_{\rm env} = 0.018^{+0.007}_{-0.006}$ in the stellar prior case. The posterior on $f^{\prime}_{\rm core}$ returns the prior, reflecting the degeneracy inherent to this inference problem. A low mean molecular weight envelope is susceptible to atmospheric mass loss over time through photoevaporation driven by stellar irradiation and planetary internal heat. To provide a complementary analysis, we simulated the internal structure of TOI-210\,b in the gas-dwarf scenario using the \texttt{JADE} code \citep{Attia_2021,Attia_2025}, which considers the thermal evolution and atmospheric mass loss of planets under evolving stellar irradiation. A description of this modeling is provided in Appendix~\ref{appendix:jade} and we discuss our photoevaporation simulation in the next Sect.~\ref{sec:photoevaporation}. The results remains broadly unchanged: we derived $f_{\rm env} = 0.009^{+0.005}_{-0.007}$ using free chemistry prior and $f_{\rm env} = 0.005^{+0.003}_{-0.004}$ with $f^{\prime}_{\rm core}$ constrained by stellar abundances with the \texttt{JADE} retrievals.

For the water world scenario, we get 2$\sigma$ lower limits $f_{\rm H_{2}O} > 0.29$ for the free chemistry and $f_{\rm H_{2}O} > 0.29$ for stellar prior case, again with $f^{\prime}_{\rm core}$ constrained by the priors. In both cases, these limits reflect the imposed upper bound of 50\% water enrichment. In other words, extreme $f_{\rm H_{2}O}$ are required to explain the low density of TOI-210\,b. We conclude it most likely possesses an extended envelope of hydrogen and helium totaling $\sim$1\% of the mass, $\sim$25\% of the radius.

\subsection{Atmospheric evolution under photoevaporation} \label{sec:photoevaporation}

We simulated the atmospheric evolution of TOI-210\,b with \texttt{JADE} in the gas dwarf scenario to assess whether an atmosphere with $f_{\rm env} \approx 0.01$ could have survived to present-day. We followed closely the procedure detailed in \cite{Bourrier_2025}. 

First, a rotating stellar model representative of TOI-210 was computed with the Geneva stellar evolution code (GENEC; \citealt{Eggenberger_2008}) using stellar radius and effective temperature constraints given in Table~\ref{table:stellar_params}. This model accounts for the internal transport of angular momentum by hydrodynamic and magnetic instabilities (see e.g., \citealt{Eggenberger_2022}). Based on the structural and rotational profile from this model, the evolution of the high energy fluxes emitted by the host star is then computed (see \citealt{Pezzotti_2021}). The initial velocity of the star being unknown, we considered the case of a moderate rotator with an initial rotation rate of $5 \times \Omega_{\odot}$ (see \citealt{Eggenberger_2019}). This rotational history then determines the evolution of X-ray and the corresponding extreme ultraviolet (EUV) luminosities, assuming the X-ray to EUV scaling relation provided by \cite{Sanz-Forcada_2011}.

\begin{figure}
    \centering
    \includegraphics[width=\linewidth]{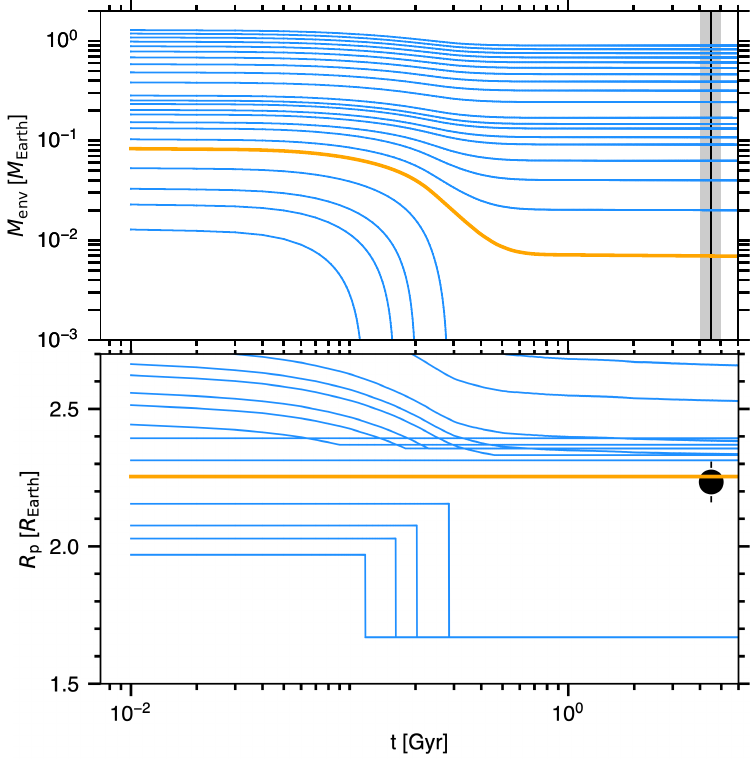}
    \caption{Atmospheric evolution of TOI-210\,b under photoevaporation simulated with \texttt{JADE}, highlighting the envelope mass (top panel) and the radius (bottom panel). The orange simulation best matches the present-day radius (black point) at the system age ($\sim$4.5\,Gyr), in an intermediate regime between full erosion of light envelopes and more massive envelopes swelling the radius.}
    \label{fig:JADE_evol}
\end{figure}

We then ran \texttt{JADE} simulations over a grid of initial planet envelope mass ($f_{\rm env,0}$), from the expected dissipation of the protoplanetary disk (10\,Myr) up to 5\,Gyr. System properties were fixed to their present-day values from Table~\ref{table:derived_params} and to the results of the internal structure retrieval (Sect.~\ref{sec:internal_structure}). The stellar luminosity curve was set to the model computed with \textsc{GENEC}, controlling the temperature profile of the planetary envelope together with internal heating, and its erosion under EUV-driven photoevaporation. The results of the simulations are presented in Fig.~\ref{fig:JADE_evol} where we observe three different regimes of evolution. Low-mass initial envelope ($f_{\rm env,0} \approxinf 1$\%) are fast eroded, leaving a bare solid core too small to reproduce the present planet radius. More massive envelopes ($f_{\rm env,0} \approxsup 1$\%) are resilient to erosion over the life of the planet, but swell the radius beyond the observed present-day value. We found that it was possible to explain the observed radius with an intermediate envelope mass of $f_{\rm env,0}\approx1$\%, as long as the atmosphere is not metal-rich to prevent the envelope expanding too much under bolometric irradiation. In these conditions, the envelope loses about half of its mass to photoevaporation, retaining $\sim$0.1\% of the total mass, while its radius remains stable across its lifetime. These results suggest that TOI-210\,b is at the transition between sub-Neptunes with gas-rich and water-rich envelopes.

\subsection{Prospects for atmospheric characterisation}

The James Webb Space Telescope (JWST; \citealt{Gardner_2023}) has already provided a first glimpse into the atmospheres of small M-dwarf planets across a wide range of planetary sizes, from sub-Earth objects like L~98-59\,b that may host a volcanic SO$_2$-rich atmosphere \citep{Bello-Arufe_2025}, to super-Earths including L 98-59\,d with hints of sulfur-rich atmosphere driven by photochemistry \citep{Gressier_2024} and LHS~1140\,b with a potentially N$_2$-rich atmosphere \citep{Cadieux_2024b, Damiano_2024}. In the sub-Neptune regime, most planets characterised so far either show strong molecular features (e.g., \citealt{Madhusudhan_2023, Benneke_2024, Rigby_2025}) or muted spectra consistent with high-altitude aerosols (e.g., \citealt{Kempton_2023, Ahrer_2025}), with a tantalising trend emerging that temperate planets with $T_{\rm eq} \lesssim 400$\,K may preferentially host clearer atmospheres \citep{Yu_2021, Brande_2024}. However, the recent discovery that the temperate sub-Neptune LP~791-18\,c ($T_{\rm eq} \approx 355$\,K) has a hazy atmosphere \citep{Roy_2026} suggests that temperature is not the sole driver that can shift a planet from a clear to a haze-dominated state.

TOI-270\,d ($R_{\rm p}$\,$\approx$\,2.00\,R$_{\oplus}$, $M_{\rm p}$\,$\approx$\,4.20\,M$_{\oplus}$, $T_{\rm eq}$\,$\approx$\,383\,K; \citealt{Kaye_2022}) and LP~791-18\,c ($R_{\rm p}$\,$\approx$\,2.49\,R$_{\oplus}$, $M_{\rm p}$\,$\approx$\,7.16\,M$_{\oplus}$, $T_{\rm eq}$\,$\approx$\,355\,K; \citealt{Greklek-McKeon_2025}) provide two key reference points in this emerging dichotomy. Despite broadly similar bulk properties and irradiation levels, they exhibit strikingly different transmission spectra. TOI-270\,d shows strong signatures from CH$_4$, CO$_2$, and likely H$_2$O, indicative of substantial metal enrichment relative to a primordial H/He envelope \citep{Benneke_2024, Constantinou_2026}. LP~791-18\,c is characterised by a prominent Rayleigh scattering slope from photochemical hazes that mute all spectral features other than CH$_4$ \citep{Roy_2026}. These contrasting atmospheric outcomes may reflect different formation pathways, with TOI-270\,d potentially originating from beyond the ice line as a water-rich world, and LP~791-18\,c likely forming under drier conditions, resulting in a less oxidised envelope that favors H$_2$ photochemistry \citep{Roy_2026}. An alternative explanation lies in the stellar environments: TOI-270 is an earlier M2V dwarf, whereas LP~791-18 is a fully-convective, more active, M6V star, suggesting that higher high-energy fluxes may push a planet toward a hazy state \citep{Morley_2015, Gao_2023}. Given the small number of well-characterised temperate sub-Neptunes with available transmission spectroscopy, each new system contributes to our understanding of the observed atmospheric diversity.

\begin{figure}
    \centering
    \includegraphics[width=1\linewidth]{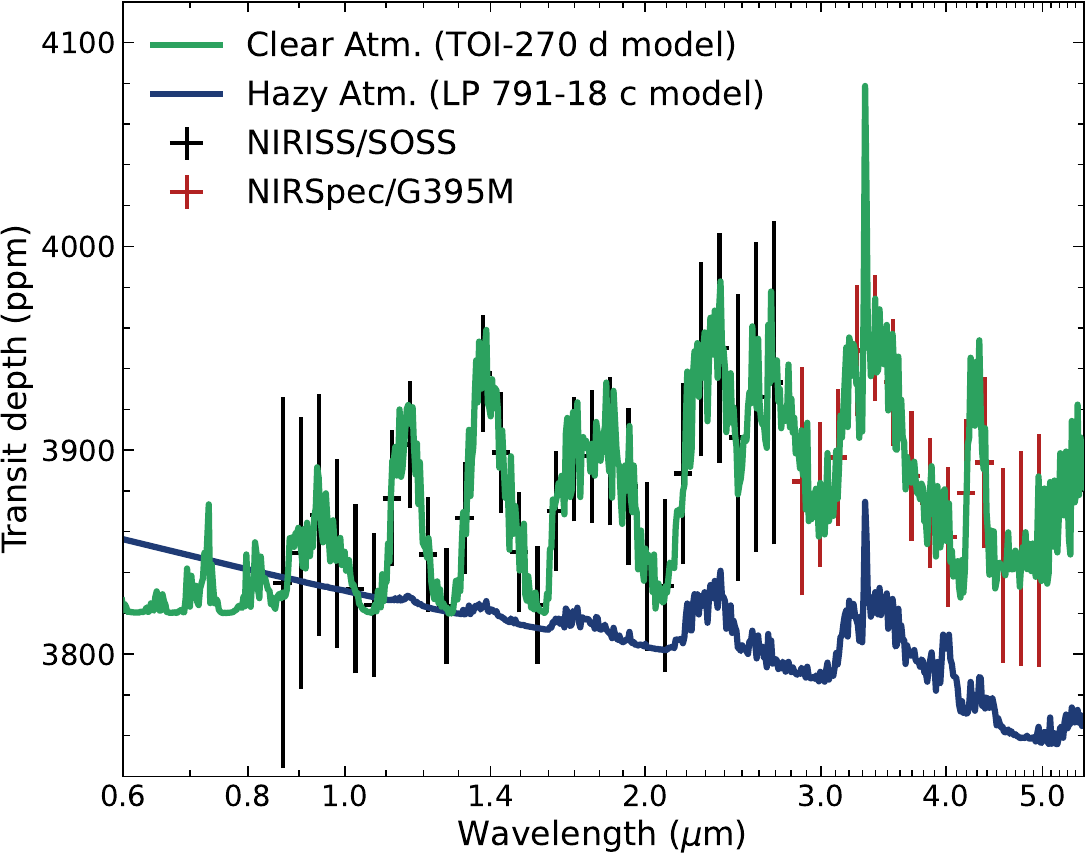}
    \caption{Synthetic transmission spectrum of TOI-210\,b with \texttt{SCARLET} based on the maximum a posteriori models for TOI-270\,d (in green; \citealt{Benneke_2024}) and LP~791-18\,c (in blue; \citealt{Roy_2026}). The single-transit precision at $R\approx 25$ computed with \texttt{PandExo} \citep{Batalha_2017} is shown as 1$\sigma$ error bars for NIRISS/SOSS (black) and NIRSpec/G395M (red).}
    \label{fig:transmission_spectrum}
\end{figure}

We evaluated the prospects for atmospheric characterisation of TOI-210\,b with JWST. TOI-210\,b shares similar mass, radius, and temperature with TOI-270\,d and LP~791-18\,c, with an intermediate M4V stellar type. We simulated realistic transmission spectra for TOI-210\,b based on the published maximum a posteriori atmospheric models of TOI-270\,d and LP~791-18\,c. The forward model spectra were generated with \texttt{SCARLET} \citep{Benneke_2012, Benneke_2013, Benneke_2015, Benneke_2019a, Benneke_2019b, Roy_2022, Roy_2023, Coulombe_2023, Piaulet-Ghorayeb_2024} by taking the best-fitting chemical volume-mixing ratios and clouds/hazes properties from these models and using TOI-210\,b's planetary parameters (Table~\ref{table:derived_params}). In Fig.~\ref{fig:transmission_spectrum}, we compare the synthetic spectra with the expected single-transit precision of NIRISS/SOSS \citep{Doyon_2023} and NIRSpec/G395M \citep{Boker_2023}, computed with \texttt{PandExo} \citep{Batalha_2017}. At a spectral resolution of $R\approx 25$, the peak precision of NIRISS (27\,ppm at 1.3\,$\mu$m) and NIRSpec (31\,ppm at 3.5\,$\mu$m) is sufficient to detect spectral features in both scenarios. JWST transmission spectroscopy of TOI-210\,b would constrain its atmospheric composition and provide a direct comparison with TOI-270\,b and LP~791-18\,c, testing whether this temperate sub-Neptune falls in the clear or hazy aerosol regime.

\subsection{RV challenges around faint M dwarfs}

Until now, TOI-210\,b was one among thousands of transiting exoplanet candidates identified by TESS awaiting a mass measurement to confirm its planetary nature. This situation reflects a broader observational challenge for faint M dwarfs ($V>14$\,mag, $H>9$\,mag), for which optical spectrographs often struggle to achieve the RV precision required for planet confirmation and characterisation. NIRPS is specifically designed to probe this regime, as exemplified by the detection of TOI-756\,b \citep{Parc_2025}, TOI-4666\,b \citep{Frensch_2026}, TOI-672\,b \citep{Osborn_2026}, and TOI-4552\,b \citep{Srivastava_2026a}.

\begin{figure}[h]
    \centering
    \includegraphics[width=1\linewidth]{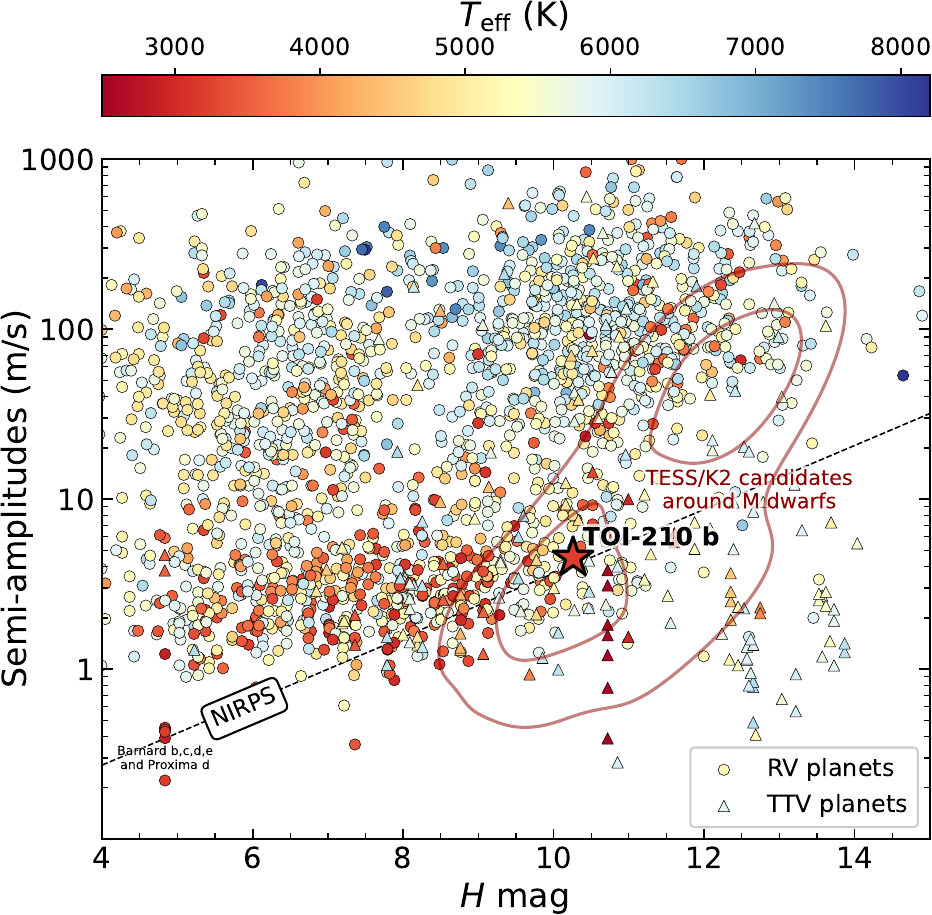}
    \caption{RV semi-amplitude ($K$) of confirmed exoplanets as a function of the $H$ magnitude of the host star and color-coded by stellar effective temperature. Circles and triangles denote RV and TTV detections, respectively. The red contours trace the 1- and 2-$\sigma$ levels of TESS/K2 candidates around M dwarfs without mass measurements, with their $K$ estimated using the mass--radius relations of \cite{Parc_2024}. TOI-210\,b lies in a sparsely populated region of this parameter space, where only a few M-dwarf planets have mass constraints. The black dashed line indicates the NIRPS precision for a 1800\,s exposure \citep{Bouchy_2025}; systems above this threshold are well suited for NIRPS follow-up, while those along it require intensive RV campaigns ($N \gtrsim 50$) to achieve $\sim$20\% mass precision \citep{Cloutier_2018}.}
    \label{fig:nirps_parameter_space}
\end{figure}

Figure~\ref{fig:nirps_parameter_space} places TOI-210\,b within the known exoplanet population from the NASA Exoplanet Archive \citep{Christiansen_2025} by showing RV semi-amplitude as a function of host-star $H$ magnitude and stellar effective temperature. TOI-210\,b lies near the peak of the distribution of K2 and TESS M-dwarf planet candidates that still lack mass measurements because of these observational challenges. This figure illustrates the capability of NIRPS to characterise transiting sub-Neptunes and giant planets around faint mid-to-late M dwarfs that remains largely inaccessible to current optical spectrographs.


The PLATO mission \citep{Rauer_2025}, currently scheduled for launch in early 2027, will target bright Sun-like stars in a long-pointing field near the southern ecliptic pole (LOPS2; \citealt{Nascimbeni_2025}) to detect Earth analogs. A dedicated sample of M dwarfs, the P4 sample, was also selected within the same field \citep{Prisinzano_2026}. Simulations predict that this sample alone could yield up to 464 new planets with $R_{\rm p} < 4\,R_{\oplus}$ after two years of observations \citep{Cabrera_2026}. These M dwarfs in the LOPS2 have a $V < 16$\,mag, corresponding approximately to $H < 12$ depending on spectral subtype \citep{Pecaut_2013}. As shown in Fig.~\ref{fig:nirps_parameter_space}, NIRPS will therefore be well-positioned for the RV follow-up of PLATO candidates orbiting M dwarfs.

\section{Conclusions} \label{sec:conclusions}

In this paper, we presented the discovery and characterisation of a new planetary system around the M4V dwarf TOI-210. The star hosts a transiting sub-Neptune (TOI-210\,b) on a 9.01-day orbit that was first identified by TESS in 2019. The mass determination was enabled by the NIRPS spectrograph working in the near-infrared, where cool dwarfs like TOI-210 emit more light. As a comparison, simultaneous HARPS observations obtained with the same exposure time and RV extraction technique were 4$\times$ less precise (23\,m\,s$^{-1}$ compared to 5.25\,m\,s$^{-1}$ for NIRPS).

An analysis of 40 sectors from TESS combined with follow-up ground-based transit photometry from LCOGT and ExTrA constrained the radius of TOI-210\,b to \radiusb. The modeling of the NIRPS RVs including a multidimensional GP to account for stellar activity yielded a planetary mass of \massb. The preferred RV solution is one with two additional Keplerian signals at 2.15 and 3.76\,days, with minimum masses 3.2$\pm$0.8\,M$_{\oplus}$ and 4.4$\pm$1.0\,M$_{\oplus}$, respectively, increasing moderately the Bayesian evidence by $\Delta \ln \mathcal{Z} = 5.9$ compared to a single planet solution. These signals are considered candidate planets, as they do not reach statistical significance to be confirmed individually, nor do they emerge robustly through a blind search with broad log-uniform prior between 0.3 and 9\,days. No transits at the candidate periods are found in TESS photometry, despite sufficient sensitivity.

From the mass--radius observations of TOI-210\,b, we investigated two plausible internal structures: an Earth-like interior surrounded by either an H/He envelope (gas dwarf) or an H$_2$O layer (water world). Using recent advancements in planetary interior models that account for the cooling and contraction of supercritical water envelopes \citep{Skinner_2026}, we showed that the density of TOI-210\,b (3.3 $\pm$ 0.7\,g\,cm$^{-3}$) is best explained by the gas dwarf scenario. We measured an H/He envelope mass fractions of $f_{\rm env} \approx 1$\% regardless of assumptions for the core mass fraction, whether free or constrained by stellar abundances of refractory elements measured by NIRPS. We verified through photoevaporation simulation with \texttt{JADE} that such envelope can survive to present-day given the modest irradiation received by TOI-210\,b ($T_{\rm eq} = 368 \pm 9$\,K).

Recent JWST observations of temperate sub-Neptunes with $T_{\rm eq} \lesssim 400$\,K have revealed a diversity of atmospheric properties, ranging from relatively clear atmospheres to those obscured by high-altitude aerosols. With its well-constrained mass, radius, and equilibrium temperature, TOI-210\,b is a prime target for follow-up atmospheric characterisation with JWST to extend the currently limited survey in the temperate regime. The host star lies near JWST's Southern Continuous Viewing Zone facilitating the scheduling of time-critical transit observations at almost any moment throughout the year. Exposure time calculations predict that a single transit with NIRISS/SOSS or NIRSpec/G395M would be sufficient to detect molecular features, even in the presence of aerosols.\\

\noindent \textit{Acknowledgements can be found in Appendix~\ref{sec:acknowledgements}}

%

\bibliographystyle{bibtex/aa.bst}
\bibliography{TOI210}


\begin{appendix}

\section{TESS light curves}

TESS has monitored the star TOI-210 over seven years, spanning 40 sectors so far (Sect.~\ref{sec:tess}). Figure~\ref{fig:tess_gp} presents a mosaic of the full \texttt{PDCSAP} light curve of TOI-210, with each panel corresponding to an individual sector.

\section{SOAR imaging contrast curve}

\begin{figure}[h!]
    \centering
    \includegraphics[width=1\linewidth]{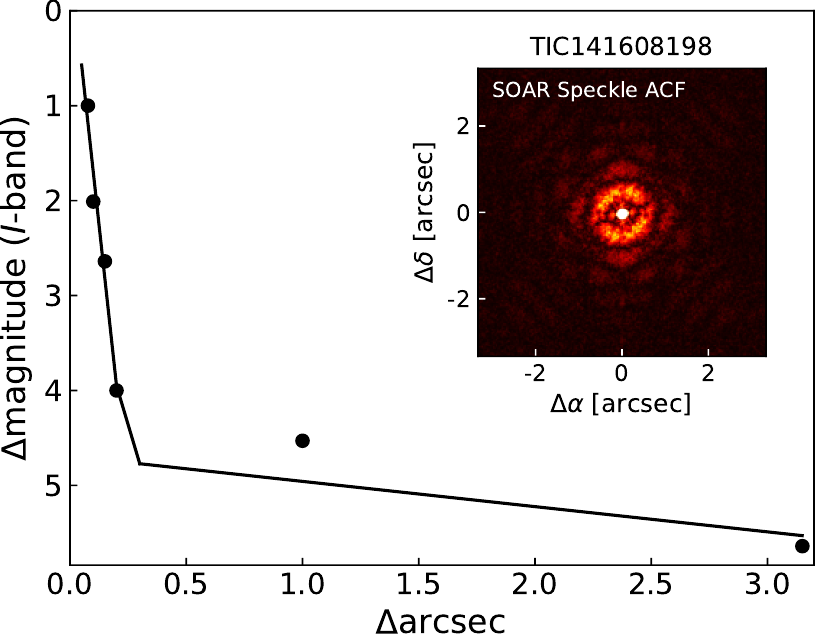}
    \caption{High-resolution speckle imaging 5$\sigma$ contrast curve of the star TIC~141608198 (TOI-210) from SOAR/HRCam (4.1\,m, $I$ band). We rule out close companions with a contrast below $\Delta I = 4.00$\,mag at $0\farcs2$ and 4.53\,mag at $1\arcsec$.}
    \label{fig:imaging}
\end{figure}

The 5$\sigma$ contrast curve derived from high-resolution SOAR imaging (Sect.~\ref{sec:soar}) is presented in Fig.~\ref{fig:imaging}.

\renewcommand{\thefigure}{A.\arabic{figure}}
\setcounter{figure}{0}
\begin{figure*}[h!]
  
  \minipage{0.225\textwidth}
  \centering
  \vspace{-0.05cm}
  \includegraphics[width=1\linewidth]{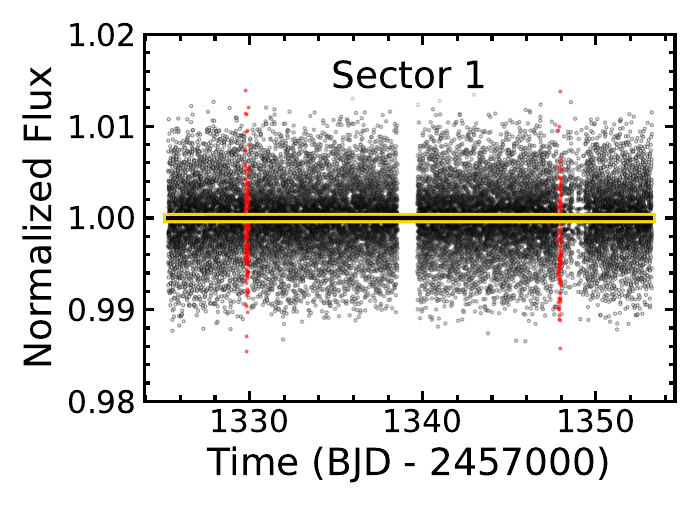}
  \endminipage\hfill
  \minipage{0.193\textwidth}
  \centering
  \includegraphics[width=1\linewidth]{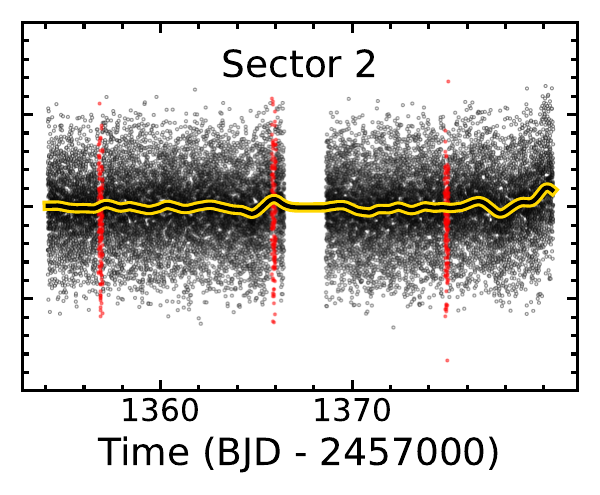}
  \endminipage\hfill
  \minipage{0.193\textwidth}
  \centering
  \includegraphics[width=1\linewidth]{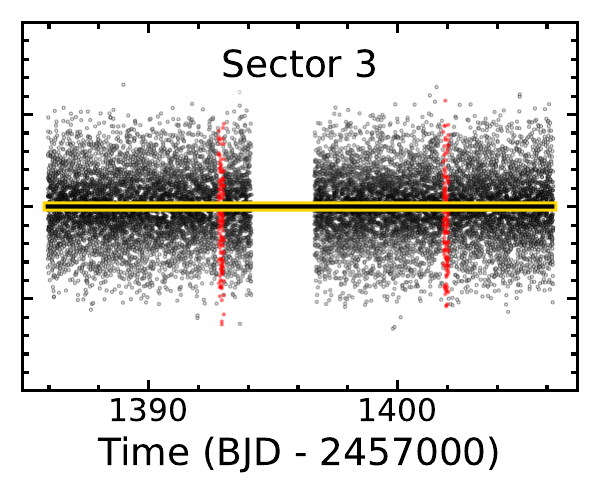}
  \endminipage\hfill
  \minipage{0.193\textwidth}
  \centering
  \includegraphics[width=1\linewidth]{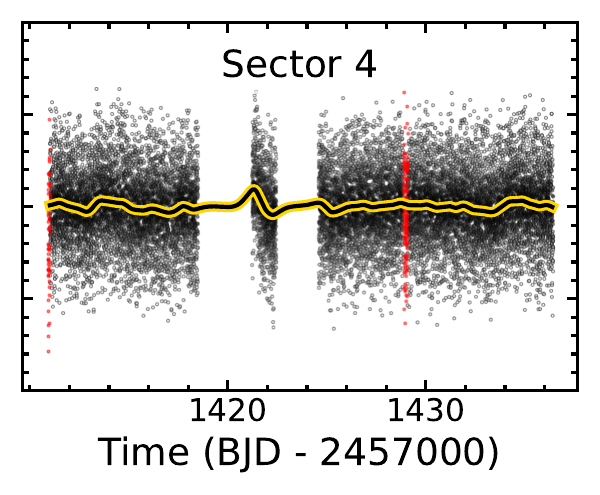}
  \endminipage\hfill
  \minipage{0.193\textwidth}
  \centering
  \includegraphics[width=1\linewidth]{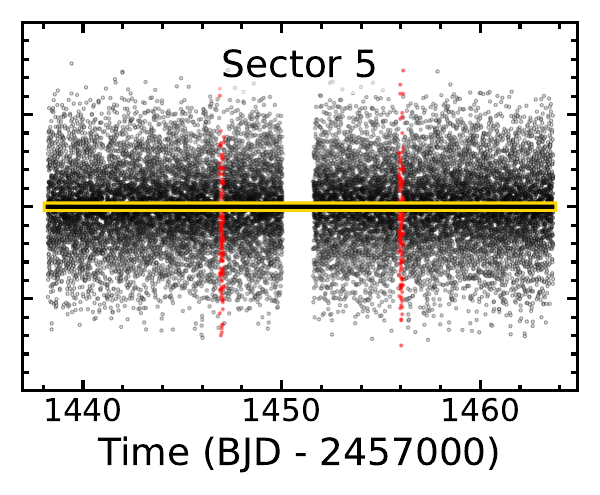}
  \endminipage\hfill\\

  \minipage{0.225\textwidth}
  \centering
  \vspace{-0.05cm}
  \includegraphics[width=1\linewidth]{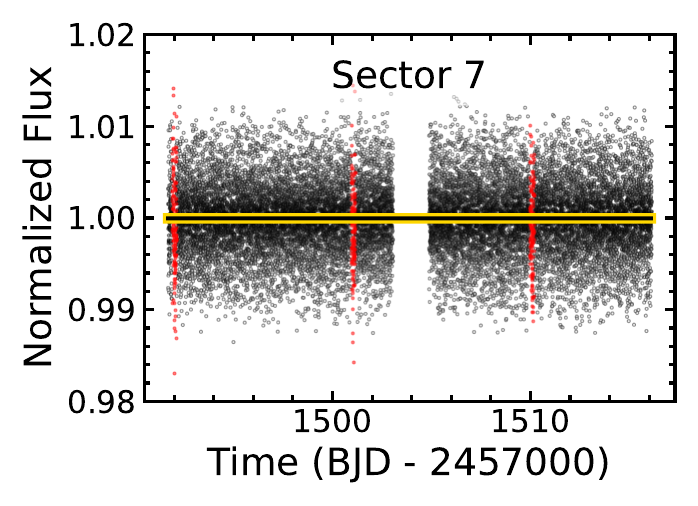}
  \endminipage\hfill
  \minipage{0.193\textwidth}
  \centering
  \includegraphics[width=1\linewidth]{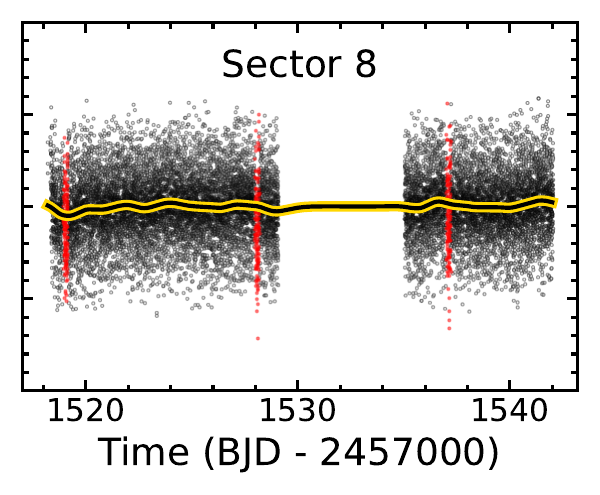}
  \endminipage\hfill
  \minipage{0.193\textwidth}
  \centering
  \includegraphics[width=1\linewidth]{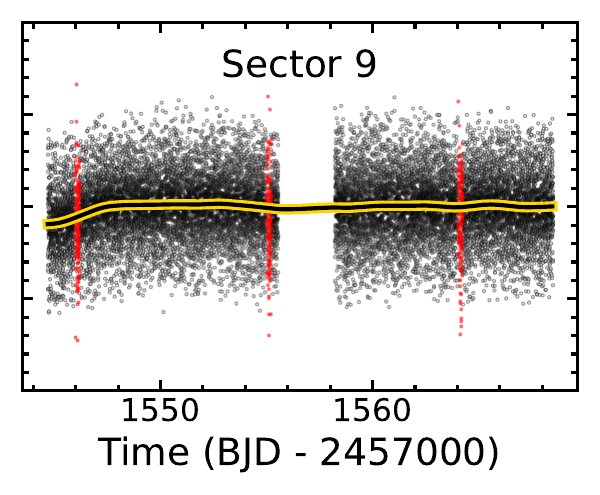}
  \endminipage\hfill
  \minipage{0.193\textwidth}
  \centering
  \includegraphics[width=1\linewidth]{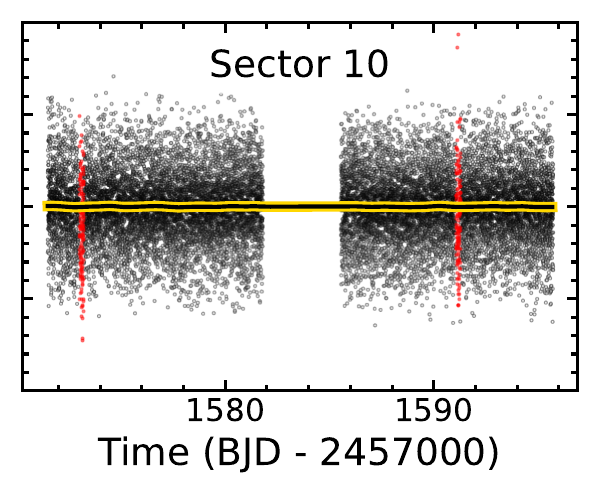}
  \endminipage\hfill
  \minipage{0.193\textwidth}
  \centering
  \includegraphics[width=1\linewidth]{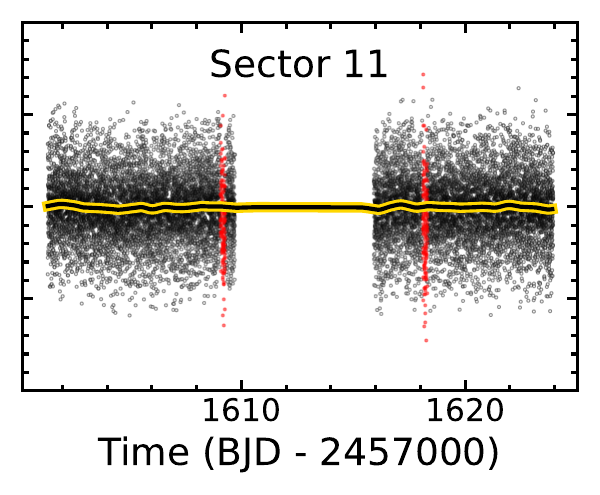}
  \endminipage\hfill\\

  \minipage{0.225\textwidth}
  \centering
  \vspace{-0.05cm}
  \includegraphics[width=1\linewidth]{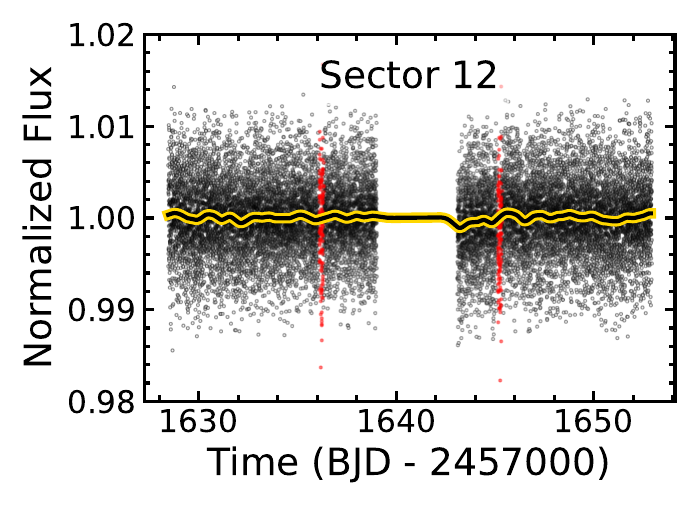}
  \endminipage\hfill
  \minipage{0.193\textwidth}
  \centering
  \includegraphics[width=1\linewidth]{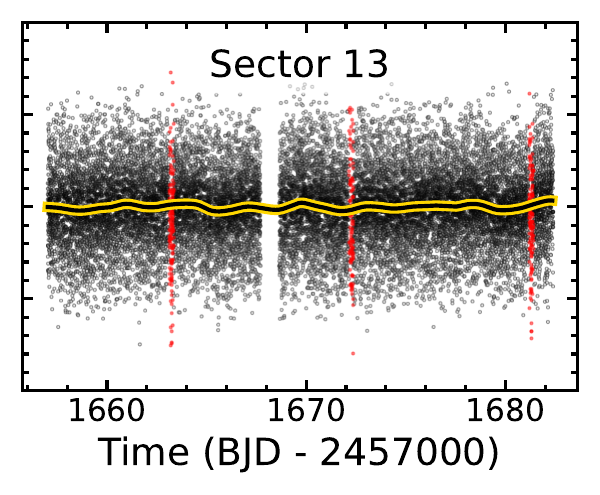}
  \endminipage\hfill
  \minipage{0.193\textwidth}
  \centering
  \includegraphics[width=1\linewidth]{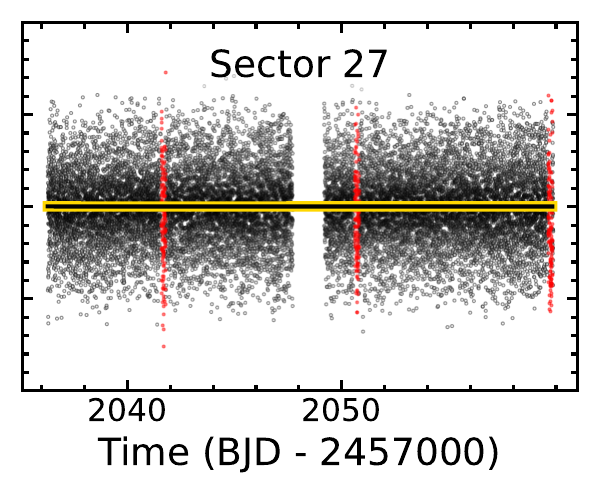}
  \endminipage\hfill
  \minipage{0.193\textwidth}
  \centering
  \includegraphics[width=1\linewidth]{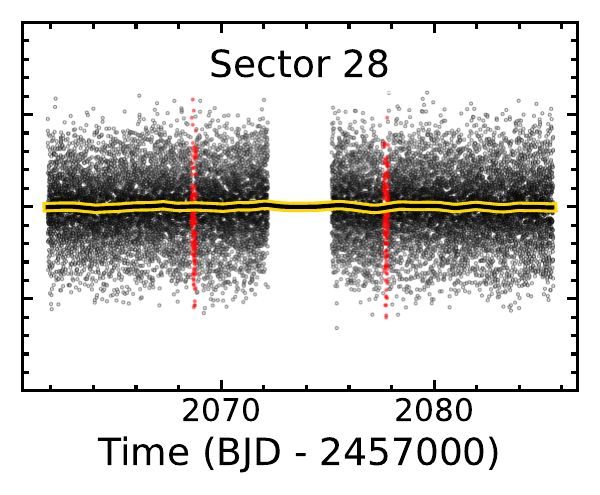}
  \endminipage\hfill
  \minipage{0.193\textwidth}
  \centering
  \includegraphics[width=1\linewidth]{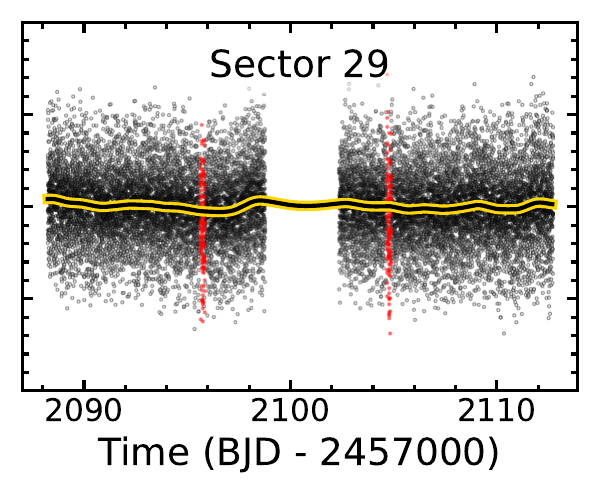}
  \endminipage\hfill\\

  \minipage{0.225\textwidth}
  \centering
  \vspace{-0.05cm}
  \includegraphics[width=1\linewidth]{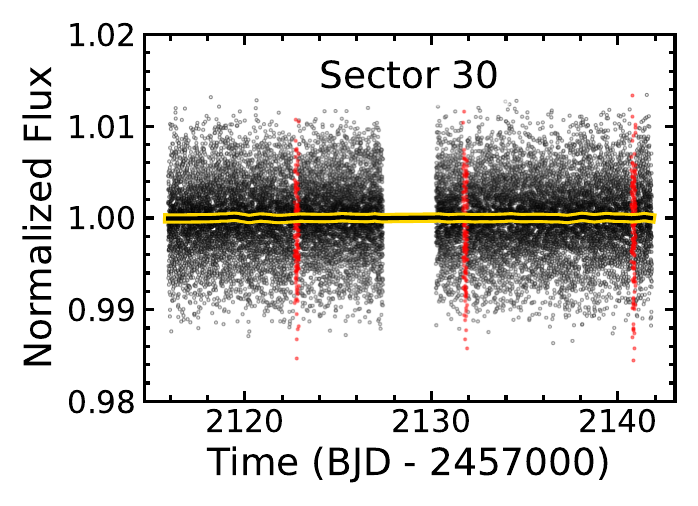}
  \endminipage\hfill
  \minipage{0.193\textwidth}
  \centering
  \includegraphics[width=1\linewidth]{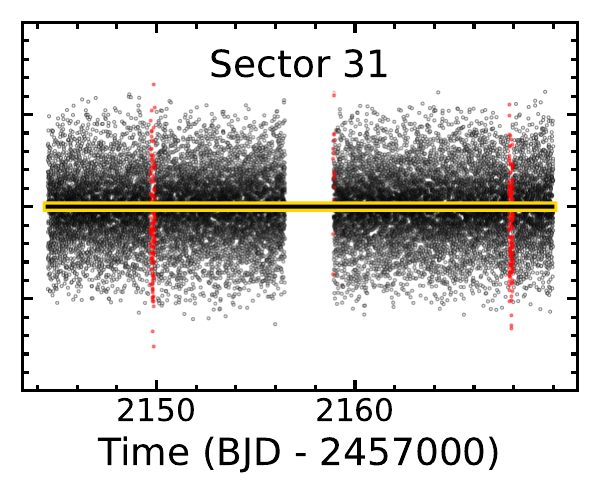}
  \endminipage\hfill
  \minipage{0.193\textwidth}
  \centering
  \includegraphics[width=1\linewidth]{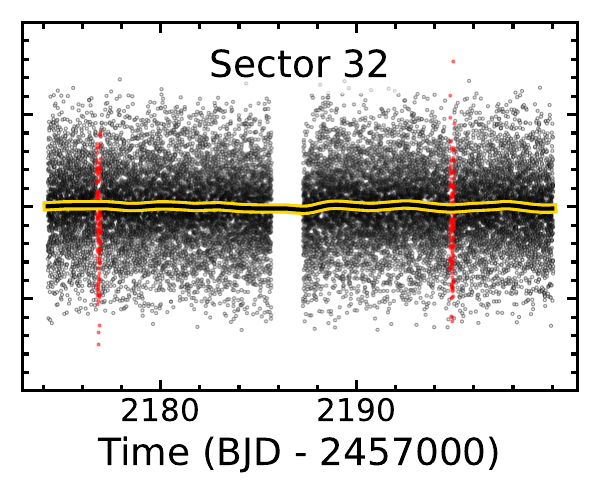}
  \endminipage\hfill
  \minipage{0.193\textwidth}
  \centering
  \includegraphics[width=1\linewidth]{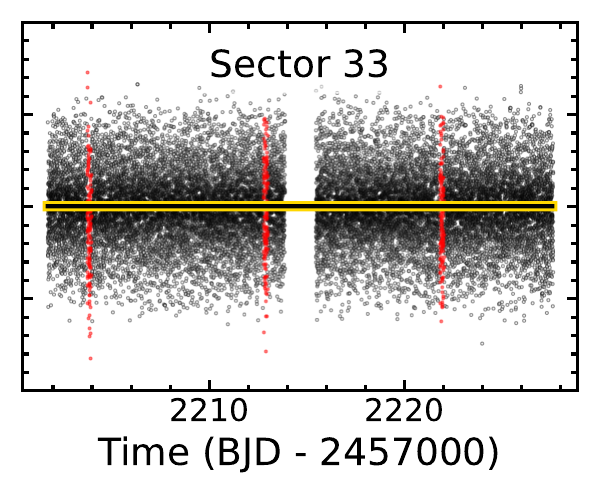}
  \endminipage\hfill
  \minipage{0.193\textwidth}
  \centering
  \includegraphics[width=1\linewidth]{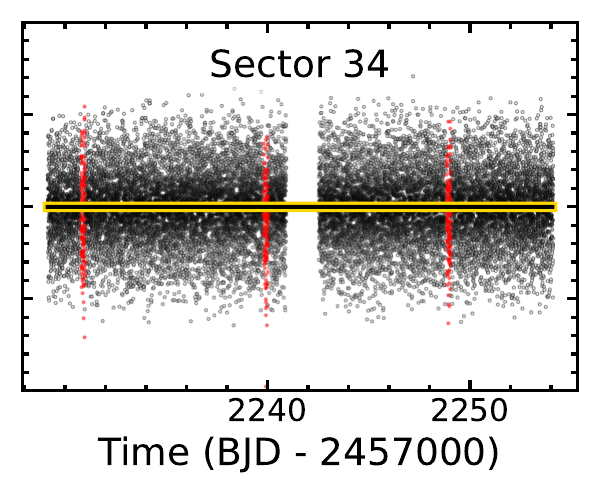}
  \endminipage\hfill\\

  \minipage{0.225\textwidth}
  \centering
  \vspace{-0.05cm}
  \includegraphics[width=1\linewidth]{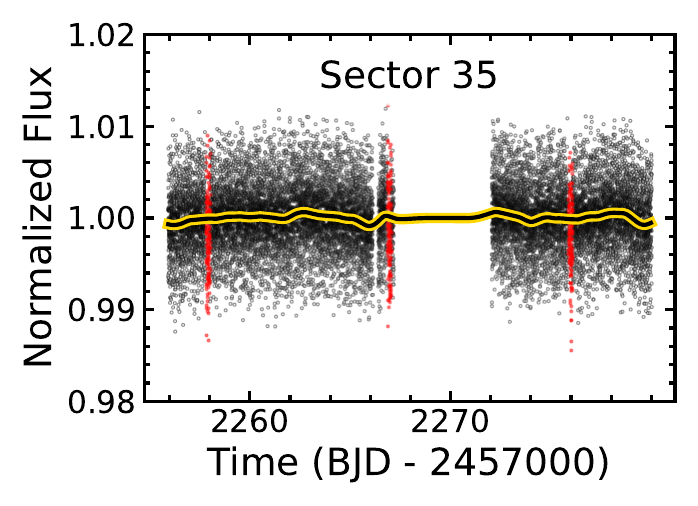}
  \endminipage\hfill
  \minipage{0.193\textwidth}
  \centering
  \includegraphics[width=1\linewidth]{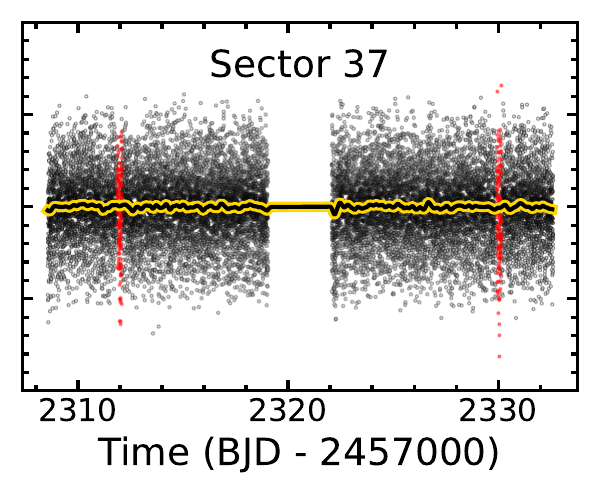}
  \endminipage\hfill
  \minipage{0.193\textwidth}
  \centering
  \includegraphics[width=1\linewidth]{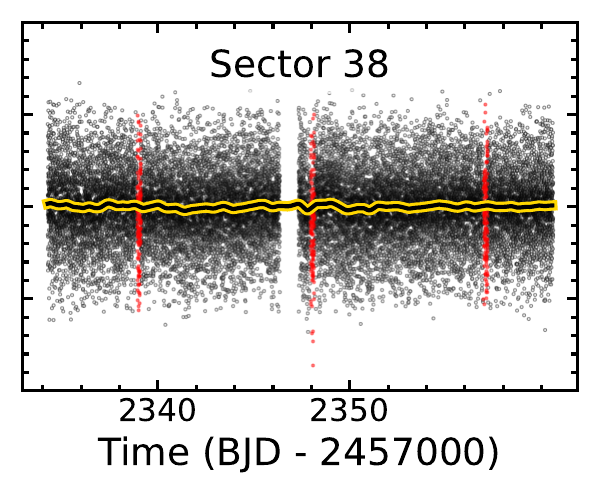}
  \endminipage\hfill
  \minipage{0.193\textwidth}
  \centering
  \includegraphics[width=1\linewidth]{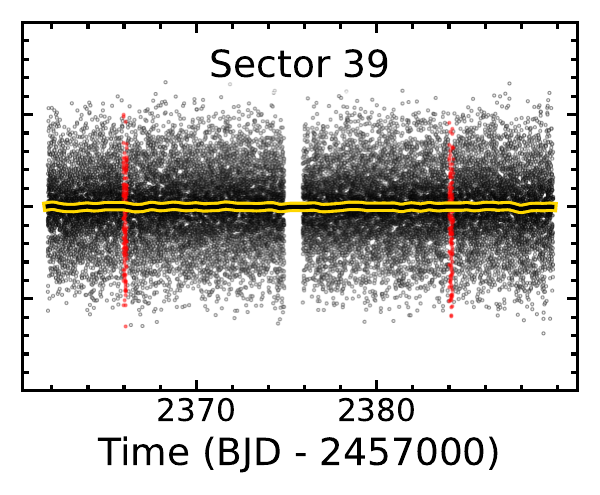}
  \endminipage\hfill
  \minipage{0.193\textwidth}
  \centering
  \includegraphics[width=1\linewidth]{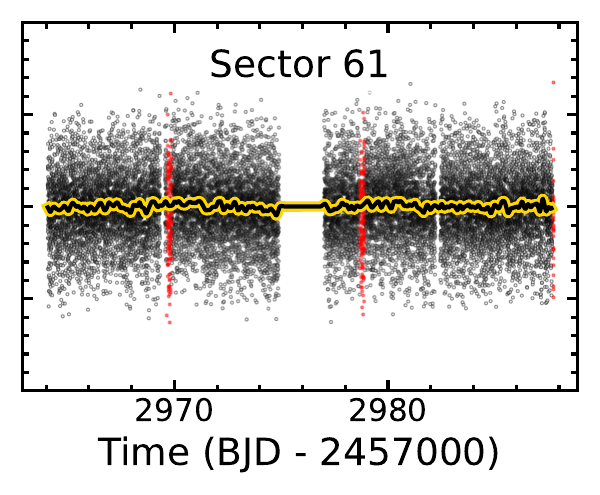}
  \endminipage\hfill\\

  \minipage{0.225\textwidth}
  \centering
  \vspace{-0.05cm}
  \includegraphics[width=1\linewidth]{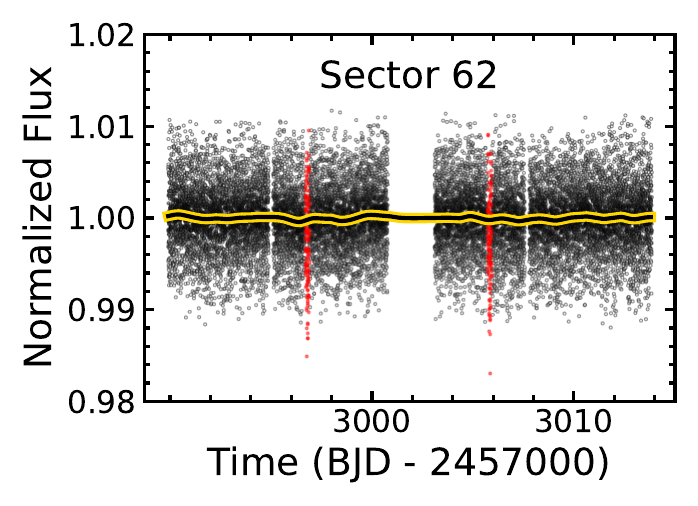}
  \endminipage\hfill
  \minipage{0.193\textwidth}
  \centering
  \includegraphics[width=1\linewidth]{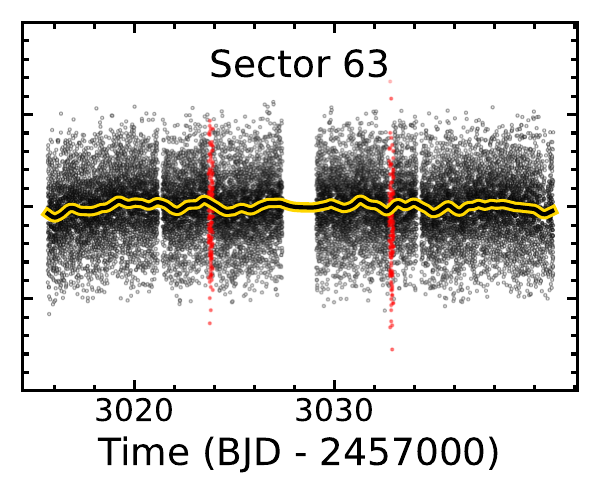}
  \endminipage\hfill
  \minipage{0.193\textwidth}
  \centering
  \includegraphics[width=1\linewidth]{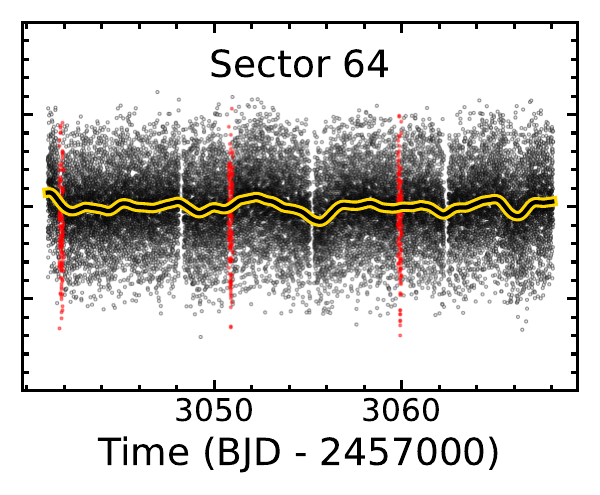}
  \endminipage\hfill
  \minipage{0.193\textwidth}
  \centering
  \includegraphics[width=1\linewidth]{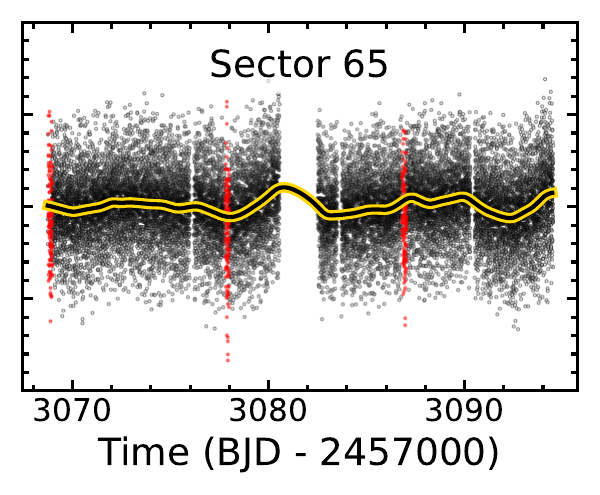}
  \endminipage\hfill
  \minipage{0.193\textwidth}
  \centering
  \includegraphics[width=1\linewidth]{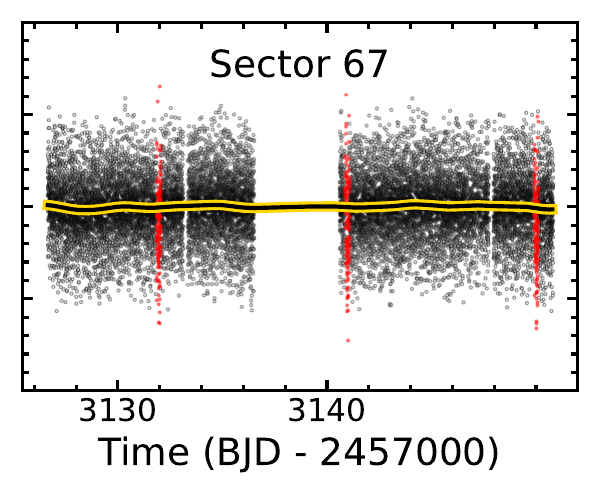}
  \endminipage\hfill\\

  \minipage{0.225\textwidth}
  \centering
  \vspace{-0.05cm}
  \includegraphics[width=1\linewidth]{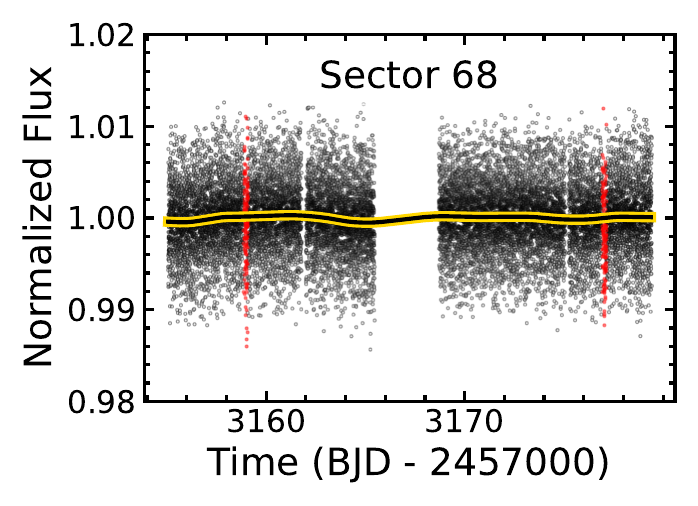}
  \endminipage\hfill
  \minipage{0.193\textwidth}
  \centering
  \includegraphics[width=1\linewidth]{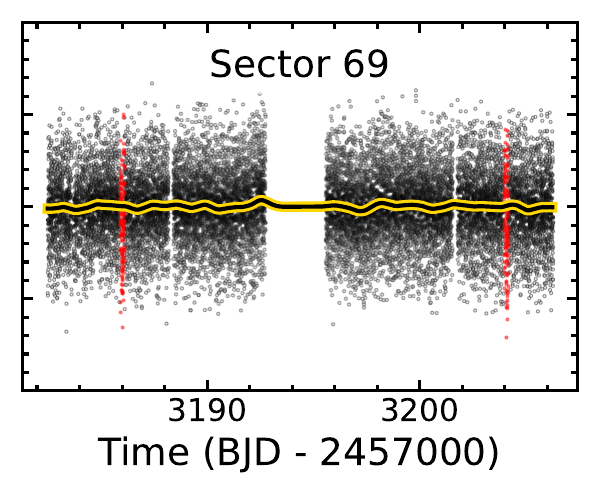}
  \endminipage\hfill
  \minipage{0.193\textwidth}
  \centering
  \includegraphics[width=1\linewidth]{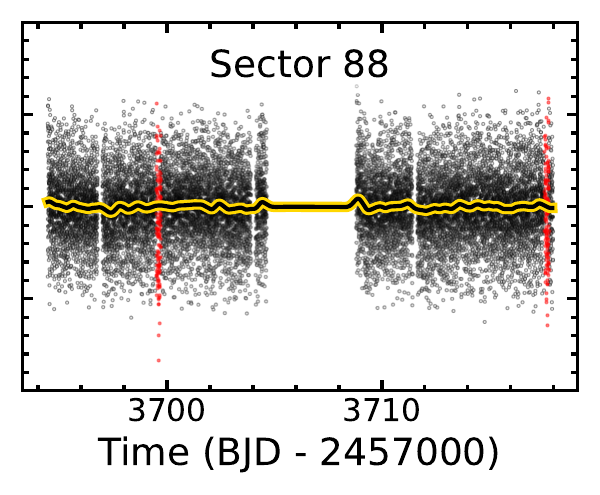}
  \endminipage\hfill
  \minipage{0.193\textwidth}
  \centering
  \includegraphics[width=1\linewidth]{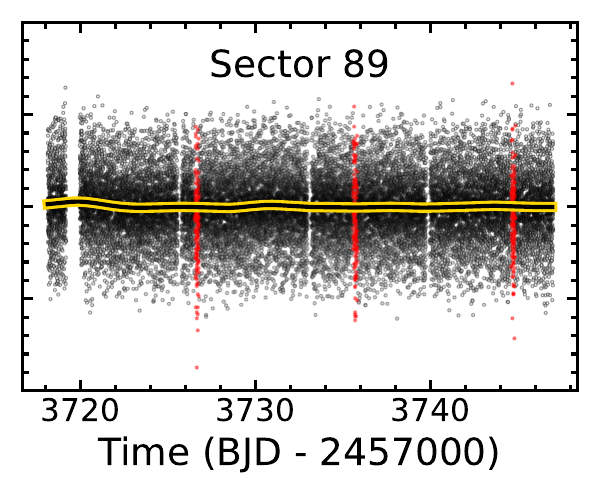}
  \endminipage\hfill
  \minipage{0.193\textwidth}
  \centering
  \includegraphics[width=1\linewidth]{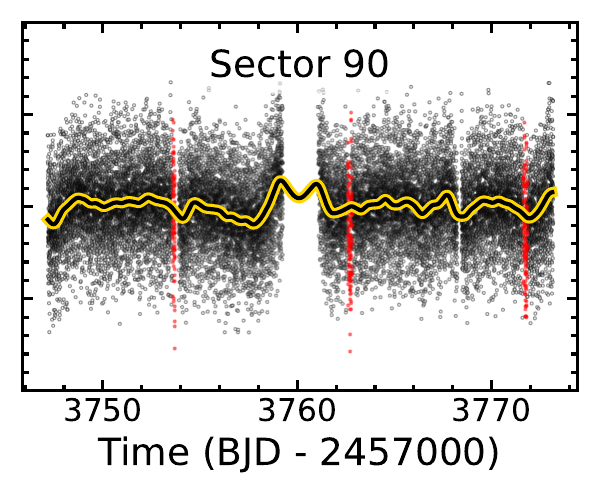}
  \endminipage\hfill\\

  \minipage{0.225\textwidth}
  \centering
  \vspace{-0.05cm}
  \includegraphics[width=1\linewidth]{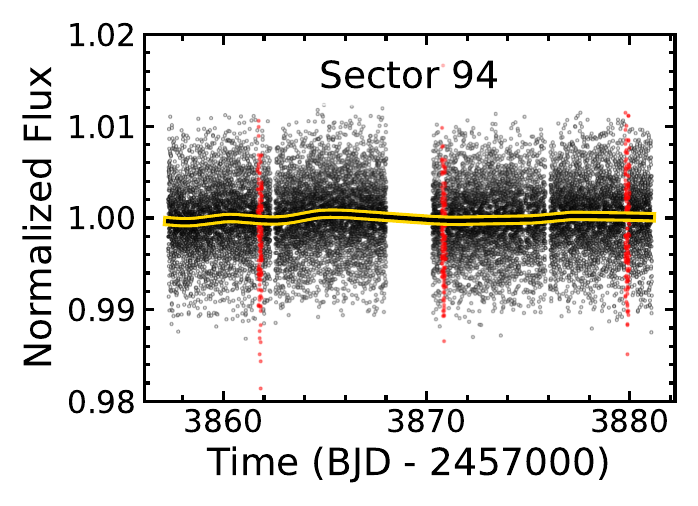}
  \endminipage\hfill
  \minipage{0.193\textwidth}
  \centering
  \includegraphics[width=1\linewidth]{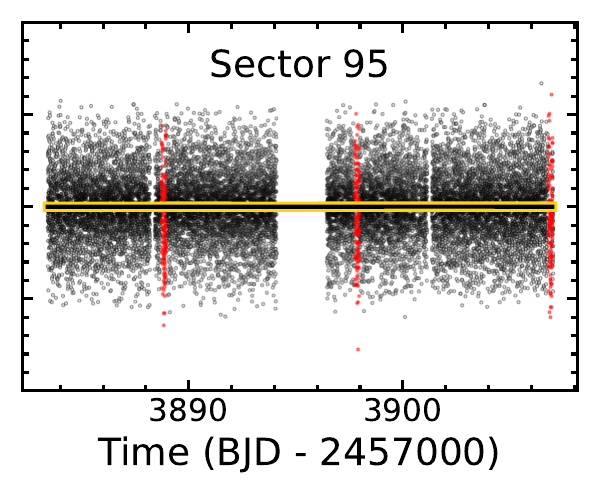}
  \endminipage\hfill
  \minipage{0.193\textwidth}
  \centering
  \includegraphics[width=1\linewidth]{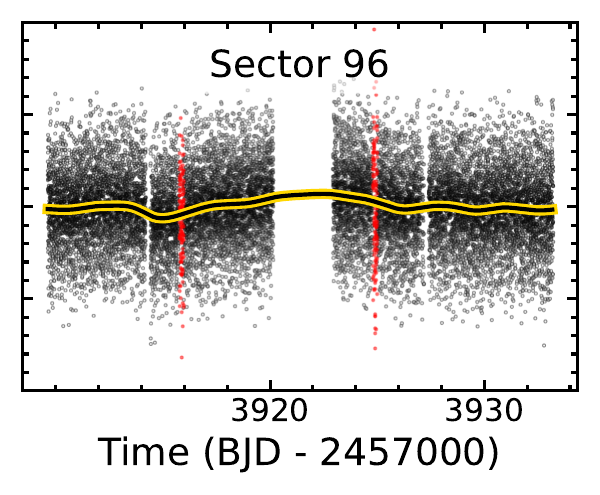}
  \endminipage\hfill
  \minipage{0.193\textwidth}
  \centering
  \includegraphics[width=1\linewidth]{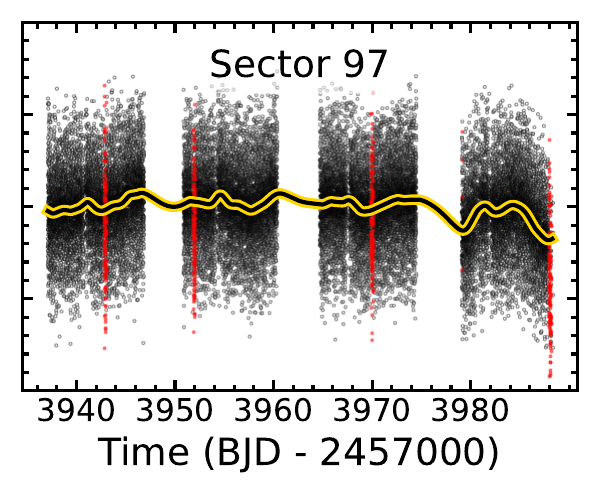}
  \endminipage\hfill
  \minipage{0.193\textwidth}
  \centering
  \includegraphics[width=1\linewidth]{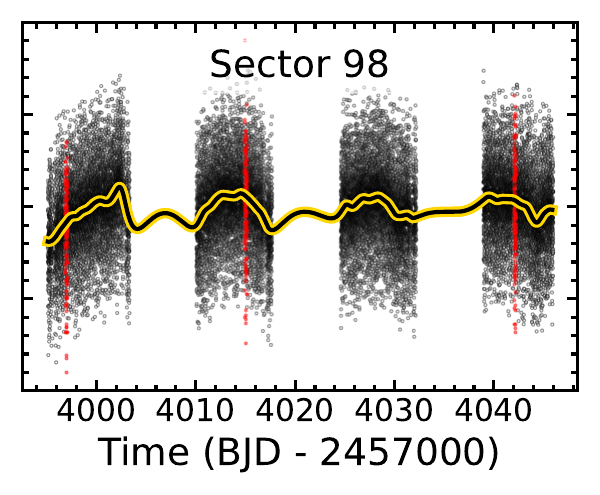}
  \endminipage\hfill
  
  \caption{Normalised \texttt{PDCSAP} flux of TOI-210 from TESS over 40 sectors from July 2018 to January 2026. The transits of TOI-210\,b are highlighted in red. The detrending model is shown as a yellow curve (details in Sect.~\ref{sec:tess}).}
\label{fig:tess_gp}
\end{figure*}

\section{NIRPS data and pipeline comparison}
\renewcommand{\thefigure}{C.\arabic{figure}}
\setcounter{figure}{0}

We provide the NIRPS d\textit{Temp}, RV, and dLW time series from the \texttt{NIRPS-DRS} in Table~\ref{table:rv}.

\begin{figure*}[b]
    \includegraphics[width=1\linewidth]{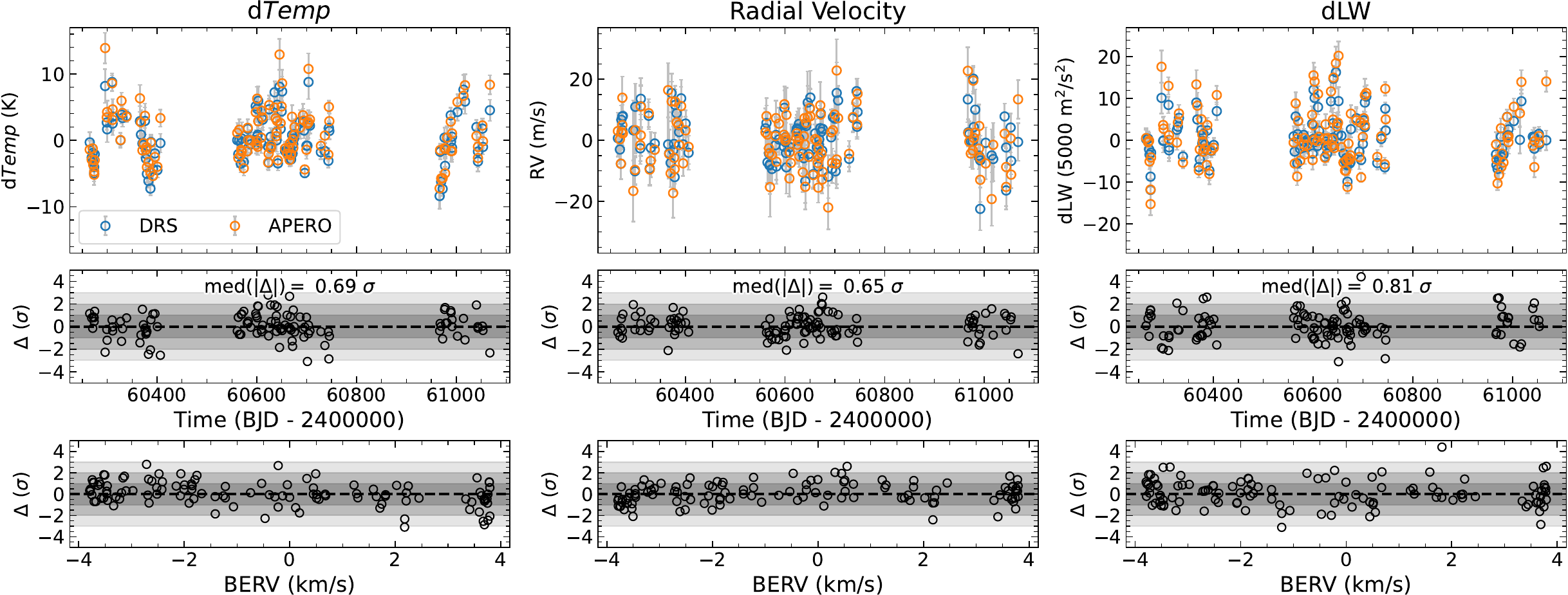}
    \caption{NIRPS d\textit{Temp}, RV, and dLW time series of TOI-210 extracted with two independent data reduction pipelines: \texttt{NIRPS-DRS} (blue circles; \citealt{Pepe_2021}) and \texttt{APERO} (orange circles; \citep{Cook_2022}). The middle panels show the difference (\texttt{NIRPS-DRS} $-$ \texttt{APERO}) in units of error, taken as the smallest between the two pipelines. The pipelines produce measurements in agreement with a median absolute deviation of 0.69$\sigma$, 0.65$\sigma$, and 0.81$\sigma$, respectively, for d\textit{Temp}, RV, and dLW. Residual correlations with BERV (bottom panels) indicate some left over systematics from the telluric correction.}
    \label{fig:drs_comps}
\end{figure*}

Pipeline-dependent systematic errors are rarely quantified in precision RV works, as multiple reduction software are rarely developed for a given spectrograph. Cross-validating the results has now become routine in the JWST era of atmospheric characterisation where low significance results (1--2\,$\sigma$) can easily vanish by changing the reduction pipeline. This cross validation is particularly important for near-infrared RV affected by telluric contamination. We compared the d\textit{Temp}, RV, and dLW of TOI-210 from the \texttt{NIRPS-DRS} and \texttt{APERO} pipelines in Fig.~\ref{fig:drs_comps}. The pipelines produced measurements in agreement with a median absolute deviation (MAD) of 0.69$\sigma$, 0.65$\sigma$, and 0.81$\sigma$ for the three observables, respectively. The telluric corrections implemented in \texttt{NIRPS-DRS} and \texttt{APERO} appear to leave residuals at different levels, which manifest as correlations in BERV space (see bottom panels of Fig.~\ref{fig:drs_comps}). After subtracting a second-degree polynomial trend with BERV, the agreement between the reductions improves further, with MAD of 0.63$\sigma$, 0.62$\sigma$, and 0.78$\sigma$. This motivated the inclusion of a BERV-detrending component during the analysis (Sect.~\ref{sec:rv_analysis}).

\clearpage

\begin{table*}[h]
\caption{\label{table:rv}Radial velocity, differential temperature (d\textit{Temp}), and differential line width (dLW) measurements of TOI-210 with NIRPS}
\centering
\renewcommand{\arraystretch}{1.25} 
\begin{tabular}{ccccccc}
\hline\hline
BJD - 2\,400\,000 & d\textit{Temp} (K) & $\sigma_{\rm d\textit{Temp}}$ (K) & RV (m\,s$^{-1}$) & $\sigma_{\rm RV}$ (m\,s$^{-1}$) & dLW (5000\,m$^2$\,s$^{-2}$) & $\sigma_{\rm dLW}$ (5000\,m$^2$\,s$^{-2}$)\\
\hline
2460264.805593 & -1.820 & 2.145 & -7.152 & 8.484 & 0.621 & 3.208 \\
2460264.816102 & 1.604 & 2.153 & 7.535 & 8.494 & 1.900 & 3.215 \\
2460268.728144 & -2.245 & 1.715 & -0.733 & 6.844 & 0.917 & 2.582 \\
2460268.738664 & -2.550 & 1.644 & -6.137 & 6.598 & 1.766 & 2.487 \\
2460270.775027 & -3.058 & 1.631 & -12.587 & 6.495 & -2.515 & 2.460 \\

\ldots & \ldots & \ldots & \ldots & \ldots & \ldots & \ldots\\
\hline
\end{tabular}
\vspace{0.1cm}
\caption*{\footnotesize {\bf Notes.} This Table is made available in its entirety in a machine-readable format.}
\end{table*}

\clearpage

\section{Supplementary material of the transit analysis}
\renewcommand{\thefigure}{D.\arabic{figure}}
\setcounter{figure}{0}

\subsection{Posterior results}

The prior and posterior results of the multi-instrument transit fit detailed in Sect.~\ref{sec:transit_analysis} are available in Table~\ref{table:transit_params}.

\begin{table}[h]
\caption{\label{table:transit_params}TOI-210\,b transit parameters from combined data set from TESS, LCOGT, and ExTrA}
\centering
\renewcommand{\arraystretch}{1.25} 
\begin{tabular}{lcr}
\hline\hline
Parameter & Prior & Posterior\\
\hline
\multicolumn{3}{c}{\textit{Stellar density}}\\[0.2cm]
$\ln \rho_{\star}$ & $\mathcal{N}\left(9.41, 0.11^2\right)$ & 9.44 $\pm$ 0.10 \\[0.2cm]
\multicolumn{3}{c}{\textit{TOI-210\,b}} \\[0.2cm]
$P$ (days) & $\mathcal{U}$(\textit{ExoFOP} $\pm$ 0.1) & 9.0105563 $\pm$\\[-0.1cm]
& & 2.8$\times$10$^{-6}$ \\
$t_{\rm 0}$ (\footnotesize{BJD\,-\,2457000}) & $\mathcal{U}$(\textit{ExoFOP} $\pm$ 0.1) & 2339.00523 $\pm$\\[-0.1cm]
& & 0.00023\\
$r_{\rm 1}$ & $\mathcal{U}\left(0, 1\right)$ & 0.54$^{+0.07}_{-0.09}$ \\
$r_{\rm 2}$ & $\mathcal{U}\left(0, 1\right)$ & 0.0619 $\pm$ 0.0008 \\[0.2cm]
\multicolumn{3}{c}{\textit{Baseline flux, limb darkening coefficients, and jitters}} \\[0.2cm]
$M_{\rm TESS}$ (ppm) & $\mathcal{N} \left(0, 10000\right)$ & $-16 \pm 46$ \\
$q_{\rm 1,TESS}$ & $\mathcal{TN}\left(0.36, 0.2, 0, 1\right)$ & 0.37$^{+0.12}_{-0.10}$ \\
$q_{\rm 2,TESS}$ & $\mathcal{TN}\left(0.17, 0.2, 0, 1\right)$ & 0.25$^{+0.14}_{-0.12}$ \\
$\sigma_{\rm TESS}$ (ppm) & $\mathcal{LU}$(1, 10000) & 12$^{+46}_{-9}$\\[0.1cm]
$M_{\textrm{LCO}g^{\prime}}$ (ppm) & $\mathcal{N} \left(0, 10000\right)$ & 278 $\pm$ 413\\
$q_{\textrm{1,LCO}g^{\prime}}$ & $\mathcal{TN}\left(0.74, 0.2, 0, 1\right)$ & 0.73$^{+0.15}_{-0.18}$ \\
$q_{\textrm{2,LCO}g^{\prime}}$ & $\mathcal{TN}\left(0.35, 0.2, 0, 1\right)$ & 0.34$^{+0.18}_{-0.17}$ \\
$\sigma_{\textrm{LCO}g^{\prime}}$ (ppm) & $\mathcal{LU}$(1, 10000) & 45$^{+398}_{-41}$\\[0.1cm]
$M_{\textrm{LCO}i^{\prime}}$ (ppm) & $\mathcal{N} \left(0, 10000\right)$ & 78 $\pm$ 161 \\
$q_{\textrm{1,LCO}i^{\prime}}$ & $\mathcal{TN}\left(0.43, 0.2, 0, 1\right)$ & 0.41$^{+0.17}_{-0.16}$ \\
$q_{\textrm{2,LCO}i^{\prime}}$ & $\mathcal{TN}\left(0.18, 0.2, 0, 1\right)$ & 0.22$^{+0.16}_{-0.13}$ \\
$\sigma_{\textrm{LCO}i^{\prime}}$ (ppm) & $\mathcal{LU}$(1, 10000) & 808$^{+202}_{-231}$\\[0.1cm]
$M_{\rm ExTrA}$ (ppm) & $\mathcal{N} \left(0, 10000\right)$ & 16 $\pm$ 73\\
$q_{\rm 1,ExTrA}$ & $\mathcal{TN}\left(0.17, 0.2, 0, 1\right)$ & 0.43$^{+0.14}_{-0.12}$ \\
$q_{\rm 2,ExTrA}$ & $\mathcal{TN}\left(0.12, 0.2, 0, 1\right)$ & 0.18$^{+0.14}_{-0.11}$ \\
$\sigma_{\rm ExTrA}$ (ppm) & $\mathcal{LU}$(1, 10000) & 19$^{+132}_{-16}$\\[0.1cm]
\hline
\end{tabular}
\end{table}

\subsection{Transit depth across photometric filters}

\begin{figure}[h]
\centering
\includegraphics[width=1\linewidth]{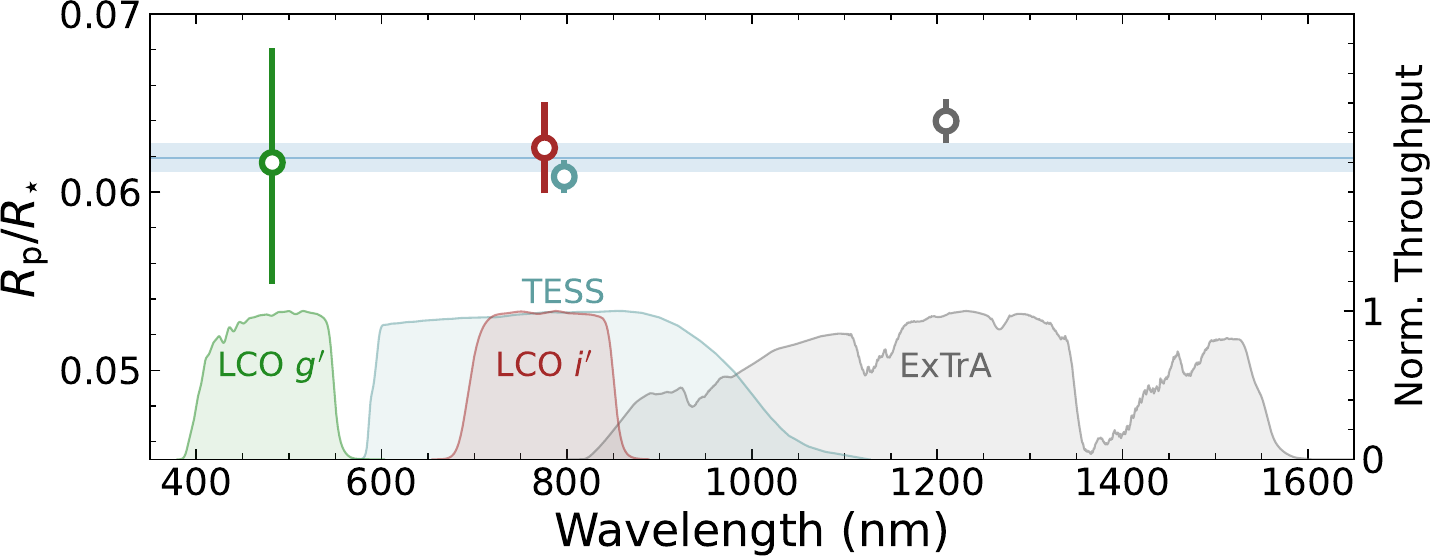}
\caption{Consistency of the planet-to-star radius ratio ($R_{\rm p} / R_\star$) of TOI-210\,b across photometric filters. The combined measurement is shown as a horizontal blue line with its 1$\sigma$ uncertainty envelope. The normalised bandpass profiles (from optical blue to the near-infrared) from LCOGT $g^{\prime}$, LCOGT $i^{\prime}$, TESS, and ExTrA are shown under the measurements. For ExTrA, the low transmission of Earth's atmosphere in telluric bands is incorporated into the throughput curve. The transit depth of TOI-210\,b shows no evidence of chromaticity.}
\label{fig:transit_chromaticity}
\end{figure}

A validation test consisting of fitting the transits from each instrument individually (TESS, LCO, ExTrA) and comparing the resulting planet-to-star radius ratio is presented in Fig.~\ref{fig:transit_chromaticity}. The transit of TOI-210\,b appears achromatic across the different photometric filters.

\subsection{Search for additional transiting planets} \label{sec:transit_search}

\begin{figure}
    \centering
    \includegraphics[width=1\linewidth]{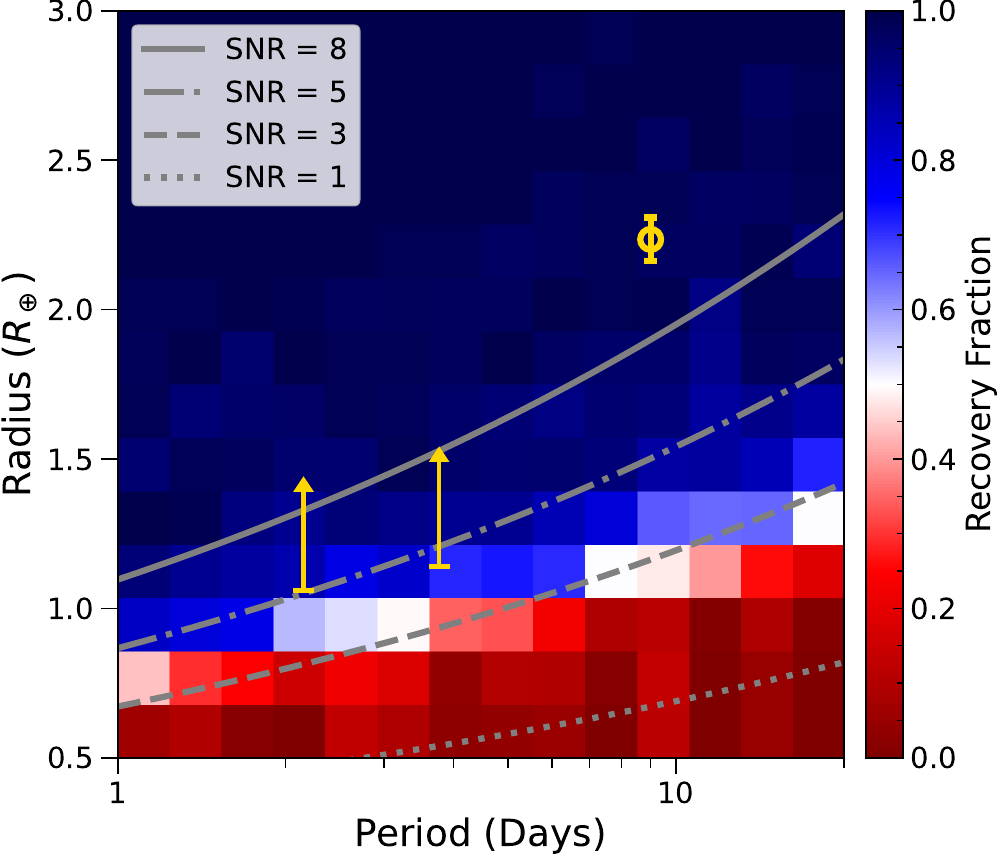}
    \caption{Transit signals recovery analysis in the TESS light curve of TOI-210. TOI-210\,b is shown in gold at its period of 9.01\,days. Candidate planets at 2.15 and 3.76\,days tentatively detected in the NIRPS RVs are indicated as lower limits, with arrows extending from minimum (iron density) to realistic (Earth density) radii based on their minimum masses. The expected transit depths fall within the TESS sensitivity, suggesting that if these RV signals are planets, they are unlikely to be transiting.}
    \label{fig:transit_search}
\end{figure}

The NIRPS RVs show moderate evidence for one or more additional Keplerian signals inner to TOI-210\,b with candidate periods at 2.15 and 3.76\,days. These potential additional planets around TOI-210 warrant a systematic search for transit signals that may have been missed by the TESS pipeline, given the faintness of the host star. We conducted an independent transit search in the \texttt{PDCSAP} light curves and performed an injection--recovery analysis to quantify the sensitivity of TESS to small planets around TOI-210.

To prepare the light curves for injection--recovery, we removed in-transit points of TOI-210\,b and only considered the out-of-transit, detrended \texttt{PDCSAP} fluxes (see Sect.~\ref{sec:tess}). 
We concatenated consecutive sectors, up to a maximum of four sectors to increase our sensitivity to transiting planets, yielding 13 light curves. This concatenation balances the sensitivity of our injection--recovery tests with the computational cost of running a large number of injections. For each injection--recovery test, we varied $P$, $R_{\rm p}$, $t_0$ and $b$. The injected period is sampled from a log-uniform distribution $\mathcal{LU} \left(1, 20 \right)$\,days, and $R_{\rm p}$, $t_0$ and $b$ are sampled from uniform distributions $\mathcal{U} \left(0.5, 3 \right)$\,R$_\oplus$, $\mathcal{U} \left(0, P \right)$\,days, $\mathcal{U} \left(0, 1 \right)$, respectively. The transiting planet signal is modeled with \texttt{batman} and injected into each light curve.

For each light curve, we ran an iterative search with Transit Least Squares \citep[TLS,][]{hippke_2019}. If the TLS periodogram recovered a significant peak with signal detection efficiency (SDE) greater than 6, it is removed from the light curve with a best-fit \texttt{batman} model. We continued this process iteratively until no significant TLS peaks were found, or a maximum of four iterations is reached. We considered a signal recovered if any of the TLS peaks from any light curve matched the injected period within a tolerance of 0.5\%.

We plot the results of our injection--recovery tests in Fig.~\ref{fig:transit_search} along with sensitivity curves in planet radius as a function of period. The sensitivity curves are computed using the standard transit S/R equation in the form $(\delta / \sigma_{\rm 0,t_{14}}) \sqrt{N_\mathrm{transits}}$, where $\delta$ is the transit depth, $\sigma_{\rm 0,t_{14}}$ is the median scatter of the light curve binned down to the transit duration, and $N_\mathrm{transits}$ the number of transits observed over all sectors. Even under the extreme assumption of pure iron compositions, the candidate planets must have radii of at least 1.06 and 1.14\,R$_{\oplus}$ for the 2.15- and 3.76-day signals, respectively. Such planets fall within the TESS detection regime shown in Fig.~\ref{fig:transit_search}, implying that transiting configurations are unlikely to have escaped detection.

\subsection{Constraints on transit timing variations} \label{sec:ttv_search}

Another powerful way to detect the presence of non-transiting planets is through transit timing variations (TTV) which their gravitational influence causes on known transiting planets. Since the NIRPS RVs are best explained by three Keplerian signals within 10\,days, we inspected the TESS and ground-based photometric data for evidence of TTVs for TOI-210\,b.

We first estimated the TTV amplitudes expected from the 2.15 and 3.76\,d candidate planets. To do so, we used the best-fitting orbital solution from the RV analysis for the three-planet model (Table~\ref{table:multi_dim_gp}) as input to \texttt{TTVFast} \citep{Deck_2014}, a numerical $N$-body integrator. The system was simulated over the entire baseline covered by TESS with time steps of 0.1\,days. Following standard practice, the longitudes of ascending node were fixed to $\Omega = 0^\circ$ for all planets. The calculated TTV amplitudes are 0.5, 0.4, and 0.04\,min for Candidate 1, 2, (assuming transiting configurations) and TOI-210\,b, respectively. The same calculations assuming significant eccentricity of 0.1 for all planets yields 115, 121, 1.4\,min instead, for an uniform distribution in $\omega$ between 0 and 2$\pi$. The large TTV amplitudes for the candidates are caused by their proximity to a 7:4 mean-motion resonance ($3.765 / 2.152 \approx 1.75$), triggering higher-order resonant effects only in the presence of some eccentricity (e.g., \citealt{Deck_2015}). If the candidate planets are real, orbital configurations that produce large TTVs could make their transits substantially more difficult to detect, potentially explaining their nondetection.

\begin{figure}
    \centering
    \includegraphics[width=1\linewidth]{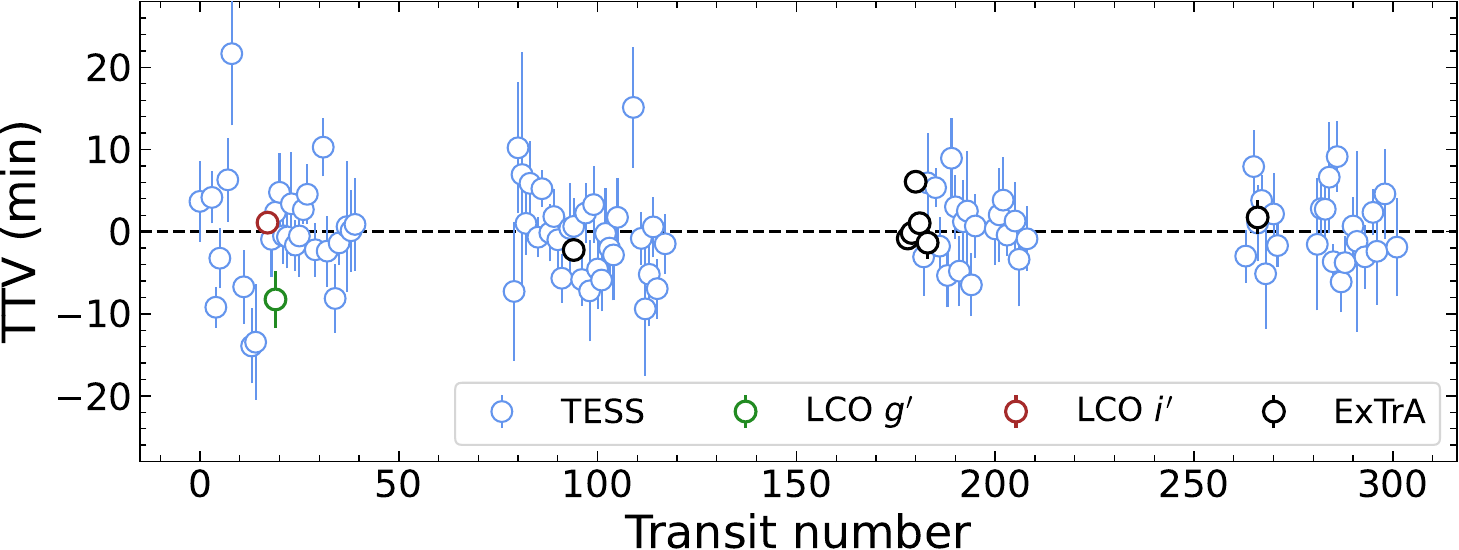}
    \caption{Transit timing variations (TTV) of TOI-210\,b from TESS, LCOGT, and ExTrA. The measurements are consistent with a linear ephemeris, with no evidence for TTVs exceeding a few minutes.}
    \label{fig:ttv}
\end{figure}

The transit timings of TOI-210\,b were obtained in \texttt{juliet} by replicating the transit analysis of Sect.~\ref{sec:transit_analysis}, this time allowing each individual transit to deviate from the linear ephemeris model within $\pm$30\,min ($P$ and $t_0$ fixed to the no TTV model). Figure~\ref{fig:ttv} presents the resulting transit timing measurements as an Observed minus Calculated ($O-C$) diagram, where TTV = 0\,min corresponds to the linear ephemeris prediction. The measured TTVs of TOI-210\,b show no obvious time correlation and their scatter of approximately 5\,min is consistent with the timing uncertainties. We find no evidence for an astrophysical TTV signal in the current data; however, as the expected TTV signal in the case of circular orbits is a factor of 100 below the timing uncertainties, this does not rule out the legitimacy of Candidates 1 and 2.

\section{Supplementary material of the radial velocity analysis}
\renewcommand{\thefigure}{E.\arabic{figure}}
\setcounter{figure}{0}

\subsection{Posterior results}
Table~\ref{table:multi_dim_gp} reports the list of priors and the posteriors median, 16$^{\rm th}$, and 84$^{\rm th}$ percentiles for the best-fit model of the NIRPS RVs, i.e., a multidimensional GP over d\textit{Temp}, RV, and dLW, a second-degree polynomial detrending with BERV for the three time series, plus three Keplerians (Sect.~\ref{sec:rv_analysis}).

\begin{table*}[h!]
\caption{\label{table:multi_dim_gp}Prior and posterior distributions of the best-fit RV model}
\centering
\renewcommand{\arraystretch}{1.25} 
\begin{tabular}{lcc}
\hline\hline
Parameter & Prior & Posterior\\
\hline
\multicolumn{3}{c}{\textit{Candidate 1}}\\[0.1cm]
$t_{0, \textrm{[c1]}}$ (BJD $-$ 2457000) & $\mathcal{U}\left(3630, 3630 + 1.2 \times2.15\right)$ & 3631.756$^{+0.092}_{-0.122}$\\
$P_{\rm [c1]}$ (days) & $\mathcal{N}\left(2.15, 0.1\right)$ & 2.1519$^{+0.0012}_{-0.0008}$\\
$K_{\rm [c1]}$ (m\,s$^{-1}$) & $\mathcal{U}\left(0, 10\right)$ & 3.49 $\pm$ 0.87\\
$e_{\rm [c1]}$ & 0 (fixed) & prior\\
$\omega_{\rm [c1]}$ & 90$^{\circ}$ (fixed) & prior\\[0.1cm]
\multicolumn{3}{c}{\textit{Candidate 2}}\\[0.1cm]
$t_{0, \textrm{[c2]}}$ (BJD $-$ 2457000) & $\mathcal{U}\left(3630, 3630 + 1.2 \times3.76\right)$ & 3632.485$^{+0.140}_{-0.180}$\\
$P_{\rm [c2]}$ (days) & $\mathcal{N}\left(3.76, 0.1\right)$ & 3.7652$^{+0.0071}_{-0.0026}$\\
$K_{\rm [c2]}$ (m\,s$^{-1}$) & $\mathcal{U}\left(0, 10\right)$ & 3.93 $\pm$ 0.93\\
$e_{\rm [c2]}$ & 0 (fixed) & prior\\
$\omega_{\rm [c2]}$ & 90$^{\circ}$ (fixed) & prior\\[0.1cm] 
\multicolumn{3}{c}{\textit{Planet b}}\\[0.1cm]
$t_{0, \textrm{b}}$ (BJD $-$ 2457000) & $\mathcal{N}\left(2339.00523, 0.00023\right)$ & prior\\
$P_{\rm b}$ (days) & $\mathcal{N}\left(9.0105563, 2.8 \times 10^{-6}\right)$ & prior\\
$K_{\rm b}$ (m\,s$^{-1}$) & $\mathcal{U}\left(0, 10\right)$ & 4.52 $\pm$ 0.84\\
$e_{\rm b}$ & 0 (fixed) & prior\\
$\omega_{\rm b}$ & 90$^{\circ}$ (fixed) & prior\\[0.1cm] 
\multicolumn{3}{c}{\textit{Multidimensional GP}}\\[0.1cm]
$\alpha_{\textrm{d}Temp}$ (K) & $\mathcal{U}\left(0, 10\right)$ & 3.08$^{+0.62}_{-0.49}$\\
$\alpha_{\textrm{RV}}$ (m\,s$^{-1}$) & $\mathcal{U}\left(-10, 10\right)$ & $-2.40^{+0.90}_{-1.00}$\\
$\alpha_{\textrm{dLW}}$ (5000\,m$^2$\,s$^{-2}$) & $\mathcal{U}\left(-10, 10\right)$ & 4.01$^{+0.89}_{-0.66}$\\
$\beta_{\textrm{RV}}$ (m\,s$^{-1}$\,d$^{-1}$) & $\mathcal{U}\left(-50, 50\right)$ & 13.97$^{+5.53}_{-4.89}$ \\
$\beta_{\textrm{dLW}}$ (5000\,m$^2$\,s$^{-2}$\,d$^{-1}$) & $\mathcal{U}\left(-50, 50\right)$ & $2.72^{+3.33}_{-2.97}$\\
$\ell$ (days) & $\mathcal{LU}\left(30, 1000\right)$ & 48$^{+18}_{-13}$ \\
$\Gamma$ & $\mathcal{LU}\left(0.01, 10\right)$ & 7.8$^{+1.5}_{-2.2}$\\
$P_{\rm rot}$ (days) & $\mathcal{LU}\left(30, 200\right)$ & 71$^{+6}_{-3}$ \\[0.1cm]
\multicolumn{3}{c}{\textit{Detrending against the BERV}}\\[0.1cm]
$a_{\textrm{d}Temp}$ & $\mathcal{N}\left(0, 0.5\right)$ & $-0.19$ $\pm$ 0.09\\
$b_{\textrm{d}Temp}$ & $\mathcal{N}\left(0, 0.5\right)$ & 0.05 $\pm$ 0.16\\
$a_{\textrm{RV}}$ & $\mathcal{N}\left(0, 0.5\right)$ & 0.05 $\pm$ 0.13\\
$b_{\textrm{RV}}$ & $\mathcal{N}\left(0, 0.5\right)$ & 0.19 $\pm$ 0.22\\
$a_{\textrm{dLW}}$ & $\mathcal{N}\left(0, 0.5\right)$ & $-0.01$ $\pm$ 0.12\\
$b_{\textrm{dLW}}$ & $\mathcal{N}\left(0, 0.5\right)$ & $0.14$ $\pm$ 0.22\\
\multicolumn{3}{c}{\textit{Offsets and Jitters}}\\[0.1cm]
$c_{\textrm{d}Temp}$ (K) & $\mathcal{U}\left(-10, 10\right)$ & 2.53 $\pm$ 0.98\\
$s_{\textrm{d}Temp}$ (K) & $\mathcal{LU}\left(0.01, 10\right)$ & 206 $\pm$ 0.21\\
$c_{\textrm{RV}}$ (m\,s$^{-1}$) & $\mathcal{U}\left(-10, 10\right)$ & $-1.19$ $\pm$ 1.18\\
$s_{\textrm{RV}}$ (m\,s$^{-1}$) & $\mathcal{LU}\left(0.01, 10\right)$ & 4.74 $\pm$ 0.78\\
$c_{\textrm{dLW}}$ (5000\,m$^2$\,s$^{-2}$) & $\mathcal{U}\left(-10, 10\right)$ & 1.58 $\pm$ 1.31\\
$s_{\textrm{dLW}}$ (5000\,m$^2$\,s$^{-2}$) & $\mathcal{LU}\left(0.01, 10\right)$ & 3.58 $\pm$ 0.30\\
\hline
\end{tabular}
\vspace{0.1cm}
\caption*{\footnotesize {\bf Notes.} [c1] and [c2] refers to the Candidates 1 and 2 identified in the NIRPS RVs (Sect.~\ref{sec:rv_analysis}). Detrending coefficients $a$ and $b$ are given in units of K for d\textit{Temp}, m\,s$^{-1}$ for RV, and 5000\,m$^2$\,s$^{-2}$ for dLW, per unit km$^2$\,s$^{-2}$ and km\,s$^{-1}$, respectively.}
\end{table*}

\subsection{Periodogram of the radial velocities}

\begin{figure*}
    \includegraphics[width=1\linewidth]{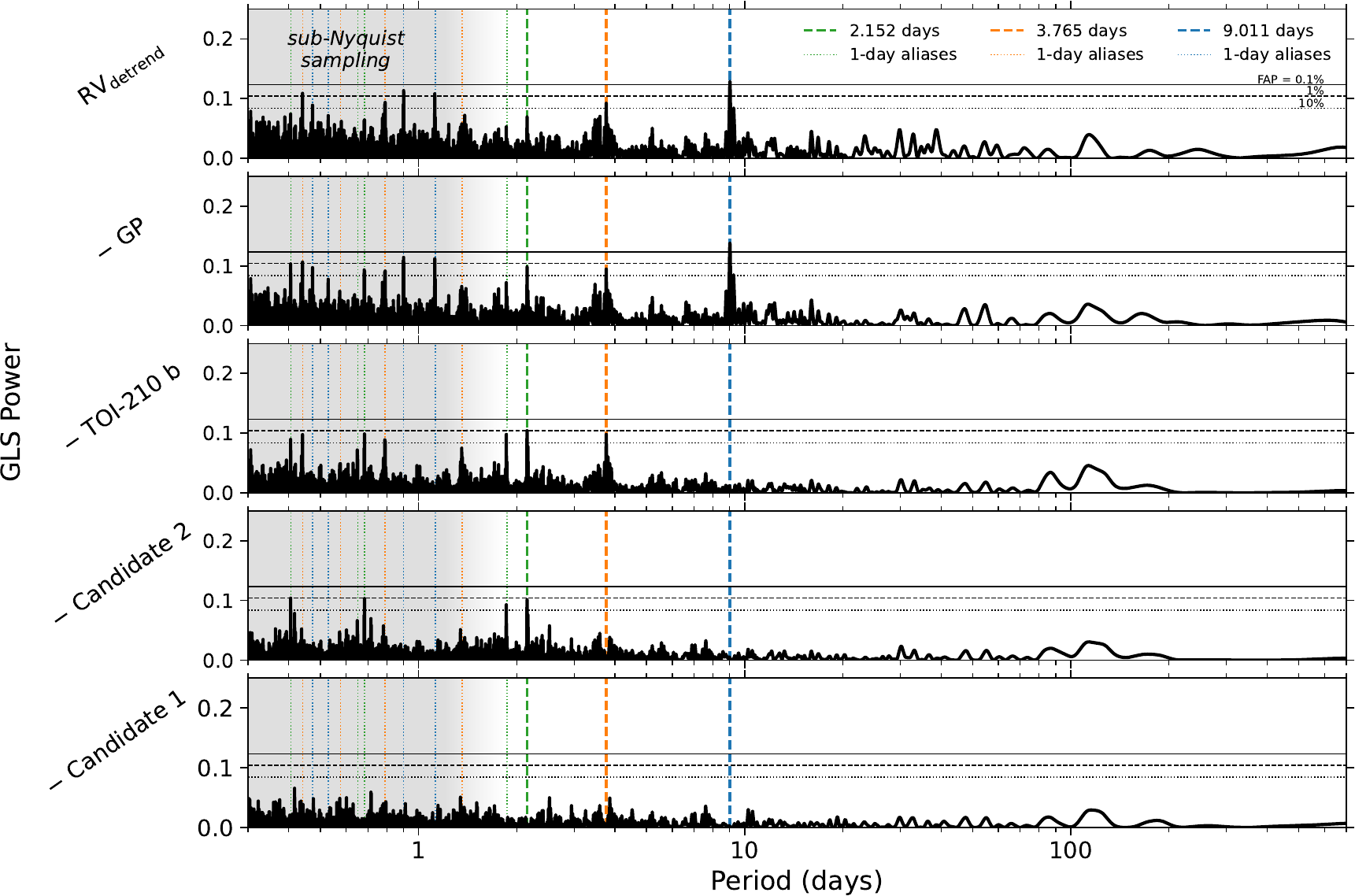}
    \caption{Generalised Lomb-Scargle periodograms of the NIRPS RVs. Each row shows the periodogram of the residual RVs after sequentially removing model components: (i) BERV systematics, (ii) stellar activity GP, (iii) TOI-210\,b at 9.01\,d, (iv) Candidate 2 at 3.76\,d, and (v) Candidate 1 at 2.15\,d. In each panel, the periods of the Keplerian signals are marked with colored vertical dashed lines, with their respective 1-day aliases (fundamental and first harmonic) shown in dotted lines of the same colors. After removing TOI-210\,b (row iii), the residuals reveal many periodicities with false alarm probability below 5\%. The full model with two additional Keplerians at 2.15 and 3.76\,days leaves no significant peaks in the residuals (row v).}
    \label{fig:periodogram_residuals}
\end{figure*}

During the analysis of the NIRPS RVs, we explored models with additional Keplerians on orbits shorter to the confirmed transiting exoplanet TOI-210\,b (Sect.~\ref{sec:rv_analysis}). The model with the highest evidence includes candidate planets at 2.15 and 3.76\,d. We show in Fig.~\ref{fig:periodogram_residuals} the Generalised Lomb-Scargle periodogram (GLS; \citealt{Zechmeister_2009}) of the RV residuals after sequentially subtracting model components. This figure highlights how the false alarm probabilities of the candidate signals evolve as stellar activity is removed and the strong 1-day aliases introduced by the observing cadence.

\subsection{Injection--recovery analysis} \label{sec:sensitivity_map}

\begin{figure}[h!]
\centering
\includegraphics[width=1\linewidth]{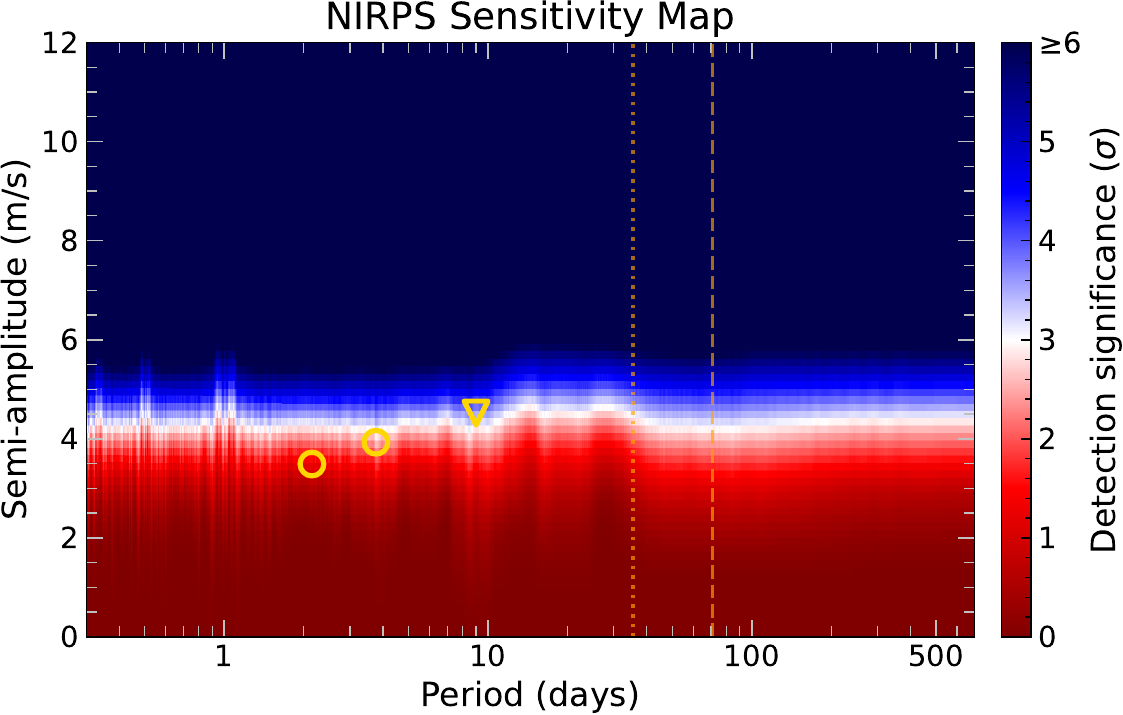}
\caption{Detection sensitivity map of planetary signals in the NIRPS data. The stellar rotation period ($P_{\rm rot}$) and its first harmonic ($P_{\rm rot} / 2$) are highlighted with vertical dashed and dotted lines, respectively. The NIRPS radial velocities can detect $\sim$4\,m\,s$^{-1}$ signals at 3\,$\sigma$ over most periods. This analysis demonstrates that the transiting planet TOI-210\,b (triangle) would be robustly recovered in a blind search, but only weakly for the candidate planets at 2.15 and 3.76\,d (circles).}
\label{fig:rv_detection_map}
\end{figure}

To assess the sensitivity of the NIRPS data to small Keplerian signals in the presence of low-level correlated stellar noise, we performed injection--recovery simulations following the methodology described by \cite{Gonzalez-Hernandez_2024} and \cite{Cadieux_2025}. We generated a simulated RV dataset using the same timestamps and individual uncertainties as the observations. We first injected a realistic activity signal by drawing a sample from the GP of the single-planet model, ensuring that the resulting scatter was consistent with the real data. This procedure ensures that the recovery tests are not biased by potential planetary signals that could be present in the observed RV residuals.
The injected signals consisted of pure sinusoids with semi-amplitudes $K$ ranging from 0 to 12\,m\,s$^{-1}$ and orbital periods between 0.3 and 1000\,days, sampled on a logarithmic grid. For each simulated dataset, we subtracted the mean GP prediction using the hyperparameters listed in Table~\ref{table:multi_dim_gp}, while keeping the original d\textit{Temp} and dLW time series unchanged. This procedure allows us to evaluate potential signal suppression or amplification introduced by the activity GP. We then computed the FAP at the injected period using the \cite{Baluev_2008} false-alarm approximation inside the Generalised Lomb-Scargle periodogram \citep{Zechmeister_2009} implementation in \texttt{astropy} \citep{Astropy_2022}. The resulting sensitivity map is presented in Fig.~\ref{fig:rv_detection_map}. Overall, the NIRPS RVs are sensitive to injected signals at the $\sim$3\,m\,s$^{-1}$ level over most orbital periods. This exercise confirms that TOI-210\,b is the only Keplerian signal that can be robustly recovered in a blind search.

\section{Interior parameters retrieval with \texttt{JADE}} \label{appendix:jade}

We briefly describe below the inference of the internal structure of TOI-210\,b with the \texttt{JADE} code \citep{Attia_2021,Attia_2025} in the gas-dwarf scenario. The \texttt{JADE} planetary structure consists in an iron core ($\alpha$-Fe), a silicate mantle (MgSiO$_3$), and an H/He-dominated envelope. The latter is divided into an upper region absorbing the stellar radiation and a lower region redistributing energy via radiation and convection. The Rosseland mean opacity of the envelope (computed from the tabulated opacities of \citealt{Ferguson_2005}) can be increased by including trace amount of metals, controlled by their mass fraction relative to the envelope. \texttt{JADE} fits for the mantle-to-planet mass fraction $f_{\rm man}$, envelope-to-planet mass fraction $f_{\rm env}$, and metal-to-envelope mass fraction $Z_{\rm env}$ with a MCMC approach, using the measured planet radius as constraint and the measured planet mass as prior. For a given set of parameters, the code integrates from the top the 1D thermodynamical structure of the envelope and the polytropic equation of state of the mantle and core, iterating over a grid of radii until finding the one that yields zero mass at planet center. The retrieval is performed at 4.5~Gyr to account for the impact of the stellar irradiation (set to the bolometric luminosity in Table~\ref{table:stellar_params}) and planetary internal luminosity on the atmospheric structure. The planet orbital properties were set to their present-day values (Table~\ref{table:derived_params}) and the atmospheric He abundance to solar (27.5\%). 

We performed a first run with uniform priors $\mathcal{U}$(0,1) on $f_{\rm env}$ and $f_{\rm man}$, and a narrower prior $\mathcal{U}$(0,0.1) on $Z_{\rm env}$. Atmospheric metallicity, and the repartition of the core and mantle mass, are poorly constrained, as expected from the degeneracies inherent to internal structure retrievals. We derive $f_{\rm env} = 0.009^{+0.005}_{-0.007}$ and $f_{\rm core} = 0.518^{+0.278}_{-0.298}$, which are consistent with the results of the grid retrieval performed in Sect.~\ref{sec:internal_structure}. Although the \texttt{JADE} retrieval allows for the absence of a volatile envelope within 2$\sigma$, the PDF on the envelope mass fraction makes this configuration highly unlikely. We then performed a second run where the core-to-mantle mass ratio is constrained with the same prior on $f^{\prime}_{\rm core}$ as above from NIRPS stellar abundances ($\mathcal{N}\left(0.27, 0.16\right)$), resulting in $f_{\rm env} = 0.005^{+0.003}_{-0.004}$ and $f_{\rm core} = 0.295^{+0.139}_{-0.144}$.







\section{Acknowledgments} \label{sec:acknowledgements}
\begin{acknowledgements}

This work has been carried out within the framework of the National Centre of Competence in Research PlanetS supported by the Swiss National Science Foundation (SNSF) under grant 51NF40\_205606.\\
CC acknowledges the support from the SNSF under the grant SPECTRE (No 200021\_215200).\\
This project has received funding from the European Research Council (ERC) under the European Union's Horizon 2020 research and innovation programme (project {\sc Spice Dune}, grant agreement No 947634).\\
CC, RD, AL, RA, \'EA, BB, NJC, PL, LMa \& JPW  acknowledge the financial support of the FRQ-NT through the Centre de recherche en astrophysique du Qu\'ebec as well as the support from the Trottier Family Foundation and the Trottier Institute for Research on Exoplanets (IREx).\\
RD, \'EA \& LMa  acknowledge support from Canada Foundation for Innovation (CFI) program, the Universit\'e de Montr\'eal and Universit\'e Laval, the Canada Economic Development (CED) program and the Ministere of Economy, Innovation and Energy (MEIE).\\
Research activities of the Board of Observational and Instrumental Astronomy at the Federal University of Rio Grande do Norte (NAOS) are supported by continuous grants from the Brazilian funding agency CNPq. This study was financed in part by the Coordena\c{c}\~ao de Aperfei\c{c}oamento de Pessoal de N\'ivel Superior -- Brasil (CAPES) -- Finance Code 001, and by the program CAPES/Print.\\
BLCM acknowledges CAPES postdoctoral fellowships and CNPq research fellowships (grant no. 305804/2022-7) and Universal (grant no. 408100/2025-7).\\
AL  acknowledges support from the Fonds de recherche du Qu\'ebec (FRQ) - Secteur Nature et technologies under file \#349961.\\
We acknowledge funding from the ERC under Grant Agreement no. 337591-ExTrA.\\
AKS, NN, RRe \& ASM  acknowledge financial support from the Spanish Ministry of Science, Innovation and Universities (MICIU) projects PID2020-117493GB-I00 and PID2023-149982NB-I00.\\
AKS  acknowledges financial support from La Caixa Foundation (ID 100010434) under the grant LCF/BQ/DI23/11990071.\\
XB, XDe \& TF  acknowledge funding from the French ANR under contract number ANR\-24\-CE49\-3397 (ORVET), and the French National Research Agency in the framework of the Investissements d'Avenir program (ANR-15-IDEX-02), through the funding of the ``Origin of Life" project of the Grenoble-Alpes University.\\
RA  acknowledges the SNSF support under the Post-Doc Mobility grant P500PT\_222212 and the support of IREx.\\
KAM  acknowledges support from the SNSF under the Post-Doc Mobility grant P500PT\_230225.\\
SCCB, EC, ED-M \& NCS acknowledge the support from FCT - Funda\c{c}\~ao para a Ci\^encia e a Tecnologia through national funds by these grants: UIDB/04434/2020, UIDP/04434/2020.\\
SCCB acknowledges the support from FCT in the form of a work contract through the Scientific Employment Incentive program with reference 2023.06687.CEECIND.\\
NCS acknowledges support from the ERC (FIERCE, 101052347).\\
RC acknowledges support from the Canada Research Chairs Program and the Natural Sciences and Engineering Research Council of Canada (NSERC).\\
NBC acknowledges support from an NSERC Discovery Grant, a Canada Research Chair, and an Arthur B. McDonald Fellowship, and thanks the Trottier Space Institute for its financial support and dynamic intellectual environment.\\
LD acknowledges financial support from the Faculty of Science at the University of Waterloo, and support from the Waterloo Centre for Astrophysics.\\
DE acknowledges support from the SNSF for project 200021\_200726.\\
ED-M acknowledges the support by the Ram\'on y Cajal contract RyC2022-035854-I funded by MICIU/AEI/10.13039/501100011033 and by ESF+ and by the PIE project 20245AT026 funded by CSIC.\\
XDu acknowledges the support from the ERC under the European Union’s Horizon 2020 research and innovation programme (grant agreement SCORE No 851555) and from the SNSF under the grant SPECTRE (No 200021\_215200).\\
PL  acknowledges financial support from the Severo Ochoa grant CEX2021-001131-S funded by MCIN/AEI/10.13039/501100011033 and from the ERC (THIRSTEE, 101164189).\\
ICL acknowledges CNPq research fellowships (Grant No. 313103/2022-4).\\
CM acknowledges the funding from the SNSF under grant 200021\_204847 “PlanetsInTime”.\\
NN acknowledges financial support by Light Bridges S.L, Las Palmas de Gran Canaria, in cooperation with the Instituto de Astrof\'isica de Canarias, and the use of Indefeasible Computer Rights (ICR) being commissioned at the ASTRO POC project in the Island of Tenerife, Canary Islands (Spain).\\
JRM  acknowledges CNPq research fellowships (Grant No. 308928/2019-9).\\
ASM   acknowledges financial support from the Spanish Ministry of Science and Innovation (MICINN) under the 2024 Ram\'on y Cajal program MICINN RYC2024-050707-I.\\
GAW is supported by a Discovery Grant from the NSERC.\\
This paper includes data collected with the TESS mission, obtained from the MAST data archive at the Space Telescope Science Institute (STScI). Funding for US Institutions for the TESS mission is provided by the NASA Explorer Program. STScI is operated by the Association of Universities for Research in Astronomy, Inc., under NASA contract NAS 5–26555.\\
KAC acknowledges support from the TESS mission via subaward s3449 from MIT.\\
This research has made use of the Exoplanet Follow-up Observation Program (ExoFOP; DOI: 10.26134/ExoFOP5) website, which is operated by the California Institute of Technology, under contract with the National Aeronautics and Space Administration under the Exoplanet Exploration Program.\\
This work makes use of observations from the LCOGT network. Part of the LCOGT telescope time was granted by NOIRLab through the Mid-Scale Innovations Program (MSIP). MSIP is funded by US National Science Foundation.\\
We thank Xinran Liu for computing support.

\end{acknowledgements}
\end{appendix}
\end{document}